\documentclass[aps,onecolumn,preprint,superscriptaddress,nofootinbib,floats]{revtex4}
\usepackage{amsmath,amssymb,color,mathrsfs, graphicx,verbatim,epsfig, bbm, wasysym, axodraw}
\usepackage[hyperfootnotes=false]{hyperref}
\usepackage{slashed}
\allowdisplaybreaks

\def\lsim{\mathrel{\rlap{\lower4pt\hbox{\hskip1pt$\sim$}}
    \raise1pt\hbox{$<$}}}
\def\gsim{\mathrel{\rlap{\lower4pt\hbox{\hskip1pt$\sim$}}
    \raise1pt\hbox{$>$}}}

\newcommand{\be}{\begin{eqnarray}}
\newcommand{\ee}{\end{eqnarray}}

\def\addresses#1#2{\hbox to \hsize{\@tablebox{#1}\hfil\@tablebox{#2}}}
\def\@tablebox#1{\vtop{\hsize=5in \begin{flushleft} #1 \end{flushleft}}}

\def\beq{\begin{equation}}
\def\eeq{\end{equation}}
\def\bit{\begin{itemize}}
\def\eit{\end{itemize}}
\def\beqa{\begin{eqnarray}}
\def\eeqa{\end{eqnarray}}

\def\met{$\displaystyle{\not}E_T$}
\def\vecmet{$\vec{\displaystyle{\not}E}_T$}

\def\mttwo{$M_{T2}$}
\def\PYTHIA{{\tt PYTHIA}}

\def\MadGraph{{\tt MadGraph}}
\def\MadGraph5{{\tt MadGraph5}}

\def\FastJet{{\tt FastJet}}
\def\abar{\bar a}
\def\bbar{\bar b}
\def\mubar{\bar\mu}
\def\ibar{\bar i}
\def\betasq{\beta^2}

\def\cossqTh{\cos^2\Theta}

\def\sinsqTh{\sin^2\Theta}

\newcommand{\bt}{\bar t}

\begin{document}

\baselineskip 0.6cm

\begin{titlepage}

\thispagestyle{empty}

\begin{flushright}
\end{flushright}

\begin{center}

\vskip 2cm

{\Large \bf A New Twist on Top Quark Spin Correlations}

\vskip 1.0cm
{\large  Matthew Baumgart$^{1,2}$ and Brock Tweedie$^3$}
\vskip 0.4cm
{\it $^1$ Department of Physics, Carnegie Mellon University, Pittsburgh, PA 15213} \\
{\it $^2$ Department of Physics and Astronomy, Johns Hopkins University, Baltimore, MD 21218} \\
{\it $^3$ Physics Department, Boston University, Boston, MA 02215} \\
\vskip 1.2cm

\end{center}

\noindent Top-antitop pairs produced at hadron colliders are largely unpolarized, but their spins are highly correlated.
The structure of these correlations varies significantly over top production phase space, allowing very detailed
tests of the Standard Model.  Here, we explore top quark spin correlation measurement from a general perspective, highlighting the 
role of azimuthal decay angles.  By taking differences and sums of these angles about the top-antitop
production axis, the presence of spin correlations can be seen as sinusoidal modulations
resulting from the interference of different helicity channels.  At the LHC, these modulations
exhibit nontrivial evolution from near-threshold production into the boosted regime, where they become sensitive to almost the entire
QCD correlation effect for centrally produced tops.  We demonstrate that this form of spin correlation measurement 
is very robust under full kinematic reconstruction, and should already be observable with high 
significance using the current LHC data set.  We also illustrate some novel ways that new physics can alter the azimuthal distributions.
In particular, we estimate the power of our proposed measurements in probing for anomalous color-dipole operators,
as well as for broad resonances with parity-violating couplings.  Using these methods, the 2012 run of the LHC may be capable 
of setting simultaneous limits on the top quark's anomalous chromomagnetic and chromoelectric dipole moments at the level of
 $3\times 10^{-18} \, {\rm cm}$ ($0.03/m_t$).

\end{titlepage}

\setcounter{page}{1}

\section{Introduction}
\label{sec:intro}

It has long been realized that top quarks should decay before they have a chance to hadronize.  As a consequence, spin effects
present in their production should be passed on to their decay products largely unobscured by the depolarizing effects of soft QCD~\cite{Barger:1988jj,Kane:1991bg}.
Several analyses at the Tevatron and LHC have been dedicated to establishing that this picture is correct by measuring the 
correlations in the spins of top-antitop pairs.  Only recently, the presence of spin correlations has been established at the 3$\sigma$ level by D0 by using powerful matrix element techniques~\cite{Abazov:2011gi}, and at the 5$\sigma$ level by ATLAS by measuring the lab-frame $\Delta\phi(l^+,l^-)$ about the beam axis in dileptonic events~\cite{ATLAS:2012ao}.

Besides verifying our basic intuition about the smallness of soft QCD effects in top quark production, it is also hoped that a detailed understanding of top quark spin correlations will provide a unique probe for new physics~\cite{Bernreuther:1993hq,Beneke:2000hk,Frederix:2007gi,Arai:2007ts,Degrande:2010kt,Cao:2010nw,Baumgart:2011wk,Barger:2011pu,Krohn:2011tw,Bai:2011uk,Han:2012fw,Fajfer:2012si}.  This becomes a particularly pressing endeavor in light of the apparent anomalies in top quark production that are currently being reported by the Tevatron experiments~\cite{Aaltonen:2012it,CDF:AFB2011dilep,Abazov:2011rq,Abazov:2012bfa}.  Nonetheless, measurement of the very rich structure of top quark spin correlations in perturbative QCD, and their highly nontrivial evolution over top production phase space, have not been considered in complete detail.  Here, we will give a more general treatment of the correlations, highlighting some effects which are missed or incompletely captured when using common observables.  We will also consider some novel ways that new physics can imprint itself on the correlations and propose a well-defined set of new measurements that can be carried out in either dileptonic or $l$+jets channels.

Measurement of the spin correlations relies on studying the decay products of the two top quarks, and there are several well-known approaches.  The matrix element method mentioned above makes use of the full information present in the six-body final state~\cite{Melnikov:2011ai}.  While this is formally the most powerful method available, it is mainly useful for addressing the binary question of whether the expected correlation is present or absent.  A classic and more physically transparent method picks a single particle from each decay -- usually a lepton, but it can also be a $b$-quark or light quark -- and looks for correlations in their polar angles defined in their parent tops' rest frames.  This method requires specifying axes from which to measure the polar angles, and there are several well-motivated choices, on which we elaborate below.  It is the standard method for most spin correlation studies, and can be optimized to pick up the Standard Model correlation at the Tevatron or LHC~\cite{Mahlon:1995zn,Uwer:2004vp,Mahlon:2010gw}.  Other options include measuring the three-dimensional opening angle between two decay products (after boosting to a common frame)~\cite{Bernreuther:1997gs}, and, as definitively demonstrated by ATLAS, measuring the azimuthal angle between leptons around the barrel of the detector~\cite{Barger:1988jj,Mahlon:2010gw,ATLAS:2012ao}.  The last strategy notably bypasses detailed reconstruction of the top production and decay kinematics.

Given these many options, and the successful measurements at the Tevatron and LHC, what remains to be measured?  The complete spin correlation is encoded in a 3$\times$3 matrix that depends on the top pair production mechanism ($q\bar q \to t \bar t$ or $gg\to t\bar t$), the partonic center-of-mass energy, and the production angle.  Assuming separate P and C conservation, the matrix is parametrized by four numbers, which represent various combinations of production helicity amplitudes.  If new C-~or P-violating physics is at play in top quark production, even more degrees of freedom can open up.  Of course, no measurement of a single variable is adequate to pin down all of these matrix entries.  Similarly, no measurement that is inclusive over top production phase space can fully reveal how the correlation matrix changes in different kinematic regions.  Nonetheless, most of this information is readily measurable.

The situation is especially interesting at the LHC, given the enormous top pair data set over a broad bandwidth of energies, as well the fact that the correlations should change significantly as we scan from $p_T < m_t$ to $p_T > m_t$~\cite{Mahlon:2010gw}.  At low $p_T$, top pairs are predominantly produced in a spin $s$-wave from $gg$ annihilation, and their spins are therefore totally anti-correlated along any axis.  At high $p_T$, production becomes chiral, and the spins lie parallel to each other (and are hence totally {\it correlated}) along the $t\bar t$ production axis.  It is common to describe this switchover in correlations purely in the language of classical spin ensembles, once an appropriate quantization axis has been defined for every point in production phase space.  However, the interference between different spin channels almost always leads to sizeable contributions to the correlation matrix, making the full evolution much more complicated.  Tracking this evolution in detail would provide us with many new opportunities to test for unexpected behavior in top production.

In order to more comprehensively characterize the spin correlations, we here explore the utility of measuring various combinations of {\it all} of the tops' decay angles.  Since much work has already been done using polar decay angles in various physically-motivated bases, the main question that we address is what can be learned by systematically including the corresponding azimuthal angles.  Though often overlooked, these by themselves probe a large portion of the correlation matrix in a very simple way.  By taking differences and sums of the azimuthal angles, we obtain distributions that are flat in the absence of spin correlations and are sinusoidally modulating in their presence.  Such effects have already been suggested for finding and categorizing $t\bar t$ resonances~\cite{Baumgart:2011wk,Barger:2011pu}, and the threshold azimuthal-difference correlation around the beams is captured by the LHC lab-frame $\Delta\phi(l^+,l^-)$ measurements~\cite{ATLAS:2012ao,CMScorr}.  Ref.~\cite{Barger:2011pu} also discussed the azimuthal-difference modulation about the production axis for $gg\to t\bar t$ in QCD, integrated over production angles.  We will generalize this result in several ways, in particular demonstrating that the azimuthal sum contains complementary information.  We will also show how both of these combinations of angles can be measured at arbitrary top production angles and velocities using complete event reconstruction.  Remarkably, it is possible to do so without inducing severe distortions in the distributions, even given the biases induced by acceptance cuts and the reconstruction difficulties associated with jets and neutrinos.

Combining these pure azimuthal correlation measurements with a pure polar correlation measurement already gives us access to most of the correlation matrix, and almost fully captures the switchover that occurs between threshold production and high-$p_T$ production at the LHC.  Much of this switchover can be seen in the azimuthal variables alone: at threshold tops exhibit mainly an azimuthal-difference modulation, while at high $p_T$ they exhibit mainly an azimuthal-sum modulation.  For a broad range of phase space at high $p_T$, the latter is in fact the dominant manifestation of the entire QCD spin correlation effect.  For completeness, we also discuss how to measure the remaining entries of the correlation matrix, which lead to mixed polar-azimuthal correlations.  These are small in QCD across the entire top production phase space, and might therefore be an interesting place to look for deviations.

New physics can affect the correlation matrix in a myriad of ways.  We will concentrate on two simple examples whose dominant effects are felt in the azimuthal correlations.  The first is a color-dipole operator, which induces an azimuthal-difference correlation whose phase directly reflects the operator's CP-violating phase.  This perspective offers the interesting possibility of probing both the chromomagnetic and chromoelectric dipole strengths with a single measurement, potentially at the level of $3\times 10^{-18} \, {\rm cm}$ ($0.03/m_t$) at 2$\sigma$ with the current run of the LHC.  The second example is a broad spin-one color-octet resonance with parity-violating couplings.  This resonance would be very difficult to discover as a peak in the $t\bar t$ mass spectrum.  However, the effects of parity violation would be evident in the azimuthal-sum spin correlations, with a strength comparable or greater than that exhibited in more traditional observables sensitive to net polarization.  The effect on the azimuthal-sum distribution represents a novel form of parity-violation in top production.

Our paper outline is as follows.  In the next section, we review the basics of top quark spin correlations.  Subsequently, in Section~\ref{sec:QCD}, we describe their expected pattern in Standard Model QCD production at leading order, and what is captured by various observables.  In Section~\ref{sec:NP}, we consider modifications to the correlations that can be induced by new physics.  In Section~\ref{sec:measurement}, we outline a possible measurement of azimuthal correlations at the LHC.  We conclude in Section~\ref{sec:conclusions}.  We also include two appendices, which contain supplementary formulas and details of our simulations.

\section{Formalism}
\label{sec:formalism}

We start with a fairly general discussion where we introduce the formalism of spin correlations, a few standard ways to measure aspects of the them using specific decay angle correlations, and our own suggestions for how to more comprehensively extract their properties and to quantify their total strength.  In subsequent sections, we show how these different manifestations of the spin correlations behave in QCD and in new physics.  We do not address detailed reconstruction issues in this section, reserving them for Section~\ref{sec:measurement}.

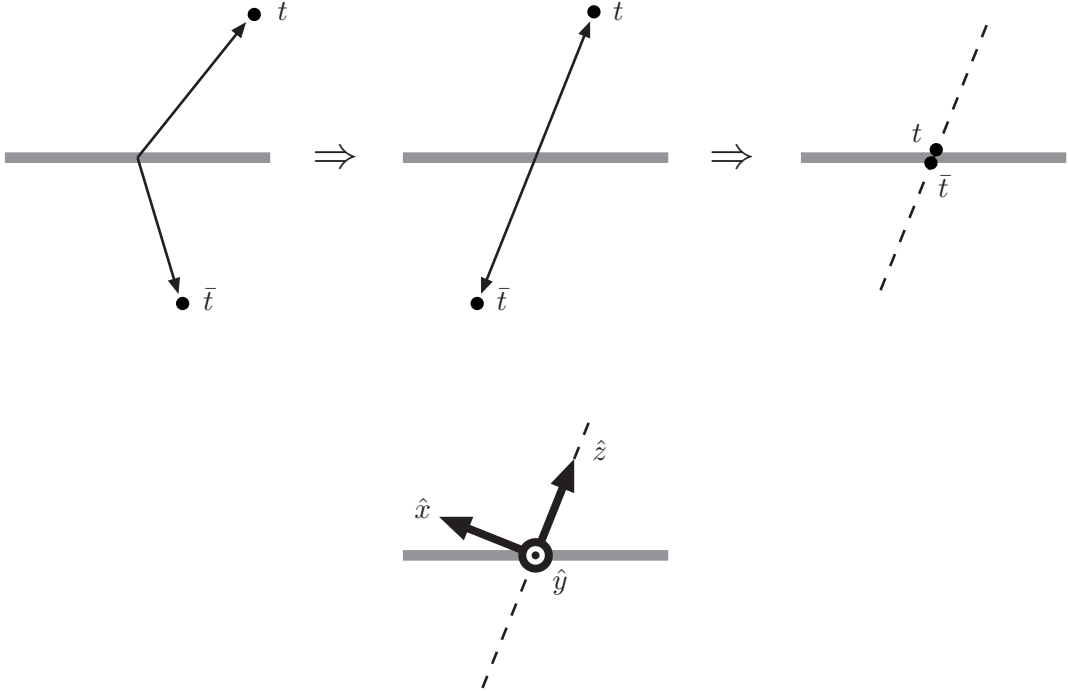
\begin{figure}[tp]
\begin{center} \begin{picture}(400,300)(0,0)
\SetColor{Gray} \SetWidth{4}  
\Line(  0,200)(100,200)
\Line(150,200)(250,200)
\Line(300,200)(400,200)
\Line(150,50)(250,50)
\SetColor{Black} \SetWidth{1}
\LongArrow( 50,200)( 90,250)   \GCirc( 94,254){2}{0} \Text(105,256)[]{$t$}
\LongArrow( 50,200)( 65,150)   \GCirc( 67,145){2}{0} \Text( 77,147)[]{$\bar t$}
\LongArrow(200,200)(220,250)   \GCirc(222,255){2}{0} \Text(232,256)[]{$t$}
\LongArrow(200,200)(180,150)   \GCirc(178,145){2}{0} \Text(188,147)[]{$\bar t$}
\DashLine(350,200)(370,250){5} \GCirc(351,203){2}{0} \Text(345,210)[]{$t$}
\DashLine(350,200)(330,150){5} \GCirc(349,198){2}{0} \Text(355,190)[]{$\bar t$}
\DashLine(200,50)(220,100){5}
\DashLine(200,50)(180,  0){5}
\Text(125,200)[]{\Large $\Rightarrow$}
\Text(275,200)[]{\Large $\Rightarrow$}
\SetWidth{3}
\LongArrow(200,50)(214,85) \Text(225,90)[]{$\hat z$}
\LongArrow(200,50)(165,64) \Text(158,68)[]{$\hat x$}
\BCirc(200,50){5} \Vertex(200,50){1.5} \Text(210,40)[]{$\hat y$}
\end{picture} \end{center}
\caption{\it Construction of a common coordinate system for the top and antitop.  The thick gray line is the beamline.  Starting from the lab frame, the $t\bar t$ CM system is actively boosted to rest, and then the individual tops are actively boosted to rest.  The decay products of the tops are measured in this frame.  The overlayed coordinate axes on the bottom figure correspond to our helicity basis.} 
\label{fig:coords}
\end{figure}

Throughout this section and the rest of the paper, we define top and antitop decay angles in a common reference frame constructed as follows.  We start in the lab frame of the colliding hadron beams, and actively boost the $t\bar t$ system to rest without rotation.  We define our $z$-axis along the $t\bar t$ production axis, pointing in the direction of the charge +2/3 top.  We define the $x$-axis within the production plane, such that it lies on the same side of the beamline as the $z$-axis.  The $y$-axis then points out of the production plane, forming a right-handed coordinate system.  We subsequently boost the tops to rest without rotation, carrying along their decay products with them.  The system is illustrated in Fig.~\ref{fig:coords}.  As is, this construction furnishes a specific realization of {\it helicity basis}, since the $z$-axis points along the top quark's CM-frame momentum vector.  (In our treatment, the top and antitop share this as a common $z$-axis, rather than constructing a separate $\bar z$-axis that points along the antitop's momentum vector.)  We can also consider other bases formed by rotating this system within the production plane, i.e. about $\hat y$.  A straightforward choice is {\it beam basis}, which orients the $z$-axis along one of the beams.  Another useful choice, to which we will return, is the {\it off-diagonal} basis of~\cite{Mahlon:1995zn}.

Since the top quark is a narrow particle, we can factorize the complete top pair production and decay process.  For a given partonic production process and a given point in production/decay phase space, the squared matrix element for the $2\to2\to6$ process (after color- and spin-summing) can be written as
\beq
 {\textstyle \Big(\, \frac{1}{3^2 \; {\rm or} \; 8^2}    \underset{\rm colors}{\sum}  \,\Big) \, \Big(\, \frac{1}{2^2} \underset{\rm spins}{\sum} \,\Big) } \,
\big|\mathcal{M}(\, q\bar q / gg \,\to\, t \bar t \,\to\, (f_1 \bar f_1' b) \, (\bar f_2 f_2' \bar b) \, )\big|^2  \; = \; \Gamma_{ab} \, \rho_{ab,\abar\bbar} \, \bar\Gamma_{\abar\bbar} \, ,
\label{eq:termdef}
\eeq
where the $f$'s are light left-handed fermions (up/down quarks or neutrino/lepton), and $\rho$ and $\Gamma$ ($\bar\Gamma$) are the production and decay spin density matrices, indexed by the top and antitop spins.  We will generally use overbars in referring to indices and properties associated with the antitop.  Repeated indices are implicitly summed unless otherwise indicated.

The production spin density matrix is defined as
\beq
\rho_{ab,\abar\bbar} \,\equiv\,  {\textstyle \Big(\, \frac{1}{3^2 \; {\rm or} \; 8^2}    \underset{\rm colors}{\sum}  \,\Big) \, \Big(\, \frac{1}{2^2} \underset{\rm initial \; spins}{\sum} \,\Big) } \, \mathcal{M}( q\bar q / gg \,\to\,  t_a \bar t_{\abar}) \, \mathcal{M}( q\bar q / gg \,\to\, t_b \bar t_{\bbar})^*.
\eeq
The study of top-antitop spin correlations, and of individual top spins, is ultimately a study of the properties of this matrix.  It is purely a function of the partonic initial state and production kinematics, and is therefore sensitive to any novel physics that affects how the tops are produced.  The matrix is hermitian by construction ($\rho^*_{ab,\abar\bbar} = \rho_{ba,\bbar\abar}$) and can be expanded in a basis of Pauli matrices:
\be
\rho_{ab,\abar\bbar} & \,=\, & \frac14 M^{\mu\mubar}\,\sigma^{\mu}_{ab}\,\sigma^{\mubar}_{\abar\bbar} \nonumber \\
                     & \,=\, & \frac14 \left( M^{00}\,\delta_{ab}\,\delta_{\abar\bbar} + M^{i0}\,\sigma^{i}_{ab}\,\delta_{\abar\bbar} + 
                               M^{0\ibar}\,\delta_{ab}\,\sigma^{\ibar}_{\abar\bbar} + M^{i\ibar}\,\sigma^{i}_{ab}\,\sigma^{\ibar}_{\abar\bbar} \right)  \nonumber \\
                     & \,\equiv\, & \frac14 M^{00}\,\left( \delta_{ab}\,\delta_{\abar\bbar} + P^{i}\,\sigma^{i}_{ab}\,\delta_{\abar\bbar} + 
                                    \bar P^{\ibar}\,\delta_{ab}\,\sigma^{\ibar}_{\abar\bbar} + C^{i\ibar}\,\sigma^{i}_{ab}\,\sigma^{\ibar}_{\abar\bbar} \right) \, ,
\label{eq:proddecomp}                                    
\ee
where $M$ is a real 4$\times$4 matrix, and we have appropriated the usual $\mu$ notation for spacetime indices, though we use a trivial metric here.  $M^{00}$ parametrizes the overall production rate, while $P^{i} \equiv M^{i0}/M^{00}$ and $\bar P^{\ibar} \equiv M^{0\bar i}/M^{00}$ characterize the net degree of top/antitop polarization, and $C^{i\ibar} \equiv M^{i\bar i}/M^{00}$ characterizes their correlations.  More explicitly, $P^{i} = \langle 2 S^i \rangle$, $\bar P^{\ibar} = \langle 2\bar S^{\ibar} \rangle$, and $C^{i\ibar} = \langle 4 S^i \bar S^{\ibar} \rangle$, where $S^i$ and $\bar S^{\ibar}$ are the top/antitop spin operators.  Given an unpolarized initial state and our choice of coordinate system, the parity and charge-conjugation symmetries characteristic of pure QCD production would individually imply
\be
{\rm P} & \,\Rightarrow\, & M^{\mu\mubar} = (-1)^{\mu+\mubar} \, M^{\mu\mubar} \nonumber \\
{\rm C} & \,\Rightarrow\, & M^{\mubar\mu} = (-1)^{\mu+\mubar} \, M^{\mu\mubar} 
\ee
(indices not summed).  These identities also hold for any other coordinates obtained by rotating within the production plane, i.e. about $\hat y$.  Parity would force much of the matrix to vanish, leaving over the diagonal spin correlations and mixed $xz$ spin correlations, as well as possible net $y$-polarizations for the tops.\footnote{Top pairs produced in QCD processes from an unpolarized initial state are in fact polarized in the $y$-direction, transverse to the production plane.  However, the effect comes in at loop-level due to the need for a complex phase between amplitudes, and is predicted to be percent-level at both the Tevatron and LHC~\cite{Kane:1991bg,Bernreuther:1995cx}.  New physics processes can also induce complex phases at tree level, significantly enhancing this polarization effect, even in perfectly P- and C-conserving theories.  We reserve exploration of this for future work~\cite{Baumgart:future}.}  By itself, charge-conjugation would require either symmetry or antisymmetry between the different correlation and polarization components, but in conjunction with parity just forces the remaining nonzero entries to be symmetric:  $P^2 = \bar P^2$ and $C^{13} = C^{31}$.  The combination CP would only require the entire $M^{\mu\mubar}$ matrix to be symmetric, independent of whether parity and charge-conjugation are individually good symmetries, and with no further constraints on the entries.  (A more extensive discussion of symmetry properties can be found in~\cite{Bernreuther:1993hq}.)

The decay spin density matrix is defined as
\beq
  \Gamma_{ab} \,\equiv\, \mathcal{M}(t_a \to f\bar f' b) \, \mathcal{M}(t_b \to f\bar f' b)^*,
\eeq
and the antitop has an exactly analogous expression.  These matrices are purely functions of the top decay kinematics.  In its fully general form, $\Gamma$ depends in a complicated way on the four-momenta of each of the top's three decay products, and may itself be affected by new physics.  However, the situation vastly simplifies if we integrate out most of the decay phase space, leaving over only the unit vector $\hat\Omega$ of one particle ($f$, $\bar f'$, or $b$) in the top's rest frame:  $\Gamma \to \tilde\Gamma(\hat \Omega)$.  As an hermitian 2$\times$2 matrix, $\tilde\Gamma$ can also be expanded in Pauli matrices, and due to rotational invariance the dependence on the remaining particle direction is fixed up to an analyzing power $\kappa$:
\beq
\tilde\Gamma(\hat\Omega)_{ab} \,\propto\, \delta_{ab} + \kappa \, \hat\Omega \cdot \vec{\sigma}_{ab} \: .
\eeq
In the Standard Model, $\kappa_{l/d} = +1$, $\kappa_{\nu/u} = -0.3$, and  $\kappa_b = -0.4$~\cite{Brandenburg:2002xr}.\footnote{The maximal analyzing power of the lepton/down is due to the $V-A$ current structure of the top decay.  It is easy to see by Fierz-transforming the decay amplitude.}  From this perspective, any new physics effects in the top decay only show up as possible rescalings of the analyzing powers, and these effects are already becoming constrained~\cite{Aaltonen:2012rz,CDF:Wpol2012,Aad:2012ky,CMS:Wpol2012}.  Since we are interested here in understanding top production, not decay, we assume the Standard Model analyzing powers.  We can also choose a single particle using means other than its flavor.  In particular, for hadronic decays of the top, we can choose a random non-$b$ quark to get $\kappa = +0.35$, or we can pick the softer of the two non-$b$ quarks in the top's rest frame to get $\kappa = +0.5$.  Assuming approximate CP conservation in the decay, the equivalent analyzing powers for the antitop should the same size as those of the top, multiplied by a minus sign (i.e., $\bar\kappa_{l/d} = -1$, $\bar\kappa_{\nu/u} = +0.3$, and  $\bar\kappa_b = +0.4$).

Picking one decay particle from each side, and contracting through the spin indices in the production and decay density matrices as in Eq.~\ref{eq:termdef}, we can see the spin structure of the production matrix elements encoded in a correlated distribution of decay angles:
\beq
\frac{d^4\sigma}{d\Omega \, d\bar\Omega} \,\propto\, 1 + \kappa\,\vec{P}\cdot\hat\Omega + \bar\kappa\,\vec{\bar P}\cdot\hat{\bar\Omega} + \kappa\bar\kappa\:\hat\Omega\cdot C \cdot \hat{\bar\Omega} \: ,
\label{eq:fullcorr}
\eeq
with $\hat\Omega \equiv (\cos\phi\sin\theta,\sin\phi\sin\theta,\cos\theta)$ and analogously for $\hat{\bar\Omega}$.  For example, it is commonly observed that the polar decay angle distribution for a single side is 
\beq
\frac{d\sigma}{d\cos\theta} \,\propto\, 1 + \kappa \, P^3 \cos\theta \: ,
\label{eq:sap}
\eeq
and therefore tells us about the top's (or antitop's) net polarization in this direction.  To begin to gain sensitivity to the correlation matrix $C$, we can instead look at the double polar decay angle distribution,
\beq
\frac{d^2\sigma}{d\cos\theta \, d\cos\bar\theta} \,\propto\, 1 + \kappa \, P^3 \cos\theta + \bar\kappa \, \bar P^3 \cos\bar\theta + \kappa\bar\kappa \, C^{33} \cos\theta\cos\bar\theta \: ,
\label{eq:polar-polar}
\eeq
or we can more directly access $C^{33}$ via
\beq
\frac{d\sigma}{d(\cos\theta\cdot\cos\bar\theta)} \,\propto\, (1 + \kappa\bar\kappa \, C^{33}\cos\theta\cdot\cos\bar\theta) \, \log\left(\frac{1}{|\cos\theta\cdot\cos\bar\theta|}\right).
\label{eq:coscos}
\eeq
By restricting to a specific axis like this, we can recover a relatively intuitive picture of a certain population of ``spin-up'' and ``spin-down'' top quarks with classical average polarizations and correlation:  $P^3$, $\bar P^3$ and $C^{33}$.  But it is clear that these polar angle distributions actually only give us a very limited view of the production density matrix.  Gaining a more complete perspective in this manner would require us to measure an extended set of polar angle distributions, utilizing a full complement of different ``$z$-axes'' chosen independently for the $t$ and $\bar t$.  Often, it is simply suggested to choose a single basis which is expected to yield a large $C^{33}$, or a $C^{33}$ which is particularly sensitive to some specific new physics effect.

Another common option is to measure the three-dimensional opening angle between the two decay products, or equivalently to take the dot product between their directions.  We call the opening angle $\chi$ (sometimes known as ``$\phi$'' in the literature).  Its distribution is
\beq
\frac{d\sigma}{d\cos\chi} \,\propto\, 1 + \frac13 \kappa\bar\kappa\:{\rm tr}[C]\,\cos\chi \: .
\label{eq:chi}
\eeq
Measuring $\chi$ gives us a very compact method for measuring the strength of the correlation independently of any net polarization, though it is only sensitive to this one specific linear combination of matrix elements.  We will see below that ${\rm tr}[C]$ captures most of the correlation at low $p_T$ at the LHC, but otherwise tends to miss most or all of the correlation.  (Though for new physics effects that contribute strictly to the trace, such as an $s$-channel pseudoscalar exchange, this variable may be quite sensitive.)

A completely general analysis of the top spins might utilize a fit of the distribution over all four of the independent decay angles, possibly invoking parity and/or charge-conjugation symmetries to limit the number of free parameters.  This is probably the ideal approach in principle, but it may be overly complicated in practice, especially if a given analysis is mainly interested in some specific aspect top spin physics.  It is also beyond the scope of the present paper to understand how effective such a fit might be under realistic measurement conditions.  

Given the formulas for polar decay angles above, an obvious question to ask is whether we could learn anything from the azimuthal decay correlations, or azimuthal-polar correlations.  The precise statement of this question of course depends on the coordinate basis used to define ``azimuthal'' versus ``polar.''  Our default choice in subsequent sections will be to use the the off-diagonal basis of~\cite{Mahlon:1995zn} for the Tevatron and the helicity basis for the LHC.  The former is motivated by the spin structure particular to $q\bar q \to t\bar t$.  In lieu of such an obvious choice for the LHC, helicity basis becomes the most physical alternative.  A large fraction of tops at the LHC are produced with relativistic velocities, and helicity-basis azimuthal correlations have a direct physical interpretation as the interference between different chiral production amplitudes.\footnote{Close to threshold, it is more natural to take one of the beams as the $z$-axis, as is done for some of the LHC measurements~\cite{ATLAS:2012ao,CMScorr} (though they do not first boost the individual top systems to rest).  However, at the LHC, the question of what is the best basis for slow tops is nominally moot, since they are produced approximately in a spin $s$-wave.  The QCD correlations should therefore be nearly basis-independent.}  Moreover, azimuthal angles measured in helicity basis should asymptotically become the least susceptible to the distorting effects of acceptance cuts, since rotating the decay system of a fast moving top about its own axis should not have a large impact on its detection efficiency.

If we just proceed to integrate out $\theta$ and $\bar\theta$ in Eq.~\ref{eq:fullcorr}, we get a formula similar to Eq.~\ref{eq:polar-polar} but which specifically probes the $x$ and $y$ parts of the net polarizations and the correlation matrix.  We can further process this into more compact one-dimensional distributions to extract useful combinations of the correlation matrix elements\footnote{Incidentally, if we instead just subsequently integrated out $\bar\phi$ or $\phi$, we would obtain convenient formulas for extracting the net transverse polarizations $(P^1,P^2)$ or $(\bar P^1,\bar P^2)$, respectively.  E.g., $d\sigma/d\phi \propto 1 + (\pi/4)\kappa\,(P^1\cos\phi + P^2\sin\phi)$.}:
\be
\frac{d\sigma}{d(\phi-\bar\phi)} & \,\propto\, & 1 + \left(\frac{\pi}{4}\right)^2 \kappa\bar\kappa \, \left[ \left(\frac{C^{11}+C^{22}}{2}\right)\cos(\phi-\bar\phi) + \left(\frac{C^{21}-C^{12}}{2}\right)\sin(\phi-\bar\phi) \right]  \nonumber \\
\frac{d\sigma}{d(\phi+\bar\phi)} & \,\propto\, & 1 + \left(\frac{\pi}{4}\right)^2 \kappa\bar\kappa \, \left[ \left(\frac{C^{11}-C^{22}}{2}\right)\cos(\phi+\bar\phi) + \left(\frac{C^{21}+C^{12}}{2}\right)\sin(\phi+\bar\phi) \right].
\ee
There are several advantages to expressing the $xy$ correlations in this manner.  An immediate one is that the measurement of the entire $xy$ part of the spin correlation is reduced to the measurement of the amplitudes and phases of these two simple distributions.  They are flat in the absence of correlations, are modulating in the presence of correlations, and, as we will see for the LHC, are left fairly intact by basic acceptance cuts and event reconstruction.  They therefore offer a simple alternative to measuring these four matrix elements one-by-one with four different ``polar-polar'' distributions as in Eqs.~\ref{eq:polar-polar} and~\ref{eq:coscos}, or extracting the sum $C^{11}+C^{22}$ via Eq.~\ref{eq:chi} (relying on other measurements to independently determine $C^{33}$ so that we can subtract it off of the trace).  The symmetry structure of the $xy$ part of the correlation matrix is also made immediately manifest.  Any net phase in the $\phi-\bar\phi$ modulation signals CP-violation ($C^{21} \neq C^{12}$, the same type discussed in~\cite{Bernreuther:1993hq}), and any net phase in the $\phi+\bar\phi$ modulation signals C-~and P-violation ($C^{12} \neq -C^{21}$ and $C^{21,12} \neq 0$).  Since these distributions are specifically sensitive to the interference between different spin configurations in a given basis, rather than their relative probabilities, they offer a complementary physical picture to that obtained with polar angle correlations in the same basis. 

Combining an azimuthal sum/difference measurement with a polar-polar correlation measurement in the same coordinate basis already gives us five of the nine entries of $C$.  What about the other four?  To access these, we can think about possible polar-azimuthal correlations, which probe the $xz$ and $yz$ parts of the matrix.  For example, Eq.~\ref{eq:fullcorr} contains terms like $C^{13}\cos\phi\sin\theta\cos\bar\theta$.  Integrating out $\theta$ and $\bar\phi$ would leave over a $d^2\sigma/d\phi\,d\cos\bar\theta$ distribution that contains  $C^{13}\cos\phi\cos\bar\theta$, as well as other terms with the coefficients $C^{23}$, $P^{1,2}$, and $\bar P^3$.  To more directly extract the elements $C^{13}$ and $C^{23}$, we can employ a simple trick:  for $\cos\bar\theta < 0$, shift our definition of $\phi$ by $\pi$.  I.e., define $\phi' \equiv \phi$ ($\phi+\pi$) for $\cos\bar\theta > 0$ ($\cos\bar\theta < 0$).  If we subsequently integrate out $\cos\bar\theta$, we are left with another simple sinusoidal distribution:
\beq
\frac{d\sigma}{d\phi'}  \,\propto\,  1 + \frac{\pi}{8} \kappa\bar\kappa \, \left( C^{13}\cos\phi' + C^{23}\sin\phi' \right),
\label{eq:PolarAz}
\eeq
with a similar expression for $d\sigma/d\bar\phi'$ (sensitive to $C^{31}$ and $C^{32}$) when the analogous $\pi$-shift is applied.  Again, the presence of P-violation in top production would be immediately obvious as a phase offset ($C^{23,32} \neq 0$), though looking for C-~or CP-violation using these distributions would require a comparison of $d\sigma/d\phi'$ and $d\sigma/d\bar\phi'$.  It is also possible to pick up the four elements $C^{13}$, $C^{23}$, $C^{31}$, and $C^{32}$ via a series of four dedicated polar-polar correlation measurements.  However, the method that we have presented requires us to reconstruct only two distributions.

Finally, we point out a novel way to characterize the total strength of the correlation effect.  It extends approaches like those in~\cite{Mahlon:1995zn,Uwer:2004vp,Mahlon:2010gw}, which diagonalize the correlation matrix assuming pure QCD production and then pick off the polar-polar correlation using the axis with the largest eigenvalue.  While our approach can in principle be used as the basis of a top spin correlation measurement in its own right, our main interest here will be to use it to quantify how much of the total spin correlation is captured up by any given measurement.  This will serve as a useful tool in our comparisons of different approaches to measuring the QCD correlations in the next section.

The top and antitop decays can be said to be correlated, in the absence of net polarization, when the angular distribution for the decay on one side is nontrivial given a {\it fixed} decay configuration on the other side.  For example, consider the case of $t\bar t$ produced in a spin $s$-wave, so that $C = {\rm diag}(-1,-1,-1)$ and $\vec{P} = \vec{\bar P} = \vec{0}$.  Using the leptonic decay on each side, and picking the lepton/antilepton as the spin analyzers ($\kappa = +1$, $\bar\kappa = -1$), the total decay distribution is $d^4\sigma/d\Omega\, d\bar\Omega \propto 1 + \hat\Omega \cdot \hat{\bar\Omega}$.  If we fix the direction of the antitop's lepton, $\hat{\bar\Omega}$, then the top's lepton will be distributed in a manner azimuthally-symmetric about this direction, but with a maximally linearly-biased polar angle distribution.  (E.g., the two leptons cannot move in exactly opposite directions.)  Similarly, if we fix the direction of the top's lepton, $\hat\Omega$, then the antitop's lepton will be seen to have a maximally linearly-biased distribution of polar angles with respect to this direction.  In this case, the entire effect is encapsulated in the distribution of the 3D opening angle $\chi$, as in Eq.~\ref{eq:chi}.

A less trivial example is $C = {\rm diag}(+1,-1,+1)$, which in fact occurs in the relativistic, central production limit of both $q\bar q \to t\bar t$ and $gg \to t\bar t$ in QCD.  Given a fixed $\hat{\bar\Omega}$, the top's lepton is not distributed in a simple way around this direction, but instead about the direction  $\kappa\bar\kappa\, C \cdot \hat{\bar\Omega}$.  The magnitude of the correlation is again maximal, but this fact is missed by all of the standard approaches to measuring it using a single variable.

To completely characterize the strength of general correlations, we suggest that instead of referring to a fixed coordinate system or measuring a simple opening angle, we measure the angle between one decay product's direction and an appropriately transformed version of the other decay product's direction.  Assuming a given correlation matrix $C$, define
\be
\cos\chi'     & \equiv & \hat\Omega       \cdot 
                         \frac{\kappa\bar\kappa\, C \cdot\hat{\bar\Omega}}{|\kappa\bar\kappa\, C \cdot\hat{\bar\Omega}|} \nonumber \\
\cos\bar\chi' & \equiv & \hat{\bar\Omega} \cdot 
                         \frac{\kappa\bar\kappa\, C^T \cdot\hat{\Omega}}{|\kappa\bar\kappa\, C^T \cdot\hat{\Omega}|}.
\ee
Note that these are generally {\it different} angles event-by-event.  Nonetheless, if we integrate out $\hat{\bar\Omega}$ or $\hat\Omega$, respectively, to get the cumulative effect of the correlation in $\cos\chi'$ or $\cos\bar\chi'$, we get the same linear coefficient:
\be
\frac{d\sigma}{d\cos\chi'}     & \propto & 1 + |\kappa\bar\kappa | \: {\cal C} \,\cos\chi' \nonumber \\
\frac{d\sigma}{d\cos\bar\chi'} & \propto & 1 + |\kappa\bar\kappa | \: {\cal C} \,\cos\bar\chi'
\ee
where
\beq
{\cal C} \,=\, \int \,\frac{d\Omega}{4\pi} \; \sqrt{\hat\Omega\cdot C^T C \cdot\hat\Omega}.
\label{eq:calC}
\eeq
Defined this way, the correlation strength ${\cal C}$ is purely a function of the eigenvalues of the matrix $C^T C$ (or equivalently $CC^T$), and can vary between zero and one.\footnote{In the case of zero expected correlation, the procedure becomes ill-defined.  But, as we will see, the correlation never vanishes in leading-order QCD, except in the extremal case of $gg \to t\bar t$ at zero-angle and infinite-boost.}  Note that the net single-side polarization effects from possible nonzero $\vec{P},\vec{\bar P}$ integrate out to zero, just as they do for the original $\chi$.  We will shortly see how this method performs in the context of QCD production.\footnote{We also point out that a similar construction can be obtained working strictly with the $xy$ block of $C$ and considering projections of decay products into this plane.  Instead of individually measuring $\phi\pm\bar\phi$ to characterize the strength of correlation within this part of the matrix, we can measure the difference in azimuthal angles between one untransformed and one transformed decay product.  This total azimuthal correlation strength can be expressed in terms of complete elliptic integrals of the second kind.  Practically, though, we find that one of the simple combinations $\phi\pm\bar\phi$ nearly saturates the correlation in QCD over most of the production phase space.}

\section{Spin Correlations in QCD}
\label{sec:QCD}

The leading-order partonic subprocesses $q\bar q \to t\bar t$ and $gg \to t\bar t$ exhibit very different patterns of spin correlations, which we now explore in detail.

The case of $q\bar q$ annihilation has been considered straightforward for a long time.  In this process, there is a natural coordinate basis for each point in production phase space for which the top and antitop spins have perfectly correlated $S^3$.  For near-threshold production, the $z$-axis is aligned with the beams (either choice works since the beams are usually unpolarized).  For relativistic production, the $z$-axis is that of helicity basis.  Intermediate boosts interpolate between the two choices via the off-diagonal basis construction~\cite{Mahlon:1995zn}.  The usual tactic is to go into off-diagonal basis and study the polar-polar correlation as in Eq.~\ref{eq:polar-polar}, either by reconstructing the distribution of $d^2\sigma / d\cos\theta\, d\cos\bar\theta$, or by reducing this down to the one-dimensional distribution $d\sigma/d(\cos\theta\cdot\cos\bar\theta)$.

To understand how this performs relative to the methods discussed above, we need to have a way to compare the strengths of the correlation's effects on different variables.  To do this, we use asymmetries of distributions obtained with the $\kappa$'s stripped off, adding events in regions where the correlated rate is larger than the uncorrelated rate, and subtracting events in regions where the correlated rate is smaller than the uncorrelated rate.  For example, the asymmetry in $d\sigma/d(\cos\theta\cdot\cos\bar\theta)$ about $\cos\theta\cdot\cos\bar\theta=0$ is directly proportional to $C^{33}$.  (It is exactly the same asymmetry that we would obtain by dividing up the 2D space of $(\cos\theta,\cos\bar\theta)$ into positive and negative quadrants.)  Similarly, the distributions $d\sigma / d(\phi-\bar\phi)$ and $d\sigma / d(\phi+\bar\phi)$, which modulate as cosines, have asymmetries between regions with $|\phi\pm\bar\phi| < \pi/2$ and $\pi/2 < |\phi\pm\bar\phi| < \pi$.  These asymmetries are proportional to $C^{11}+C^{22}$ and $C^{11}-C^{22}$.\footnote{Defined in this way, they are {\it insensitive} to possible nonvanishing $C^{21}\pm C^{12}$.  However, these combinations can be accessed either by a full sinusoidal fit or by forming asymmetries between regions of positive and negative $\phi\pm\bar\phi$.}  The distributions of $\cos\chi$ and $\cos\chi'$ ($\cos\bar\chi'$) are linearly-biased, and their asymmetries about zero are respectively proportional to tr$[C]/3$ and the total correlation $\cal C$ (Eq.~\ref{eq:calC}).  The asymmetry in $\cos\chi'$ ($\cos\bar\chi'$) is also what we would obtain by forming the asymmetry over the complete four-dimensional phase space for the two decay directions.

\begin{figure}[tp]
\begin{center}
\epsfxsize=0.44\textwidth\epsfbox{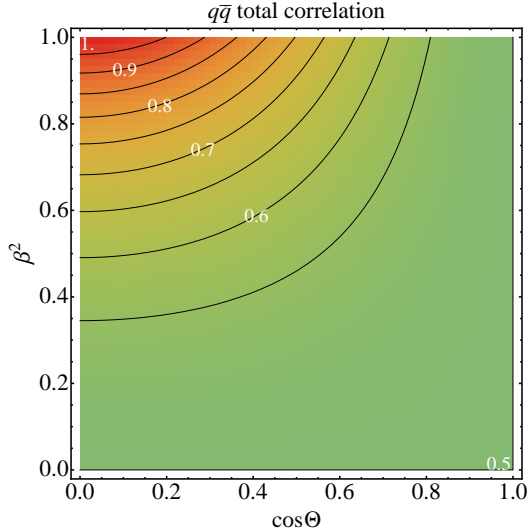}
\caption{\it Total LO spin correlation strength in $q\bar q \to t\bar t$.  Plotted versus top production angle and squared-velocity in the partonic CM frame.}
\label{fig:qqTotalCorr}
\end{center}
\end{figure}

\begin{figure}[tp]
\begin{center}
\epsfxsize=0.44\textwidth\epsfbox{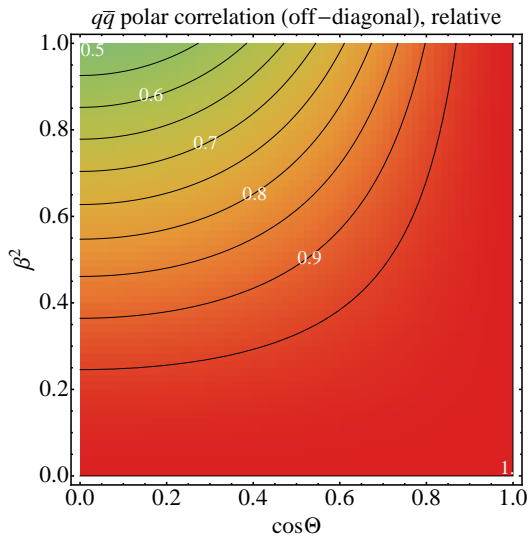}
\caption{\it LO correlation strength in off-diagonal-basis polar angles in $q\bar q \to t\bar t$, relative to total strength.  Plotted versus top production angle and squared-velocity in the partonic CM frame.} 
\label{fig:qqPolarRel}
\end{center}
\end{figure}

\begin{figure}[tp]
\begin{center}
\epsfxsize=0.44\textwidth\epsfbox{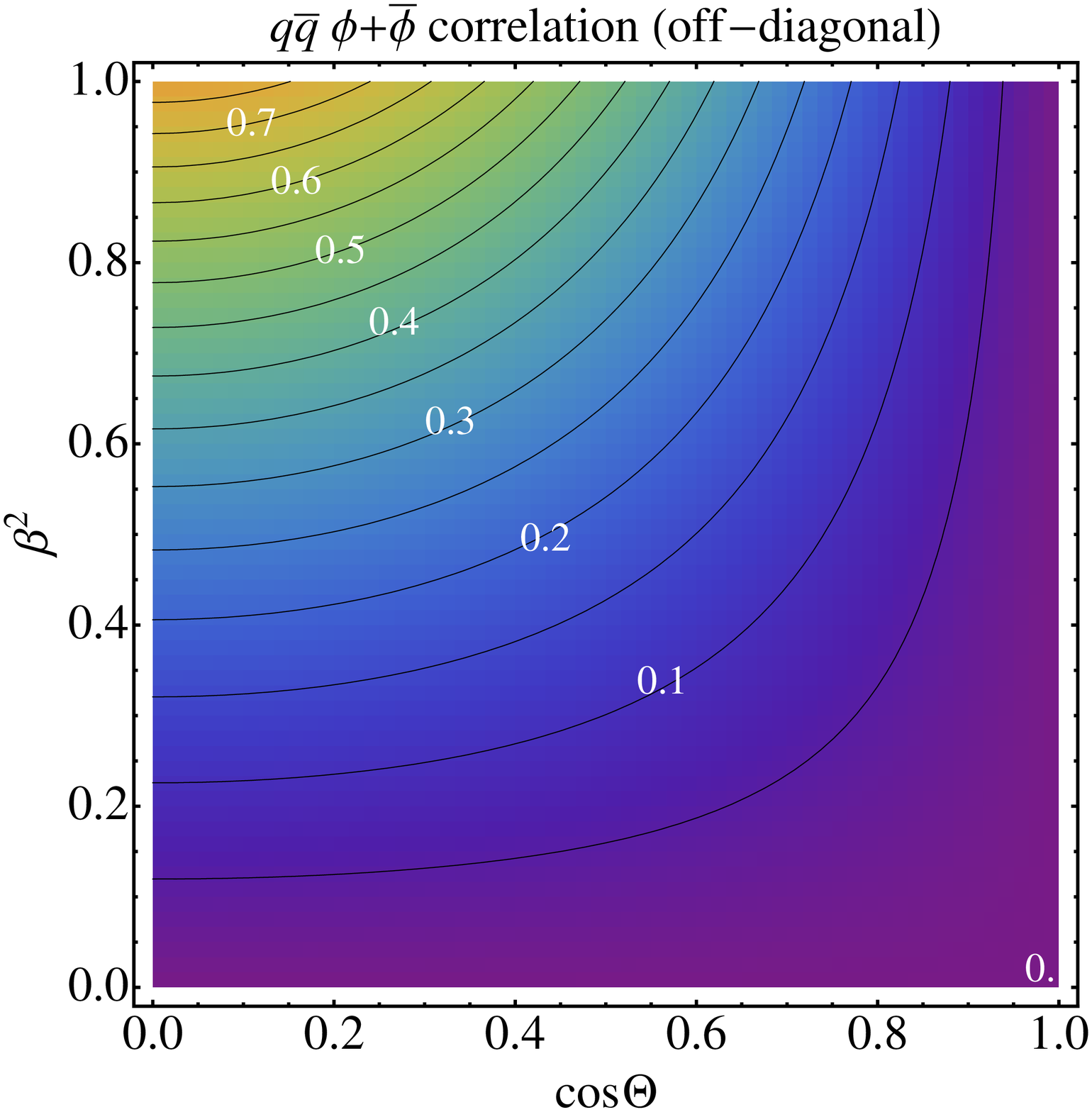}
\epsfxsize=0.44\textwidth\epsfbox{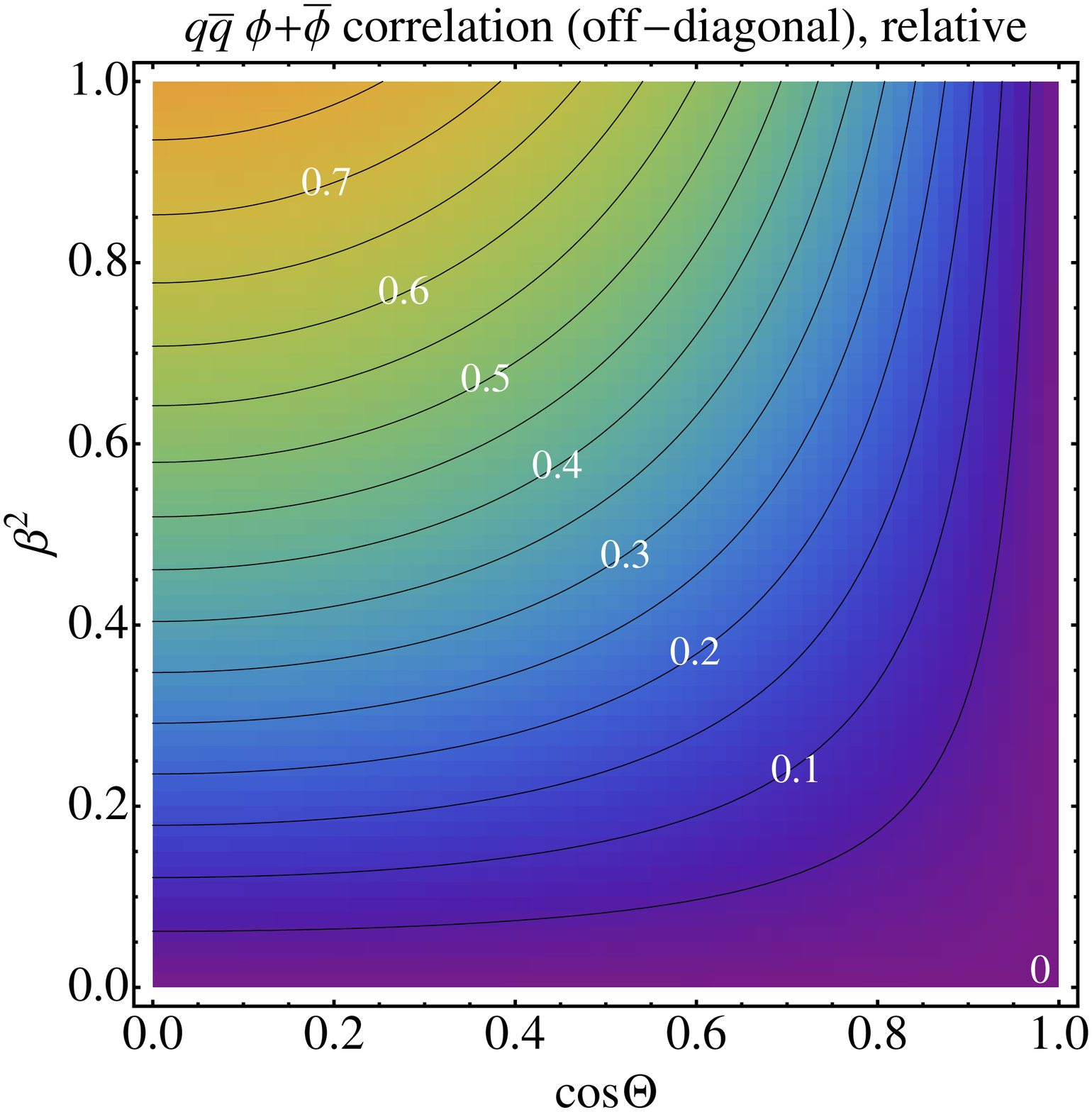}
\caption{\it  LO correlation strength in off-diagonal-basis $\phi+\bar\phi$ in $q\bar q \to t\bar t$.  Absolute (left) and relative to total strength (right).  Plotted versus top production angle and squared-velocity in the partonic CM frame.} 
\label{fig:qqAzSum}
\end{center}
\end{figure}

In Fig.~\ref{fig:qqTotalCorr} we show the total correlation $\cal C$ for $q\bar q \to t\bar t$ as a function production angle ($\Theta$) and squared-velocity ($\beta^2$) in the partonic CM frame.  The strongest possible correlation, ${\cal C} = 1$, leads to a 50\% asymmetry in $\cos\chi'$ ($\cos\bar\chi'$).  To keep a common normalization, such that ``1'' corresponds to maximal correlation, we multiply the asymmetries of all of our variables by 2.  According to this measure, the correlation measured by $\cos\theta\cdot\cos\bar\theta$ in off-diagonal basis is $0.5$ (i.e., 25\% asymmetry) over the entire production phase space.  We do not need to plot this number, but we illustrate how it compares relative to the total correlation in Fig.~\ref{fig:qqPolarRel}.  In Fig.~\ref{fig:qqAzSum}, we show the $\phi+\bar\phi$ correlation strength, also in off-diagonal basis, both absolute and relative to $\cal C$.  In this basis, the $\phi-\bar\phi$ and $xz$ correlations identically vanish at leading-order.  (The $\cos\chi$ correlation strength is a flat $1/3$, strictly weaker than $\cos\theta\cdot\cos\bar\theta$.)

Figure~\ref{fig:qqTotalCorr} indicates that the correlation in $q\bar q \to t\bar t$ production is half-maximal at low velocities and/or forward angles, but becomes maximal at relativistic velocities and central angles.  The corresponding correlation matrices are $C = {\rm diag}(0,0,+1)$ and $C = {\rm diag}(+1,-1,+1)$.  (Complete formulas can be found in Appendix~\ref{sec:formulas}.)  Consequently, the $\cos\theta\cdot\cos\bar\theta$ asymmetry does not always reflect the total strength of the correlation (Fig.~\ref{fig:qqPolarRel}), and the ``missing'' part of the correlation yields the $\phi+\bar\phi$ modulation (Fig.~\ref{fig:qqAzSum}).  This additional correlation comes from the fact that the two possible spin configurations (up-up and down-down) are not generally produced incoherently.  Their interference only vanishes when the tops are at rest or moving along the beamline, in which case their spins are directly inherited from the spins of the annihilating quarks.  At the Tevatron, where $q\bar q$ annihilation is dominant, tops are typically produced near threshold, and the spin interference contribution to the correlation is roughly at the 10\% level.  A dedicated high-$p_T$ analysis might nonetheless reveal the emergence of the $\phi+\bar\phi$ correlation, statistics permitting.

The analogous story for $gg \to t\bar t$, which is of primary concern at the LHC, is more complicated.  As pointed out by~\cite{Mahlon:2010gw}, we should really think of this as two separate processes:  the annihilation of opposite-spin and same-spin gluons (or, equivalently, same-helicity and opposite-helicity).  For an unpolarized initial state, the former dominates for $p_T < m_t$, and the latter dominates for $p_T > m_t$.  The former always produces opposite-spin tops in helicity basis, and the latter always produces same-spin tops in off-diagonal basis.  When the two processes are superimposed, the resulting pattern of spin correlations is highly nontrivial.  Ref.~\cite{Mahlon:2010gw} considered the polar angle correlations of $gg \to t\bar t$, establishing that there is generally no basis in which this correlation saturates (i.e., $C^{33} = \pm 1$).  However, they show which basis maximizes the polar correlation.  Practically, this corresponds to finding a basis that diagonalizes the matrix $C$, and then choosing as the ``$z$-axis'' that has the eigenvalue of the largest magnitude (as suggested by~\cite{Uwer:2004vp}).  We now consider what these correlations look like using our more general approach.

\begin{figure}[tp]
\begin{center}
\epsfxsize=0.44\textwidth\epsfbox{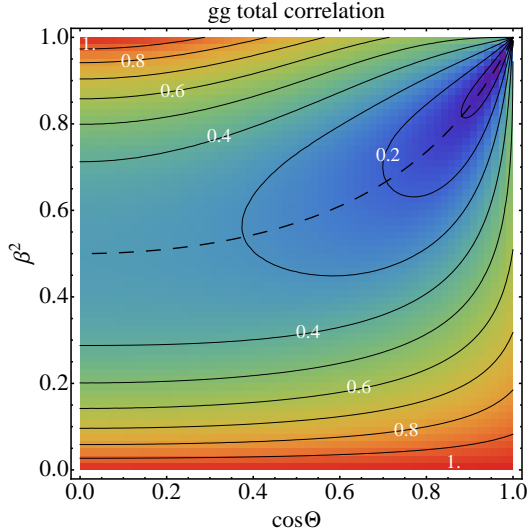}
\caption{\it Total LO spin correlation strength in $gg \to t\bar t$.  Plotted versus top production angle and squared-velocity in the partonic CM frame.  (Dashed line indicates $p_T = m_t$.)} 
\label{fig:ggTotalCorr}
\end{center}
\end{figure}

First, in Fig.~\ref{fig:ggTotalCorr}, we show the total correlation $\cal C$.  The correlation is maximal both near threshold and for relativistic central production, with a broad contour of minima at $p_T = m_t$.  In the near-threshold region, production is dominated by opposite-spin gluons annihilating into tops in a spin $s$-wave with $C = {\rm diag}(-1,-1,-1)$.  In the relativistic region with $p_T \gg m_t$, same-spin gluons dominate, and the correlations look similar to those of $q\bar q$ annihilation.  At central production angles these again approach $C = {\rm diag}(+1,-1,+1)$.  Near $p_T = m_t$, the two initial spin states contribute comparably, and their correlations combine destructively.  The cancellation becomes perfect in the limit of relativistic forward production.

\begin{figure}[tp]
\begin{center}
\epsfxsize=0.44\textwidth\epsfbox{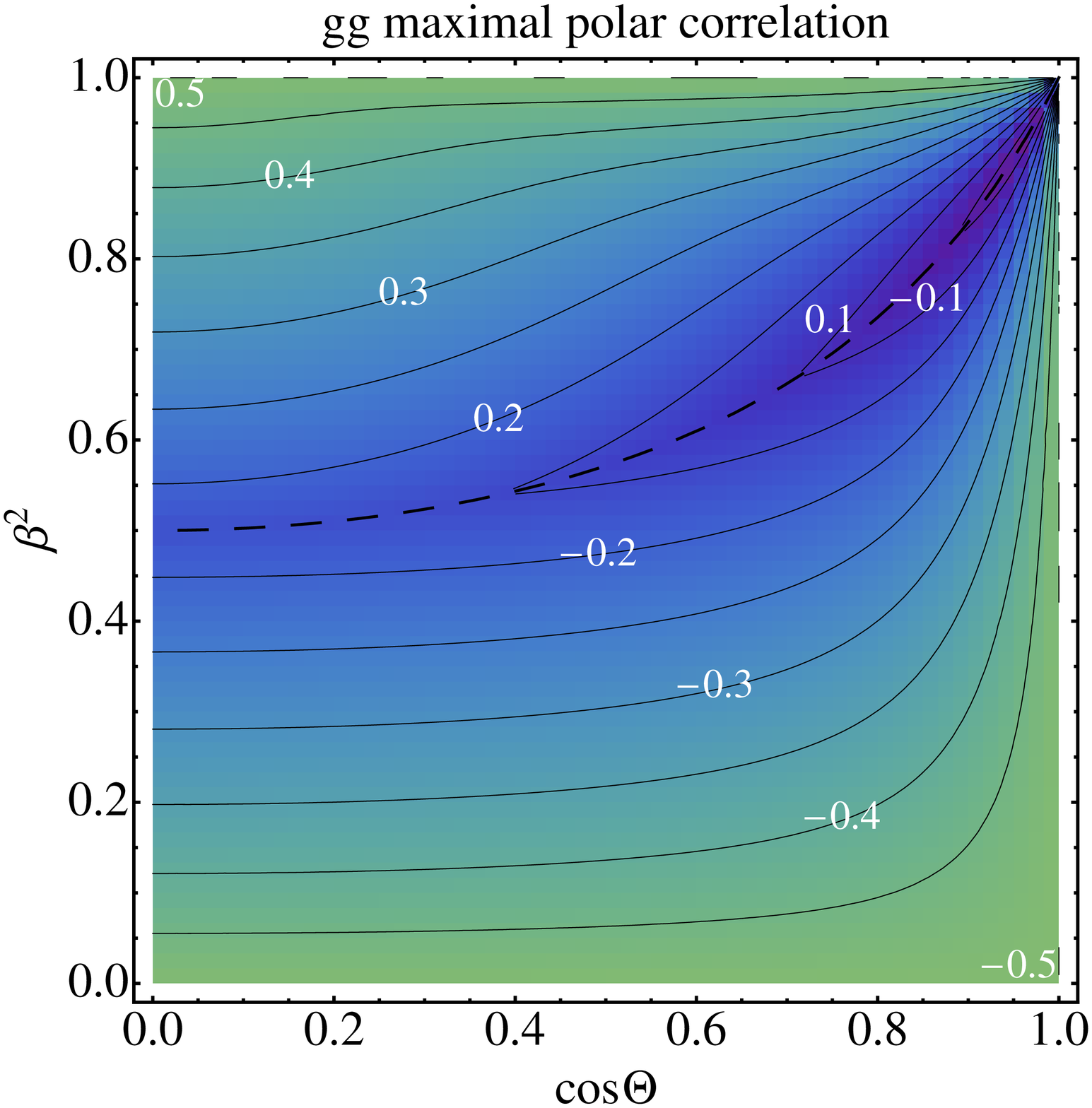}
\epsfxsize=0.44\textwidth\epsfbox{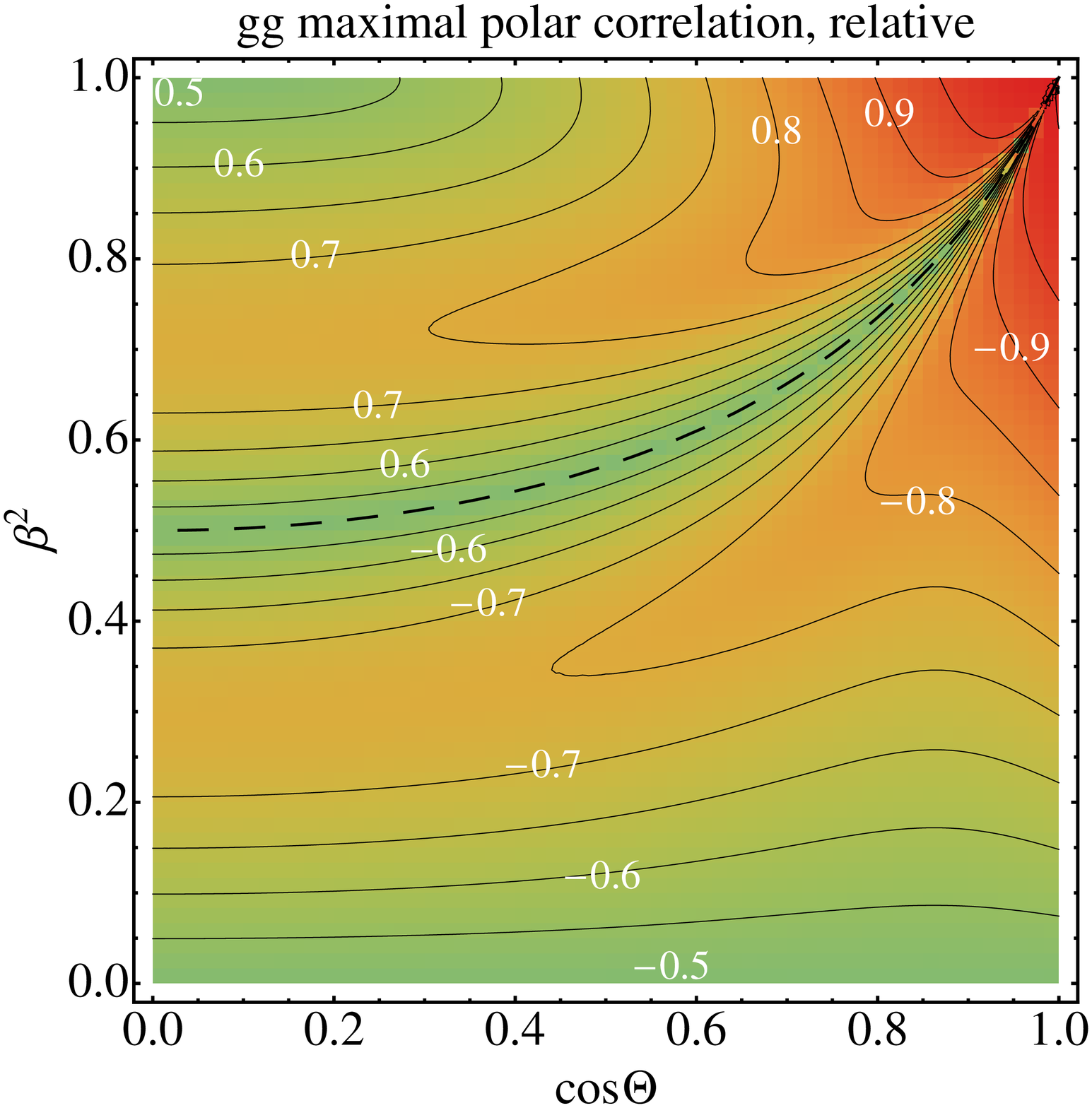}
\caption{\it Maximal LO correlation strength obtainable via polar angles in $gg \to t\bar t$.  Absolute (left) and relative to total strength (right).  Plotted versus top production angle and squared-velocity in the partonic CM frame.  (Dashed line indicates $p_T = m_t$.)} 
\label{fig:ggPolar}
\end{center}
\end{figure}

\begin{figure}[tp]
\begin{center}
\epsfxsize=0.44\textwidth\epsfbox{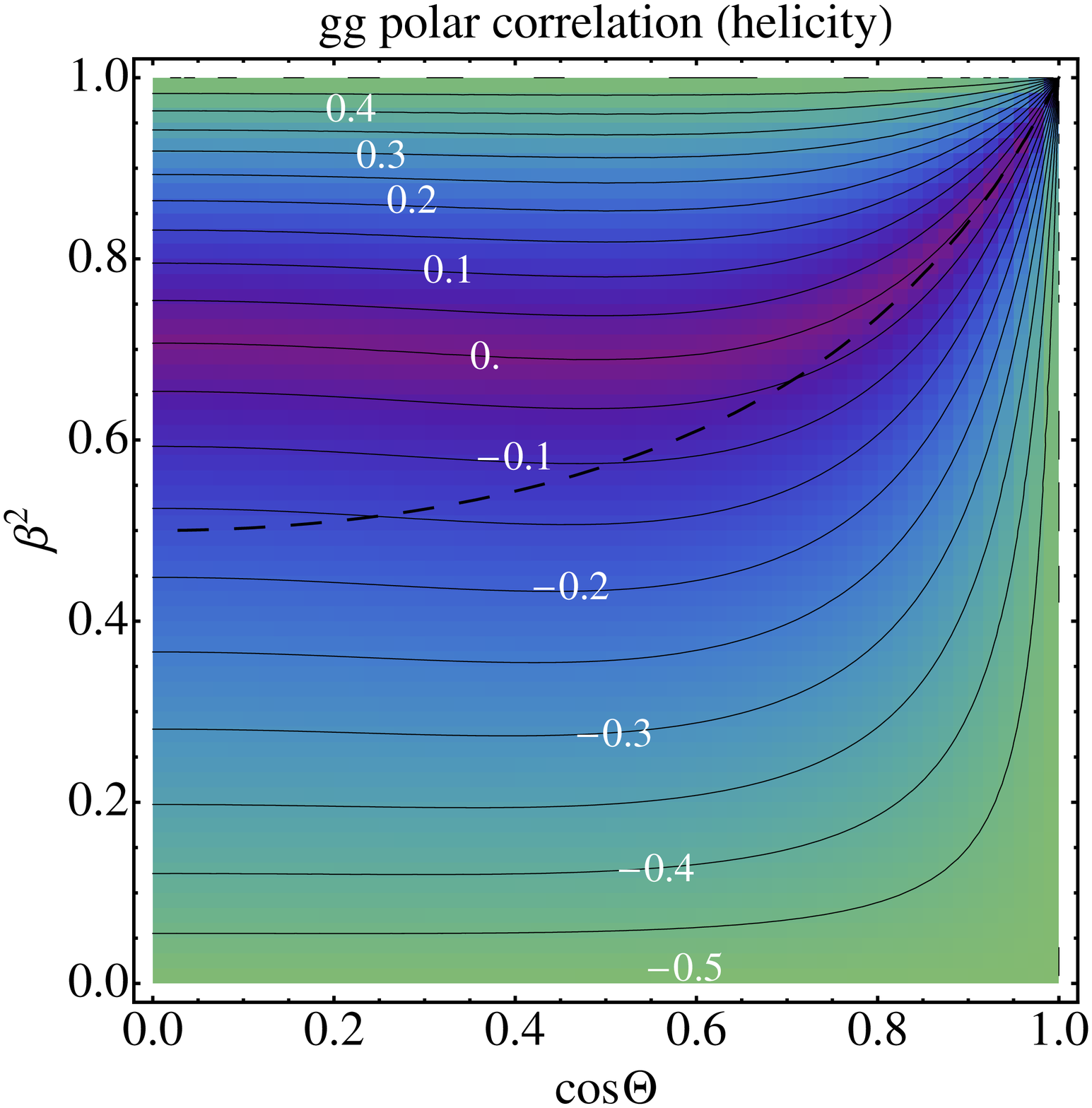}
\epsfxsize=0.44\textwidth\epsfbox{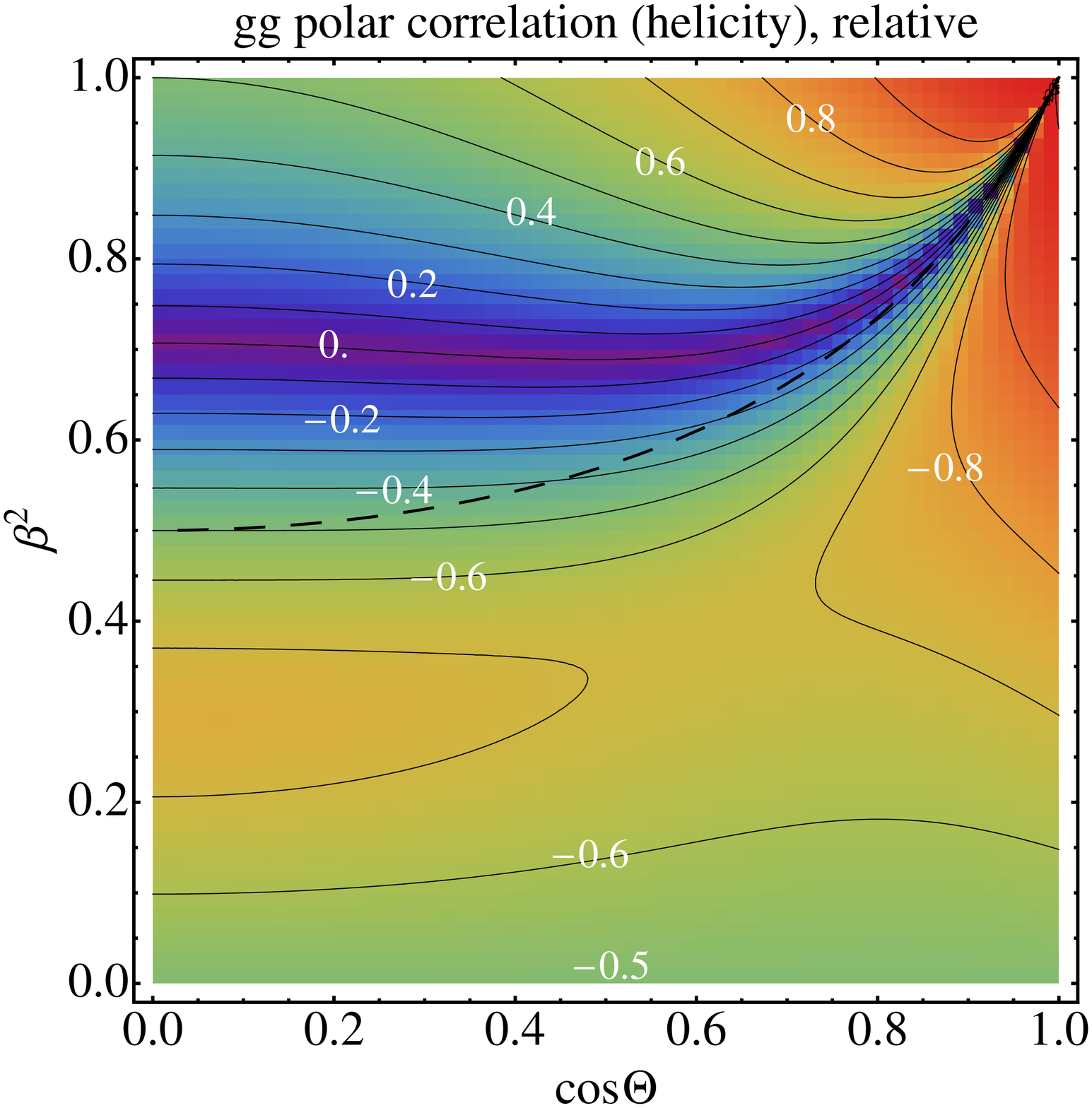}
\caption{\it LO correlation strength in helicity-basis polar angles in $gg \to t\bar t$.  Absolute (left) and relative to total strength (right).  Plotted versus top production angle and squared-velocity in the partonic CM frame.  (Dashed line indicates $p_T = m_t$.)} 
\label{fig:ggPolarHel}
\end{center}
\end{figure}

We plot the maximal polar correlation in Fig.~\ref{fig:ggPolar}, both the absolute strength and the strength relative to $\cal C$.  The situation is again clearly nontrivial.  According to our measure, this method picks up between 50\% and 75\% of the total correlation throughout the bulk of the phase space, approaching 100\% only for relativistic forward production, where the correlation is shutting off.  Note that the largest-magnitude eigenvalue of $C$ flips sign at $p_T = m_t$, from negative at low $p_T$ to positive at high $p_T$.  This also corresponds to a discrete jump in our choice of $z$-axis: for central production at low $p_T$ the ideal $z$-axis is mainly aligned with the top momentum vector (helicity basis), whereas at high $p_T$ the ideal $z$-axis is mainly aligned with the beams.  This switchover happens because the helicity-basis polar correlation necessarily passes through a zero as we transition from $s$-wave production near threshold to chiral production at high momentum.  It therefore becomes small in magnitude and cannot contribute to a large eigenvalue.  The situation is illustrated in Fig.~\ref{fig:ggPolarHel}, where we can see the zero occurring near $\beta^2 = 1/\sqrt2$ ($m_{t\bar t} \simeq (1.85)(2m_t)$) for a broad range of production angles.

\begin{figure}[tp]
\begin{center}
\epsfxsize=0.44\textwidth\epsfbox{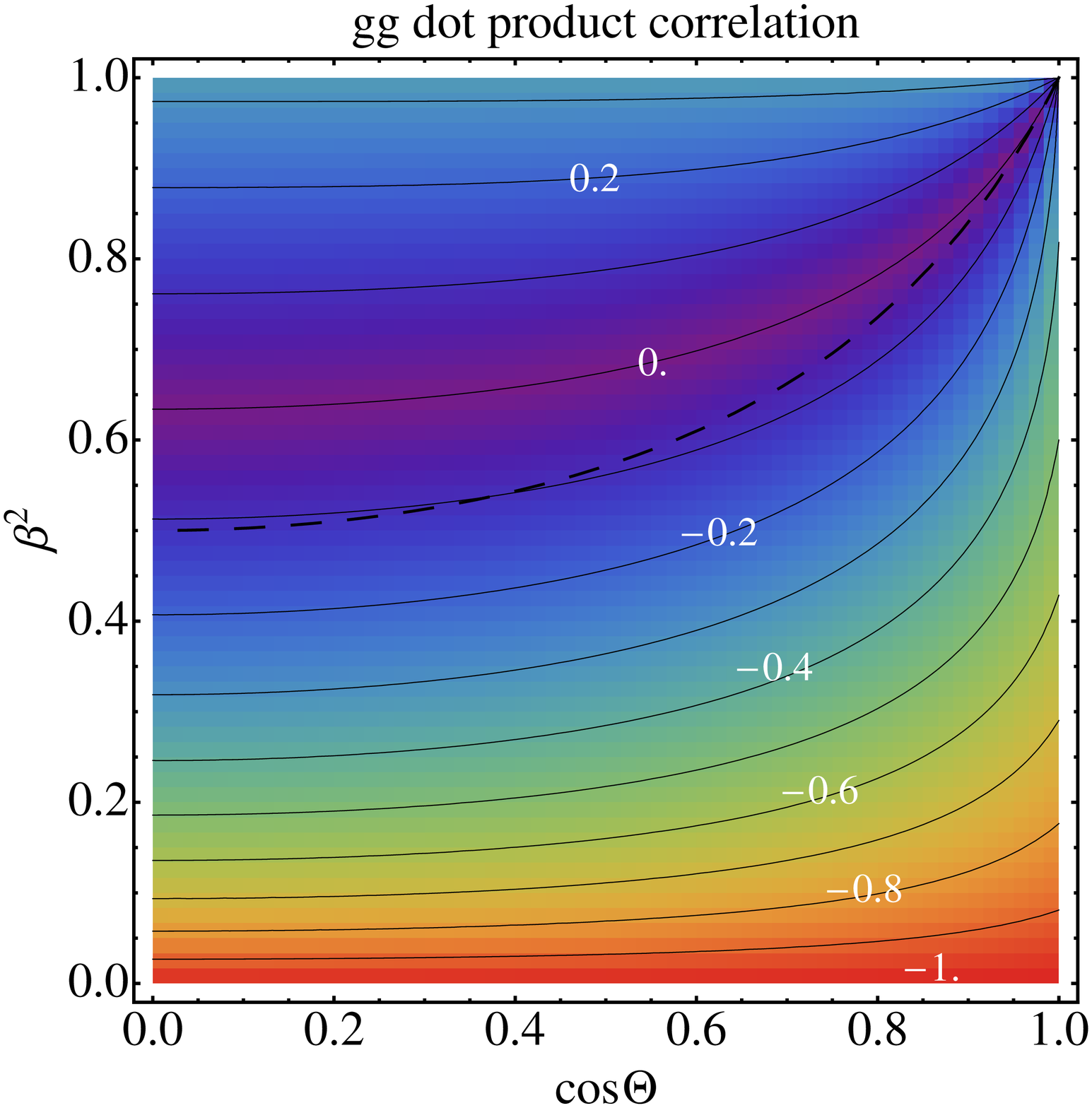}
\epsfxsize=0.44\textwidth\epsfbox{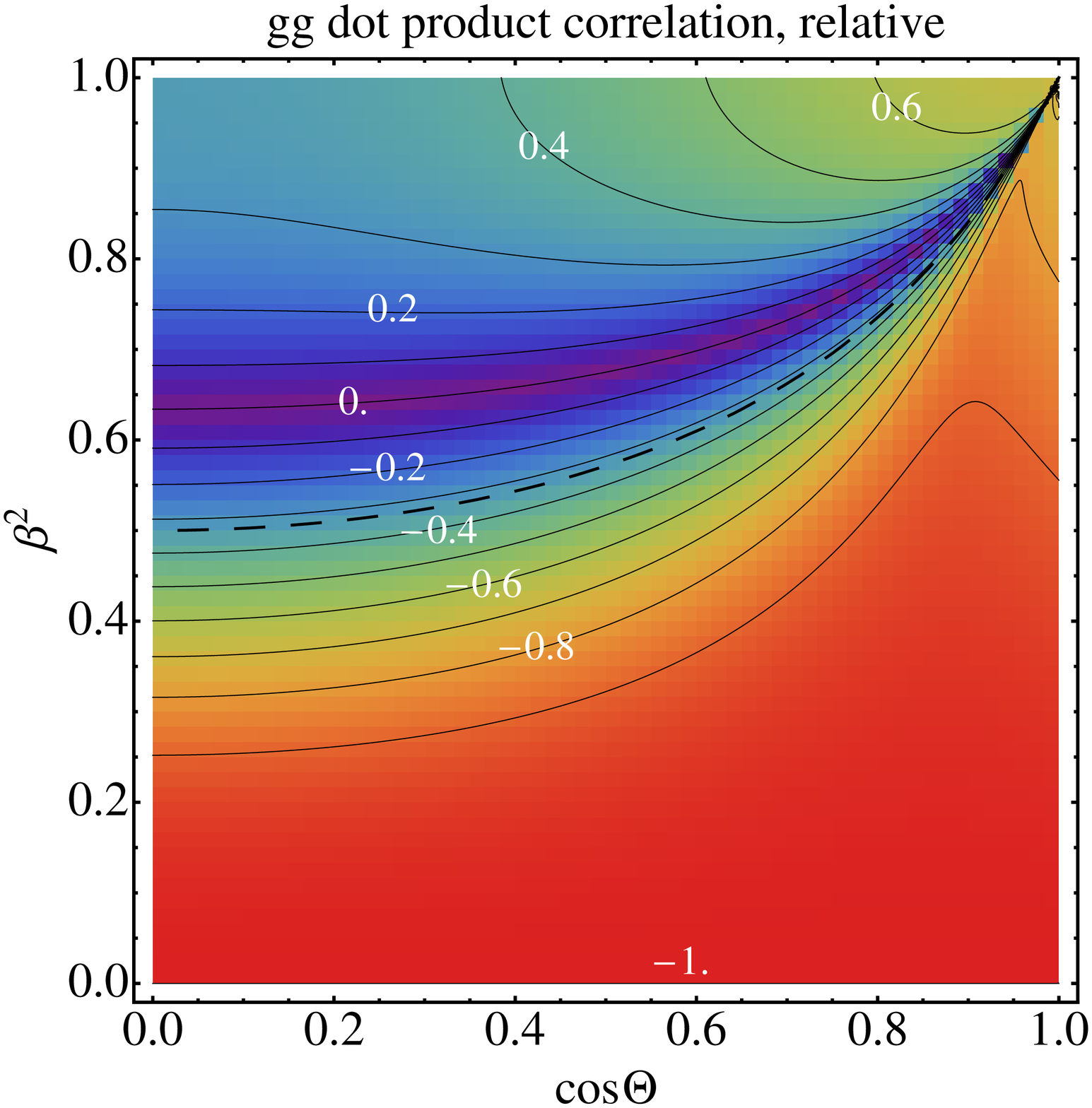}
\caption{\it LO correlation strength in the dot product ($\cos\chi$) in $gg \to t\bar t$.  Absolute (left) and relative to total strength (right).  Plotted versus top production angle and squared-velocity in the partonic CM frame.  (Dashed line indicates $p_T = m_t$.)} 
\label{fig:ggDotProduct}
\end{center}
\end{figure}

\begin{figure}[tp]
\begin{center}
\epsfxsize=0.44\textwidth\epsfbox{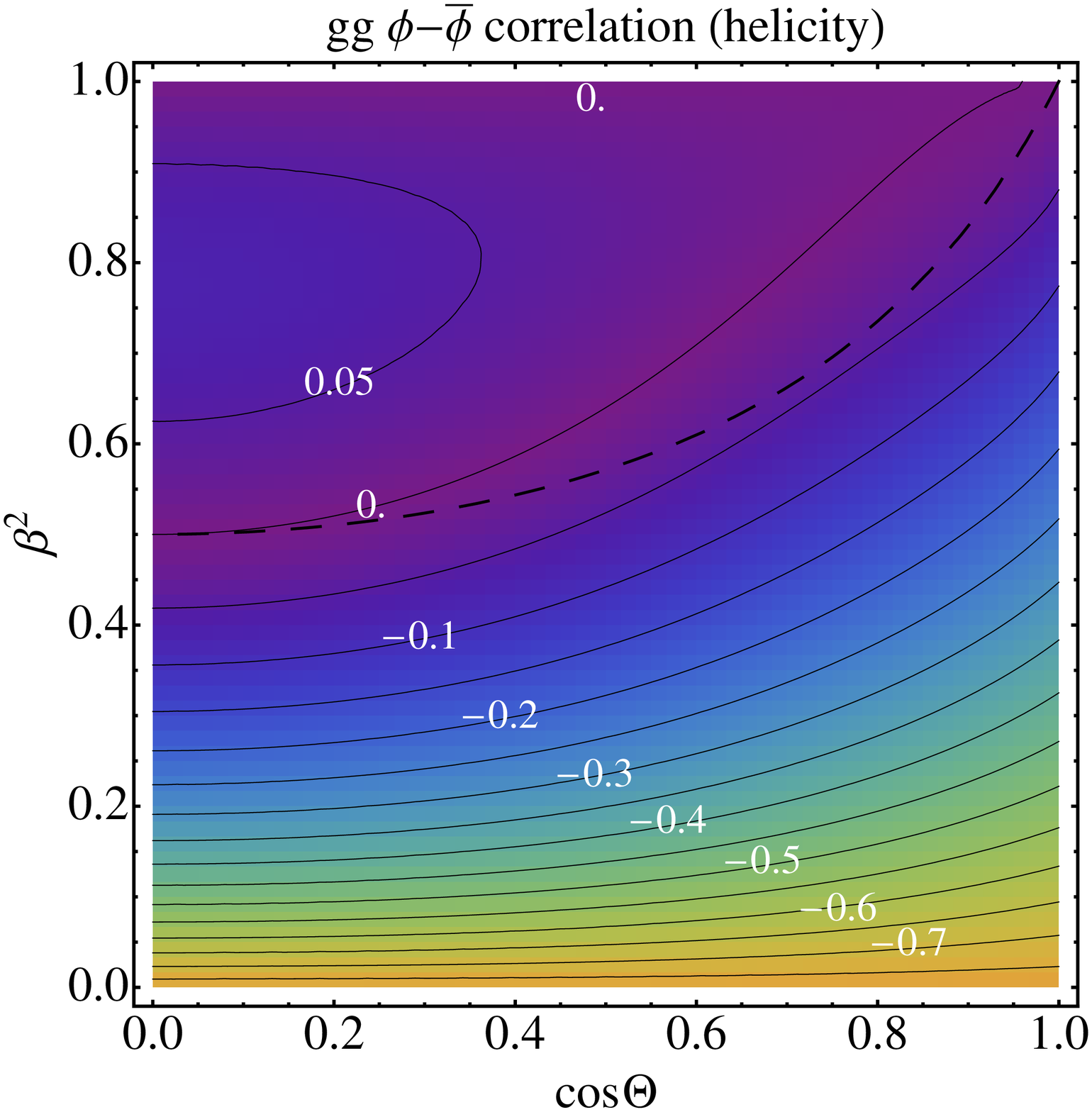}
\epsfxsize=0.44\textwidth\epsfbox{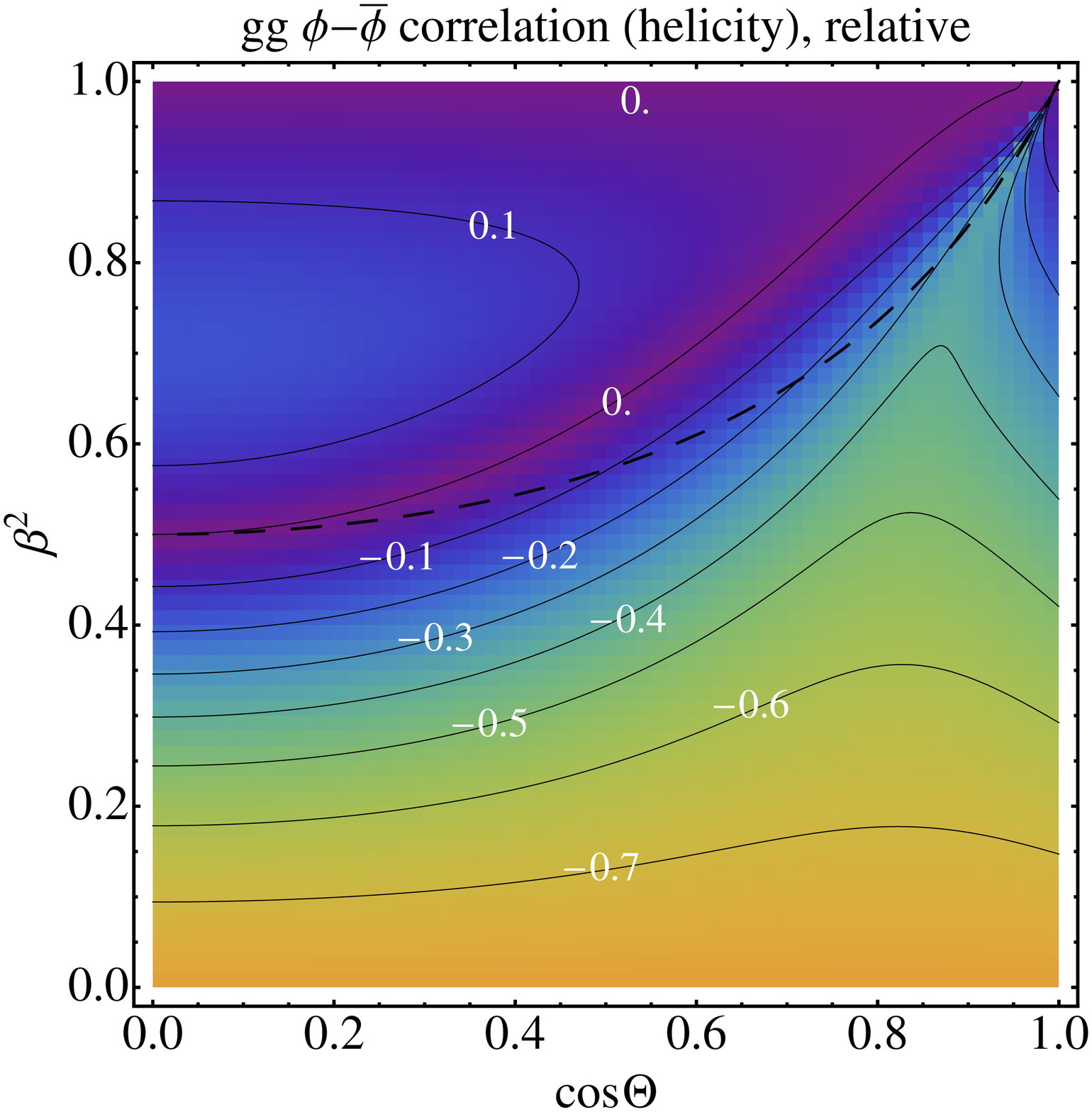}
\caption{\it LO correlation strength in helicity-basis $\phi-\bar\phi$ in $gg \to t\bar t$.  Absolute (left) and relative to total strength (right).  Plotted versus top production angle and squared-velocity in the partonic CM frame.  (Dashed line indicates $p_T = m_t$.)} 
\label{fig:ggAzDiff}
\end{center}
\end{figure}

\begin{figure}[tp]
\begin{center}
\epsfxsize=0.44\textwidth\epsfbox{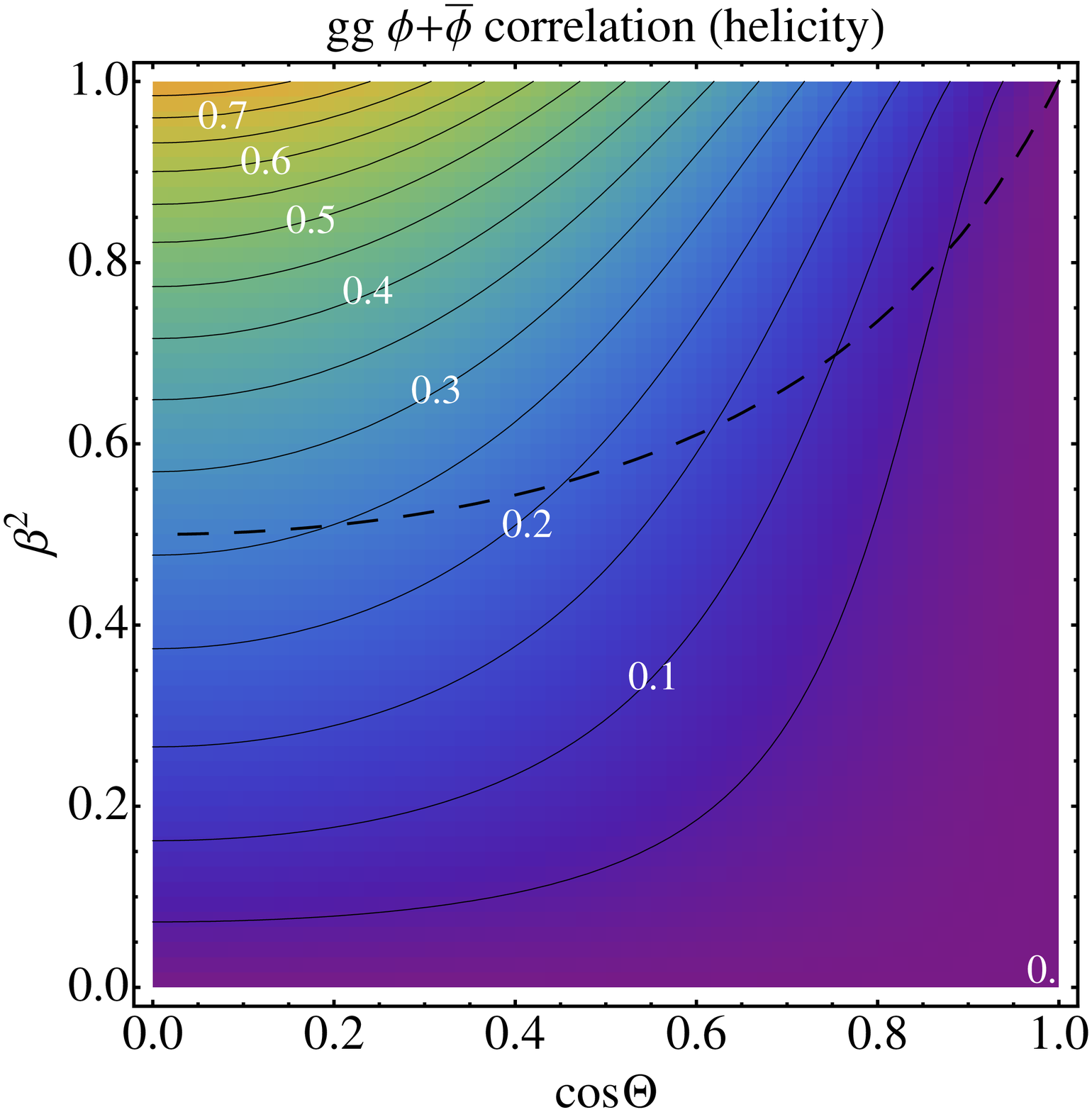}
\epsfxsize=0.44\textwidth\epsfbox{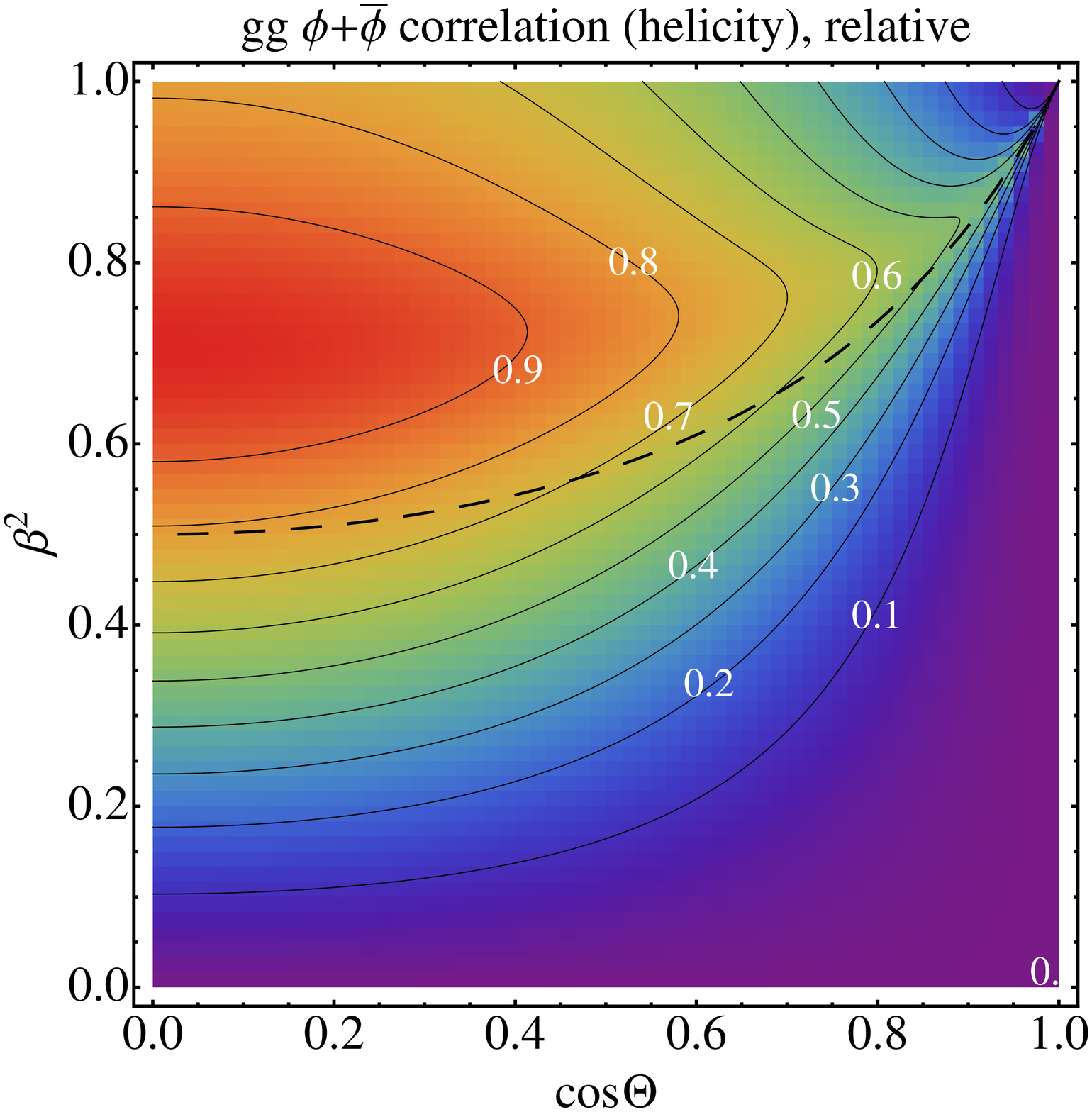}
\caption{\it LO correlation strength in helicity-basis $\phi+\bar\phi$ in $gg \to t\bar t$.  Absolute (left) and relative to total strength (right).  Plotted versus top production angle and squared-velocity in the partonic CM frame.  (Dashed line indicates $p_T = m_t$.)} 
\label{fig:ggAzSum}
\end{center}
\end{figure}

\begin{figure}[tp]
\begin{center}
\epsfxsize=0.44\textwidth\epsfbox{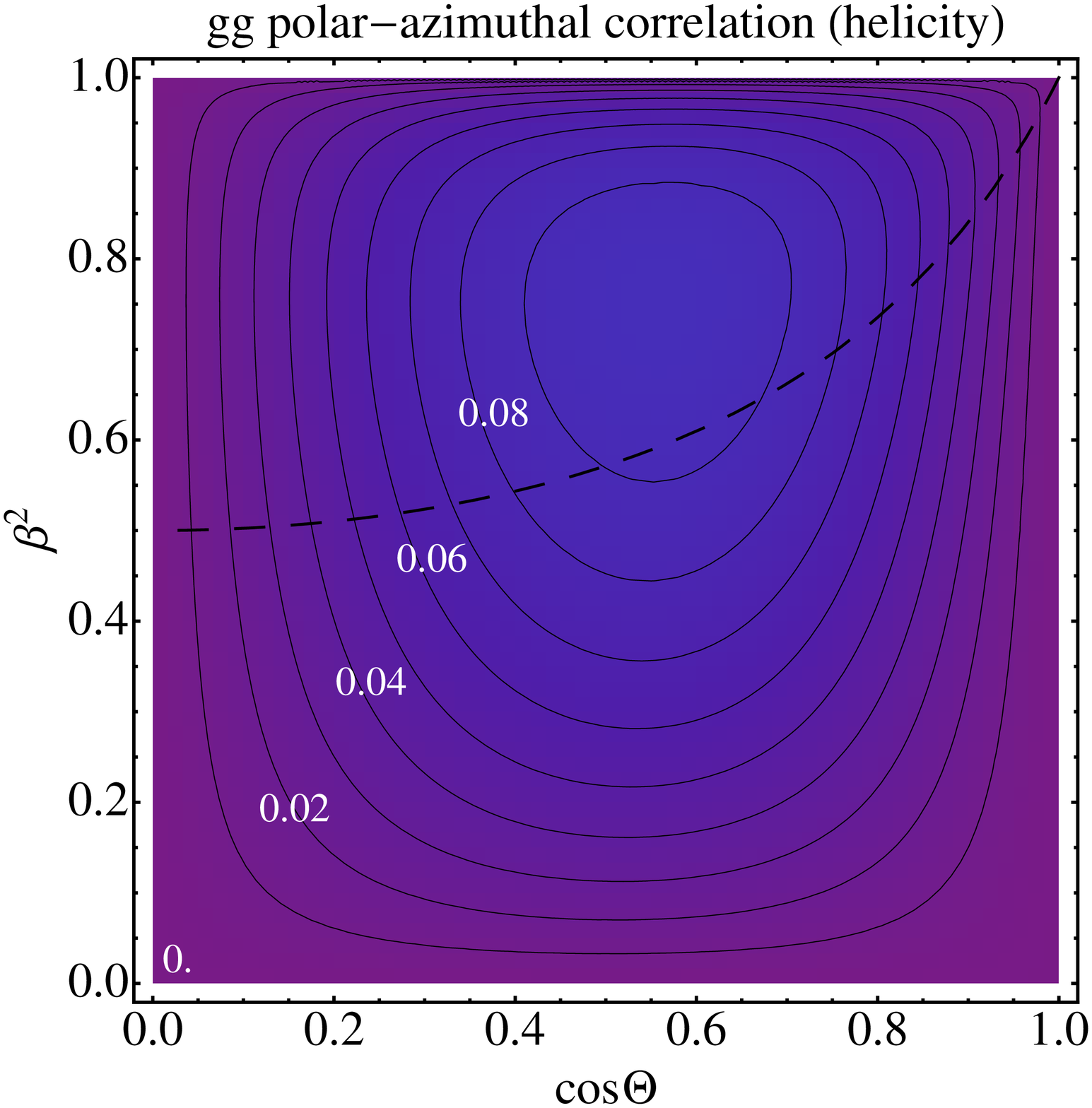}
\epsfxsize=0.44\textwidth\epsfbox{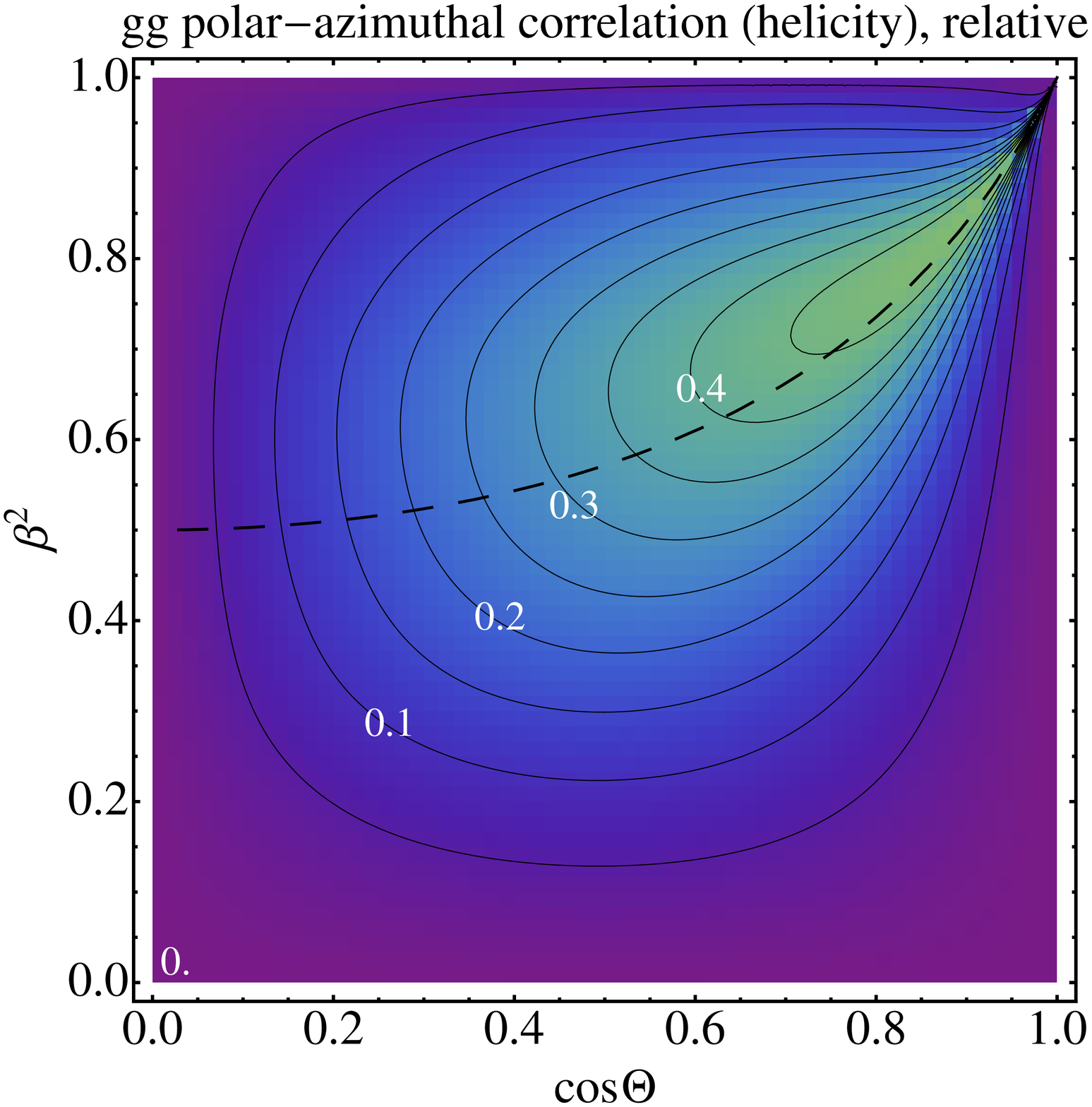}
\caption{\it LO polar-azimuthal ($xz$) cross-correlation strength in helicity-basis in $gg \to t\bar t$.  Absolute (left) and relative to total strength (right).  Plotted versus top production angle and squared-velocity in the partonic CM frame.  (Dashed line indicates $p_T = m_t$.)} 
\label{fig:ggPolarAz}
\end{center}
\end{figure}

Next, we consider the other variables which we have been discussing.  The simple dot-product $\cos\chi$ appears in Fig.~\ref{fig:ggDotProduct}.  As commonly observed, it picks up most of the total correlation close to threshold.  Near $p_T = (1.3)m_t$, it passes through a zero, and then asymptotically approaches $1/3$ at high boost.  In Figs.~\ref{fig:ggAzDiff} through~\ref{fig:ggPolarAz}, we show the variables that make dedicated use of azimuthal angles, specializing to helicity basis.  These include the azimuthal angle difference $\phi-\bar\phi$ (Fig.~\ref{fig:ggAzDiff}), the azimuthal sum $\phi+\bar\phi$ (Fig.~\ref{fig:ggAzSum}), and the polar-azimuthal, or $xz$, cross-correlation (Fig.~\ref{fig:ggPolarAz}).  Like the dot-product correlation, the azimuthal-difference correlation is mainly active near threshold (albeit with smaller strength) and passes through a zero at intermediate $p_T$.  Unlike the dot-product correlation, it largely fails to regenerate at high $p_T$, reaching back up to only about 5\% near $p_T \simeq (1.85)m_t$ before shutting off again.  Its turnoff is also more closely aligned with the $p_T = m_t$ contour.  The azimuthal-sum correlation is in some sense the inverse of the azimuthal-difference.  It is weak at low $p_T$ and strong at high $p_T$.  In fact, for $\beta^2$ near $1/\sqrt2$, where the helicity-basis polar correlation shuts off, the azimuthal-sum nearly saturates the entire correlation, surpassing 99\% of $\cal C$ for central production.  For even higher $p_T$, the correlation remains strong, typically above 80\% relative to the total.  Finally, we consider the $xz$ correlation, which can be obtained either from Eq.~\ref{eq:PolarAz} or from Eq.~\ref{eq:coscos} by measuring one of the tops' polar decay angles with respect to the $x$-axis instead of the $z$-axis.  (Both approaches yield the same asymmetry.)  This correlation is typically quite small, below 10\%, and is strictly zero on all four edges of Fig.~\ref{fig:ggPolarAz}.  It accounts for a nontrivial fraction of the total correlation only at somewhat forward angles with $p_T \simeq m_t$. 

Clearly, then, there is quite a lot that can be measured at the LHC.  At low momenta, $p_T \lsim m_t$, a measurement of a large dot-product correlation would suggest the expected $s$-wave configuration.  (The current ATLAS and CMS measurements already support this~\cite{ATLAS:2012ao,CMScorr}.)  Indeed, as illustrated by Fig.~\ref{fig:ggDotProduct}, the total correlation is almost wholly reflected in $\cos\chi$, even for momenta somewhat above threshold where the total correlation has already dropped by $O(2)$.  However, a measurement of $\cos\chi$ by itself is inadequate to conclusively establish this behavior.  We have seen how to instead break the correlation down into its individual components in helicity basis, providing a more comprehensive view.  The expectation at low $p_T$ is small correlations in $\phi+\bar\phi$ and $xz$, and comparable $O$(0.1--1) correlations in $\phi-\bar\phi$ and the polar decay angles.

At high momenta, $p_T \gsim m_t$, the the helicity-basis polar correlation passes through a zero, and the azimuthal correlations dominate.  While it is possible to capture part of this correlation via polar angles constructed with the optimized choice of ``$z$-axis'' from~\cite{Uwer:2004vp,Mahlon:2010gw} (basically the beam-axis), it is much more efficient and straightforward to simply measure $\phi+\bar\phi$, which encodes almost the entire correlation over a broad range of momenta for moderately central angles.  The smallness of all of the remaining correlations (polar, $\phi-\bar\phi$, and $xz$) can then be checked independently.  In this way, the complete and rather complicated evolution of the spin correlations can be fully mapped out using a sequence of fairly simple measurements.  (At very high $p_T$, the process $q\bar q \to t\bar t$ also becomes important, but exhibits essentially the same pattern of spin correlations as $gg \to t\bar t$.)

\section{Spin Correlations from New Physics}
\label{sec:NP}

New physics can alter the spin correlations of top pairs in a large variety of ways, many of which have been explored in the literature~\cite{Bernreuther:1993hq,Beneke:2000hk,Frederix:2007gi,Arai:2007ts,Degrande:2010kt,Cao:2010nw,Baumgart:2011wk,Barger:2011pu,Krohn:2011tw,Bai:2011uk,Han:2012fw,Fajfer:2012si}.  Polar-polar correlations in particular are practically the default probe, especially when the new physics in question does not introduce a strong net polarization or other symmetry violations.  Often an ideal basis is first identified to maximize the polar-polar effect.  Nonetheless, considering the complete correlation matrix, especially the $xy$ block containing azimuthal correlations, can be quite useful.  In some cases, azimuthal correlations even capture the majority of the correlation effect from new physics.

In~\cite{Baumgart:2011wk}, we showed how correlations in both $\phi-\bar\phi$ and $\phi+\bar\phi$ could be used to characterize resonances in the $t\bar t$ invariant mass spectrum.  Spin-0 resonances produce a $\phi-\bar\phi$ modulation, and the phase of this modulation is directly related to the phase in the scalar's Yukawa coupling to tops.  This observation has also been made in~\cite{Barger:2011pu}, which included a more comprehensive study of interference effects.  Spin-1 and spin-2 resonances produce a $\phi+\bar\phi$ modulation, which is sensitive to the signed ratio of the resonance's chiral couplings to top quarks.  This feature is particularly relevant to heavy axigluon resonance or contact-interaction models that purport to explain the top forward-backward asymmetry at the Tevatron~\cite{Frampton:2009rk}, as they will induce a ``wrong-signed'' $\phi+\bar\phi$ modulation at high-$m(t\bar t)$ compared to the SM.

In this section, we consider two additional new physics scenarios that leave strong imprints on the azimuthal correlations.  The first is the presence of dimension-five chromomagnetic and chromoelectric dipole (CMDM and CEDM) operators.  The effects of these operators have been studied extensively in the past.  The usual logic is to probe the CMDM using total rates, and the CEDM using various forms of CP-violating observables \cite{Atwood:1992vj,Atwood:1994vm,Rizzo:1996zt,Lee:1997up,Zhou:1998wz,Beneke:2000hk,Atwood:2000tu,Sjolin:2003ah,Gupta:2009wu,Zhang:2010dr,Degrande:2010kt,Kamenik:2011dk,Hioki:2012vn,Englert:2012by}.  It is also possible to see the effects of the CEDM in the neutron electric dipole moment, and this leads to a particularly powerful indirect constraint \cite{Kamenik:2011dk}.  However, because the chirality structure of these operators are similar to Yukawa couplings, one of their dominant effects is a modification of the $\phi-\bar\phi$ distribution.  A measurement of this distribution therefore provides a sensitive probe of both operators simultaneously.  

The second scenario is a broad resonance with parity-violating couplings.  A standard way to reveal parity violation is to measure the longitudinal polarization of the individual tops.  But we have seen above that parity-violation can also manifest itself as the appearance of forbidden terms in the spin correlation matrix.  We explore a simple example model of a resonance that is so broad as to be unobservable in the $t\bar t$ mass spectrum, but which modifies the $\phi+\bar\phi$ distribution at least as strongly as it modifies distributions sensitive to the net top polarizations.

\subsection{Chromomagnetic and chromoelectric dipole operators}

Our starting point is a modification to the QCD couplings to top quarks due to the following operators
\beq
\Delta {\mathcal L}  \,=\, \frac{g_s}{2} \, G_{\mu\nu}^a \, \bar t \left[ T^a \sigma^{\mu\nu} (\mu + i\gamma^5 d) \right] t \, ,
\eeq
with $\sigma^{\mu\nu} \equiv (i/2)[\gamma^\mu,\gamma^\nu]$.  These operators are dimension-five, but arise from more fundamental dimension-six operators with a Higgs field insertion.  (They also lead, for example, to a novel $tbW^+g$ coupling.)  The dimensionful couplings $\mu$ and $d$ characterize their strength.  One physical consequence of these operators are contributions to the anomalous chromomagnetic and chromoelectric dipole moments of the top as seen by soft gluon exchanges, from which the couplings derive their usual labels as the CMDM and CEDM.  They also lead to effects in high-energy processes, and can be constrained by measurements at hadron and lepton colliders.  Using direct collider constraints from the measured $t\bar t$ cross sections, Ref.~\cite{Kamenik:2011dk} obtains $\mu\times m_t \, <$ 0.05 and $d\times m_t\, <$ 0.16.  They also demonstrate a strong indirect constraint on the CEDM due to loop-induced effects on the electromagnetic EDMs of the neutron and the mercury nucleus, implying $d\times m_t\, < \, 2 \times 10^{-3}$.  We will see in Section~\ref{sec:measurement} that our independent set of constraints from spin correlations gives us competitive sensitivity to the CMDM and CEDM with respect to the direct studies.  Spin correlations alone will have the potential to probe dipole moments below 0.01 over the lifetime of the LHC, possibly even approaching the level of the indirect CEDM constraint.

The CMDM operator conserves all of the discrete Lorentz symmetries, while the CEDM operator is P-violating and C-conserving, leading to a net CP-violation.   In the language of Section~\ref{sec:formalism}, the CMDM can only affect the entries of the 4$\times$4 spin density matrix $M$ that are already populated by the SM.  The CEDM can also populate the forbidden $xy$, $yx$, $yz$, and $zy$ entries, but because of C-conservation these entries are still symmetric.  Neither operator induces a net polarization at tree-level.

Assuming that the couplings $\mu$ and $d$ are small relative to $1/m_t$, we work to linear order, and therefore pick up only the leading effect from interference with QCD.  To this order, the CMDM still appears in all C/P-allowed entries of the production matrix, but the CEDM only appears in the aforementioned entries associated with P-violation.  While the CMDM can affect the total rate and cause shifts to the spin correlations exhibited by the SM, the CEDM is ``all symmetry violation,'' inducing novel effects that never occur in QCD.

We can get a sense for the total effects of these operators on the spin correlations by applying a variation of the method described in Section~\ref{sec:formalism}.  We assume that the normalized 3$\times$3 correlation matrix $C$, in the presence of new physics, can be written as $C = C_{\rm SM} + \epsilon \, C_{\rm NP}$.  Here $\epsilon$ is an expansion parameter that characterizes the strength of the new physics.  Note that $C_{\rm NP}$ must incorporate effects not only in the spatial part of the production density matrix, but also modifications to the total rate.  Picking one fermion each from the top and antitop decay, we construct a modified opening angle from their unit vectors in their respective parents' rest frames
\beq
\cos\chi_{\rm NP}' \,\equiv\, \hat\Omega  \cdot \frac{\kappa\bar\kappa\, C_{\rm NP} \cdot\hat{\bar\Omega}}{|\kappa\bar\kappa\, C_{\rm NP} \cdot\hat{\bar\Omega}|} \, ,
\eeq
or alternatively with $\hat\Omega \leftrightarrow \hat{\bar\Omega}$.  This angle is distributed according to
\beq
\frac{d\sigma}{d\cos\chi_{\rm NP}'}  \,\propto\,  1 + |\kappa\bar\kappa | \: {\cal C} \,\cos\chi_{\rm NP}' \, ,
\eeq
where
\be
{\cal C}                     & \,=\, & \int \,\frac{d\bar\Omega}{4\pi} \; \frac{\hat{\bar\Omega}\cdot C_{\rm NP}^T C \cdot\hat{\bar\Omega}}{\sqrt{\hat{\bar\Omega}\cdot C_{\rm NP}^T C_{\rm NP} \cdot\hat{\bar\Omega}}}  \nonumber \\
\frac{d \cal C}{d\epsilon}   & \,=\, & \int \,\frac{d\bar\Omega}{4\pi} \; \sqrt{\hat{\bar\Omega}\cdot C_{\rm NP}^T C_{\rm NP} \cdot\hat{\bar\Omega}} \; .  \label{eq:slope}
\ee
This construction gives us the strongest possible slope of ${\cal C}$ versus the new physics strength $\epsilon$.  Defined in this way, positive values of $\cos\chi_{\rm NP}'$ always correspond to regions of decay phase space where the normalized rate is enhanced with respect to the SM (for positive $\epsilon$), and negative values of $\cos\chi_{\rm NP}'$ correspond to regions where it is de-enhanced.

\begin{figure}[tp]
\begin{center}
\epsfxsize=0.44\textwidth\epsfbox{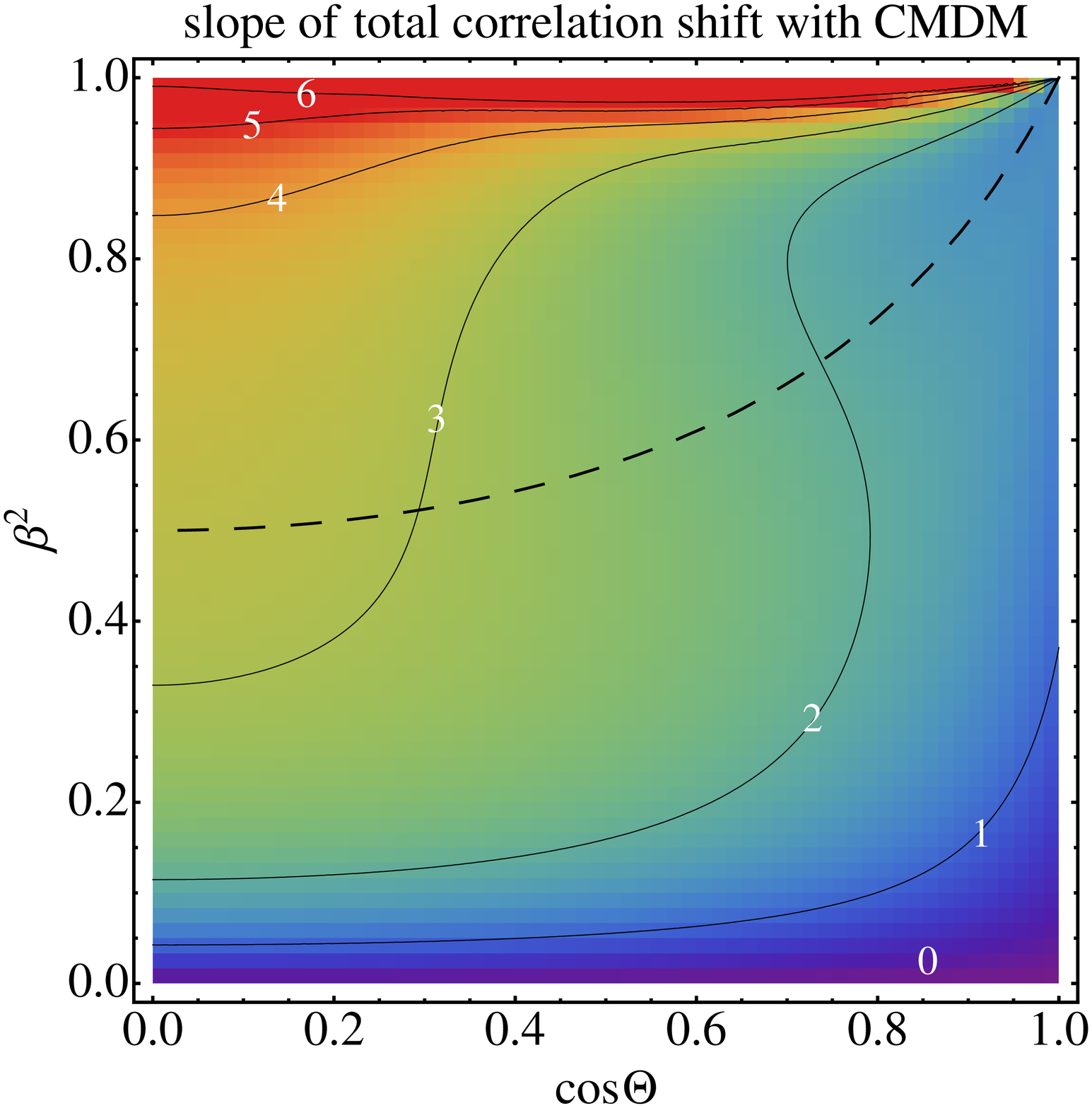}
\epsfxsize=0.44\textwidth\epsfbox{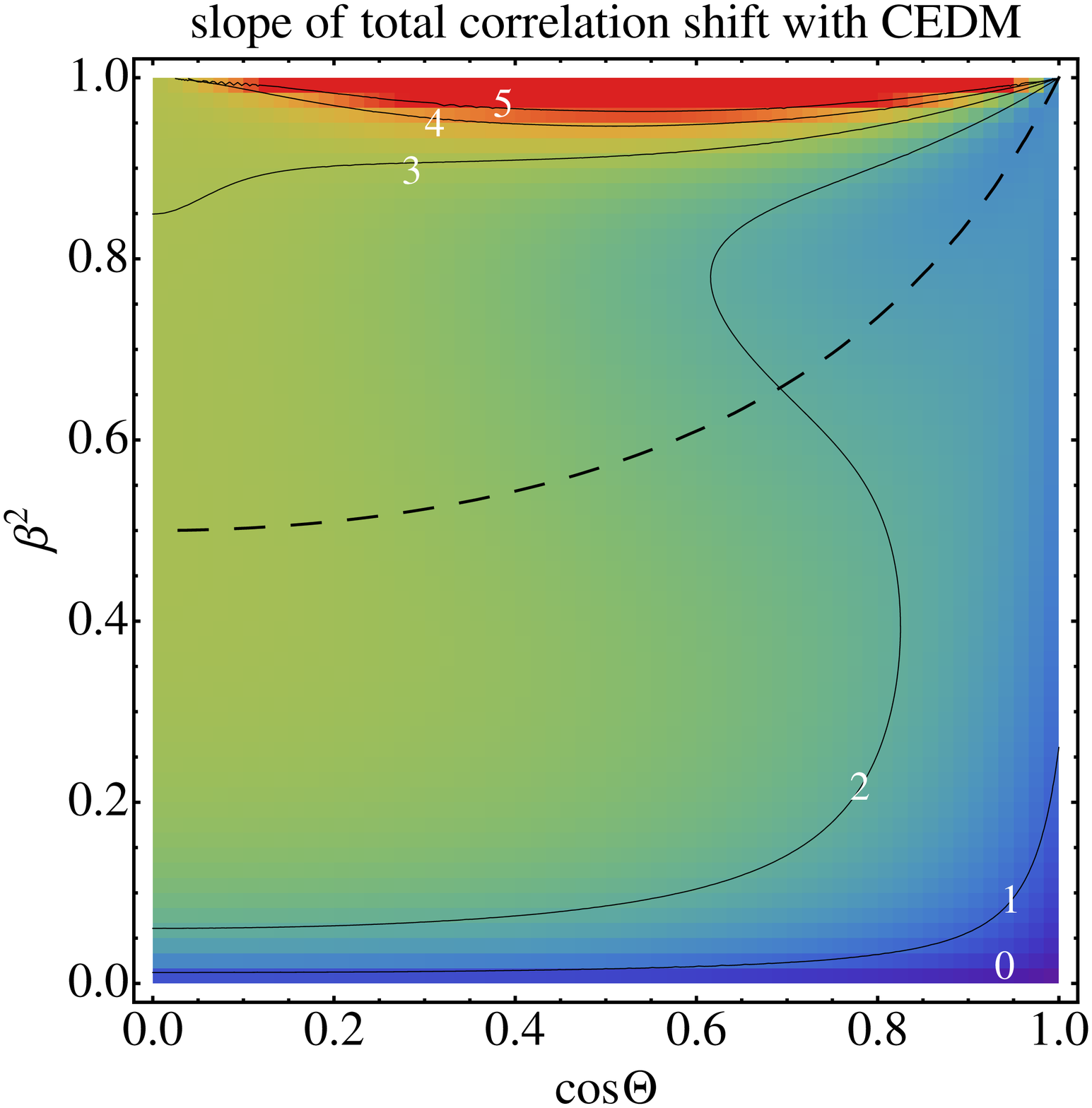}
\caption{\it Slope of the total spin correlation shift, defined in Eq.~\ref{eq:slope}, with respect to a CMDM (left) or a CEDM (right) in $gg \to t\bar t$.  The correlation is expanded to linear order in $\mu\times m_t$ or $d\times m_t$, respectively.  Plotted versus top production angle and squared-velocity in the partonic CM frame.  (Dashed line indicates $p_T = m_t$.)  High-valued contours have been cut off due to the formal divergence at high-$\beta^2$ .} 
\label{fig:ggTotalCorrCDM}
\end{center}
\end{figure}

In Fig.~\ref{fig:ggTotalCorrCDM}, we map out this slope over $gg \to t\bar t$ production phase space, independently for the CMDM and CEDM with $\epsilon \equiv \mu\times m_t$ and $\epsilon \equiv d\times m_t$, respectively.  For both the CMDM and CEDM, we see that the shifts to the correlations are maximized for boosted production.  They are fairly well-behaved over much of the phase space, though there is a formal divergence as $\beta^2 \to 1$ at intermediate production angles due to the growth of the dimension-five interaction strength with energy.\footnote{This divergence is an artifact of our linear approximation of the new physics.  Adding in the full dipole dependence regulates it, but effects sensitive to higher orders in the dipole could also receive corrections from even higher-dimensional operators.  Thus, the specific form of the correlations at very high velocity is more model-dependent.  Nonetheless, these effects are not immediately visible, due to the rapidly-falling PDF's.}  Thinking of boosted tops as approximately chiral quarks, QCD dominantly produces same-spin (opposite-helicity) tops, whereas inserting a single dipole operator leads to top production with dominantly opposite-spin (same-helicity).  The interference between these different, purely chiral, spin channels is entirely localized in the $xz$ and $yz$ off-diagonal parts of the correlation matrix, and these are the sources of the divergences.  Interference effects in the rest of the correlation matrix are everywhere finite, as one of the processes must sacrifice an $m_t/E$-suppressed helicity-flip.\footnote{The same effect occurs in the total rate interference, which is entirely due to the CMDM.  While the interference would naively grow with energy, one of the interfering amplitudes must undergo a helicity-flip.  This prevents the rate interference from blowing-up at high $t\bar t$ invariant mass.  Indeed, the rate interference is a rather mild function of the tops' production angle and velocity, typically giving a fractional contribution of $O(5\mu\times m_t)$ for both $gg \to t\bar t$ and $q\bar q \to t\bar t$ production.  Note that for both the total rate and the spin correlations, the ``new physics squared'' contributions do not suffer from $m_t/E$ suppressions, and do ultimately take over.  The effects are not strictly predictive, as interference from even higher-order operators can also become relevant.  However, the leading-order interference nominally dominates for $m(t\bar t) \lsim 2.5$~TeV for $\mu \times m_t \lsim 0.01$.}  In practice, much of the correlation effect at large but finite $\beta^2$ comes from interference with opposite-spin QCD production, inducing modifications to the $\phi-\bar\phi$ distributions.  This is fortuitous, as we have seen that the SM modulations in this variable die off at high energies.

\begin{figure}[tp]
\begin{center}
\epsfxsize=0.44\textwidth\epsfbox{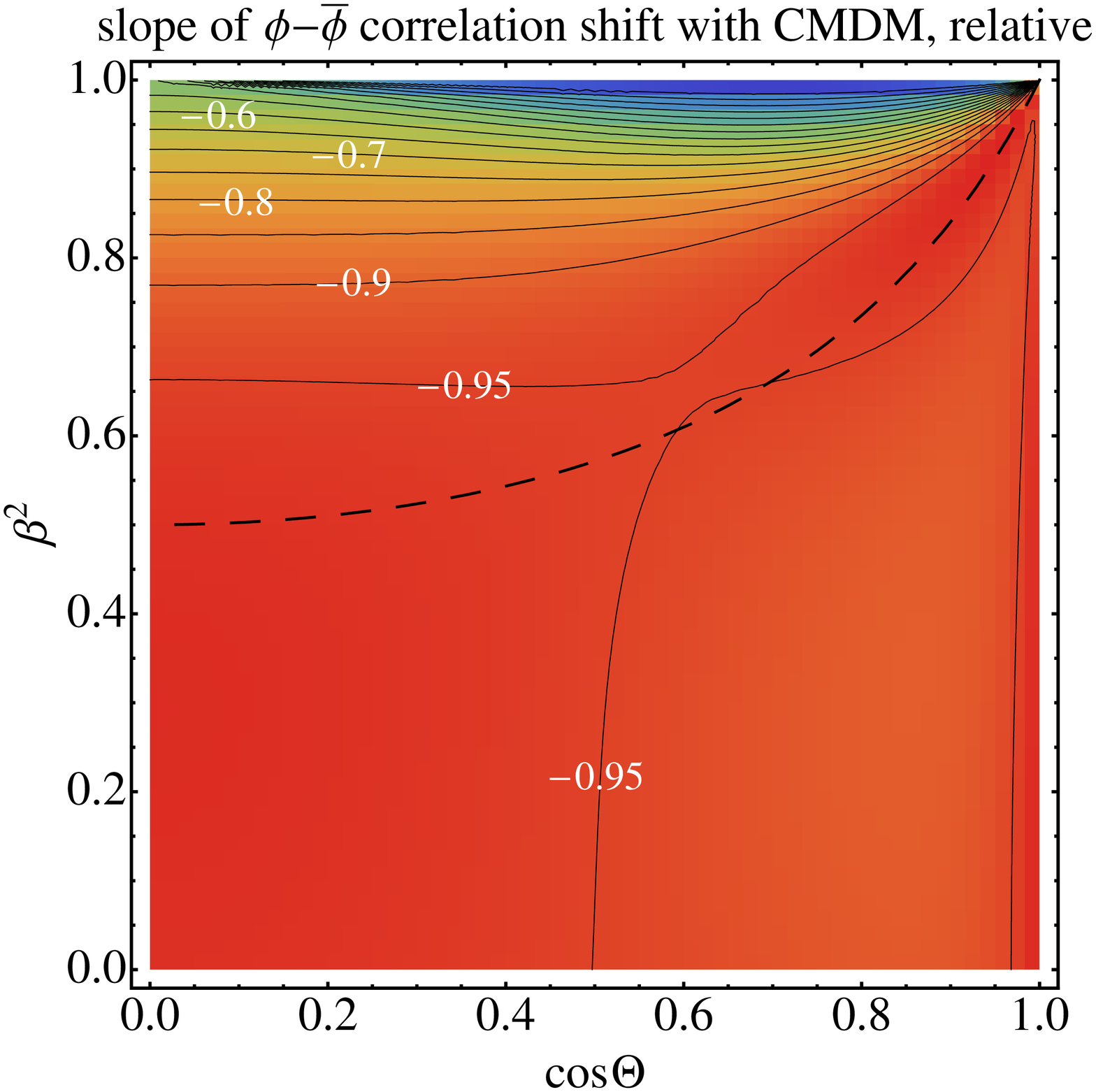}
\epsfxsize=0.44\textwidth\epsfbox{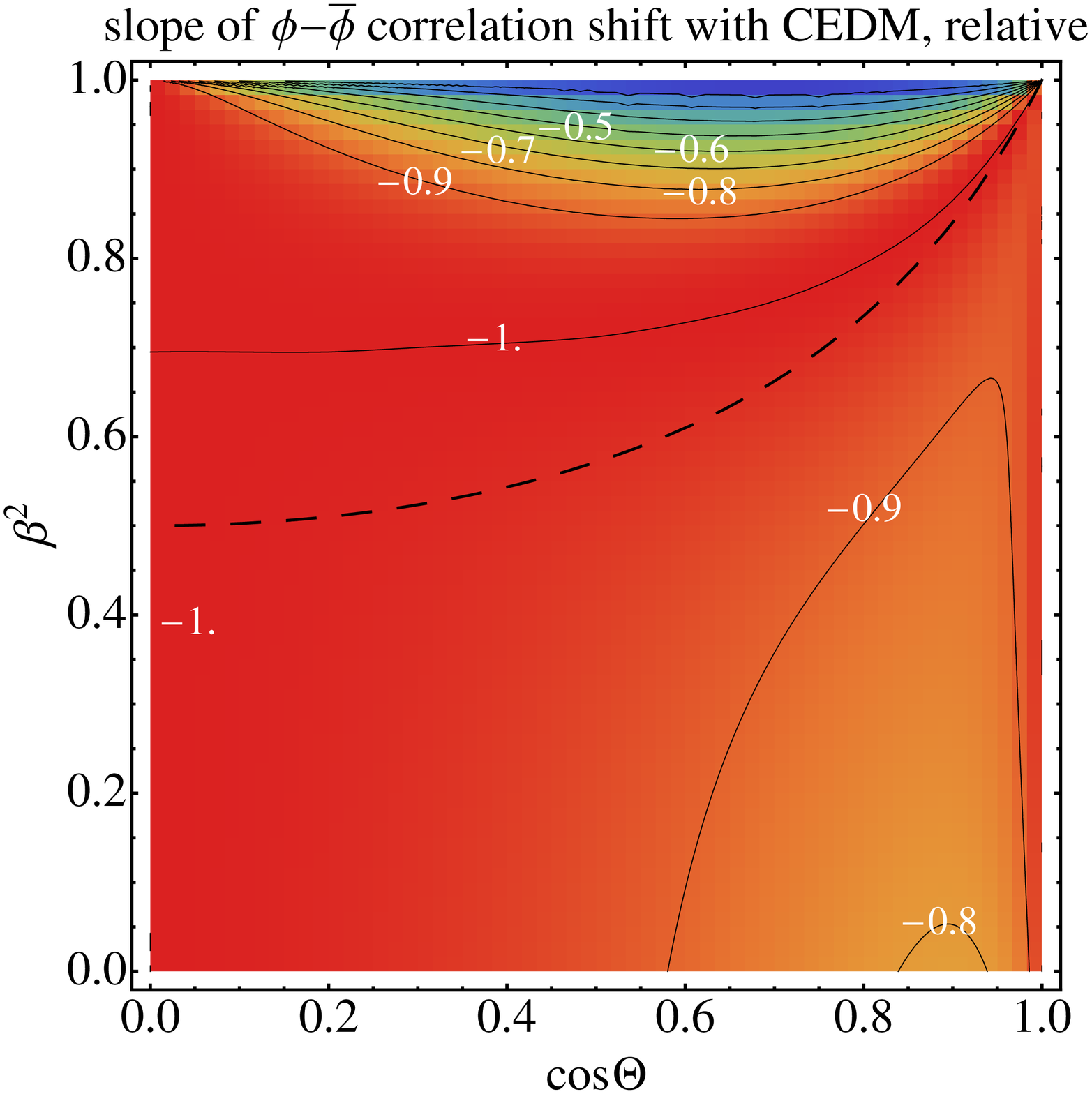}
\caption{\it Slope of the $\phi-\bar\phi$ spin correlation shift with respect to a CMDM (left) or a CEDM (right) in $gg \to t\bar t$, relative to the slope of the total shift in Fig.~\ref{fig:ggTotalCorrCDM}.  The correlation is expanded to linear order in $\mu\times m_t$ or $d\times m_t$, respectively.  Note that $\mu$ induces a pure cosine-wave modulation, and $d$ induces a pure sine-wave modulation.  Plotted versus top production angle and squared-velocity in the partonic CM frame.  (Dashed line indicates $p_T = m_t$.)} 
\label{fig:ggAzDiffCDM}
\end{center}
\end{figure}

To illustrate this point in more detail, we show in Fig.~\ref{fig:ggAzDiffCDM} the relative slope of the $\phi-\bar\phi$ modulation effect compared to the maximal slope ${\cal C}$, continuing to focus on $gg \to t\bar t$ production.  As in Section~\ref{sec:QCD}, we use asymmetries to define a common normalization, allowing us to compare the strength of sinusoidal correlations to linear correlations.  For the CMDM, we define the asymmetry between the regions $|\phi-\bar\phi| = [0,\pi/2]$ and $|\phi-\bar\phi| = [\pi/2,\pi]$.  For the CEDM, we define the asymmetry between the regions $\phi-\bar\phi = [0,\pi]$ and $\phi-\bar\phi = [-\pi,0]$.  We see that, from this perspective, the $\phi-\bar\phi$ modulations capture the vast majority of the correlation shifts over much of the production phase space.  The only exceptions are in the high-$p_T$ limit for the CMDM (above roughly $400$~GeV), and for high-$\beta^2$, intermediate-angle production for the CEDM.  These are exactly the regions where the $xz$ and $yz$ correlations are naively beginning to diverge.  While it is still possible to define a more powerful probe of the correlation shifts using the above matrix formalism (or related ``optimal observable'' methods~\cite{Atwood:1992vj,Zhou:1998wz,Atwood:2000tu,Sjolin:2003ah,Hioki:2012vn}), the extreme simplicity and high sensitivity of the single variable $\phi-\bar\phi$ strongly motivate us to consider probing both dipole moments simultaneously using azimuthal angle measurements.

While we have been focusing on $gg \to t\bar t$ production, which is the main production channel for tops at the LHC, we cannot neglect the impact of the CMDM and CEDM on $q\bar q \to t\bar t$ production.  At high-$p_T$, $q\bar q$ contributes $O$(30\%) of the total rate.  We account for this in detail when we estimate measurement prospects below.  Here, we simply comment that the effects on the $\phi-\bar\phi$ modulation are typically about half as large, and come in with opposite sign.  Adding in $q\bar q$ therefore dilutes the final modulations at the LHC by as much as a factor of $O$(2) relative to pure $gg$.  For the CMDM, this cancellation formally reduces the sensitivity of an azimuthal angle measurement compared to a total cross section measurement, as the latter's slope has the same sign (and similar magnitude) for both production mechanisms.  Nonetheless, we emphasize that a variable that undergoes a simple {\it shape} distortion may ultimately win out, especially when statistics begin to allow us to probe down to percent-level effects or smaller.  (The luminosity uncertainty at the LHC is itself a few percent.)  For the CEDM, it is also formally possible to improve the measurement, but it is not clear that the added complexity would genuinely pay off.  In particular, we note that for central production, the {\it entire} effect is captured by $\phi-\bar\phi$, and there is no way to escape the fact that $gg$ and $q\bar q$ partially cancel each other.  

For both electric and magnetic dipoles, one might hope that the Tevatron might avoid this cancellation as $t \bar t$ pairs there are produced overwhelmingly through $q \bar q$ annihilation.  However, we find the dominant effects for the CMDM are typically not fully captured by azimuthal-difference modulations about the $z$-axis, but rather by azimuthal-sum modulations about the $x$-axis.  For non-central tops, most of the CEDM effect is captured by azimuthal-difference modulations about the $x$-axis.  It would be interesting to explore a detailed measurement of these effects, but we do not undertake this here.  Still, it serves as yet another example of the perspective offered by taking a more general approach to the spin correlations.

\subsection{Broad parity-violating resonance}
\label{sec:resonance}

Here, we consider a new spin-one color-octet particle that couples to light quarks and top quarks as
\beq
\Delta {\mathcal L}  \,=\, g_s A_\mu^a \, \Big(   \bar q \left[ T^a \gamma^\mu  (v_{q} + \gamma^5 a_{q}) \right]q   + \bar t \left[ T^a \gamma^\mu (v_{t} + \gamma^5 a_{t}) \right]t \Big).
\eeq
We have studied this model before in~\cite{Baumgart:2011wk}.  However, there we assumed a narrow resonance with large $S/B$, so that interference effects with QCD were subdominant and the chirality structure of the light quark couplings was not apparent.  In that case, we get C- and P-violation when $v_t a_t \ne 0$, which leads to (identical) net polarizations of the top and antitop in the $xz$-plane.  While the correlation matrix has nontrivial functional dependence on $v_t$ and $a_t$, none of the P-violating entries become nonzero.  When we include interference with QCD, these entries can all turn on, yielding novel symmetry-violating effects in spin observables.  In particular, nonvanishing $xy$ and $yx$ entries would induce a sinusoidal component of the $\phi+\bar\phi$ modulation.  To the best of our knowledge, this type of P-violation has never been studied before for top quarks.

As a concrete example, we add a resonance with $(v_q,a_q) = (0.05,0)$, $(v_t,a_t) = (0,5)$, $M = 500$~GeV, and $\Gamma/M = 0.5$.  Such a resonance may serve to parametrize the effects of TeV-scale composite physics that couples strongly to top quarks, and only couples to light quarks through kinetic mixing with the gluon.  If the main decay channel was through top quarks, we would only expect $\Gamma/M \sim 0.2$.  Therefore, the unusually broad width requires additional decay channels, perhaps into high-multiplicity jets via new (on- or off-shell) colored particles \cite{Tavares:2011zg,Gross:2012bz}.

\begin{figure}[tp]
\begin{center}
\epsfxsize=0.44\textwidth\epsfbox{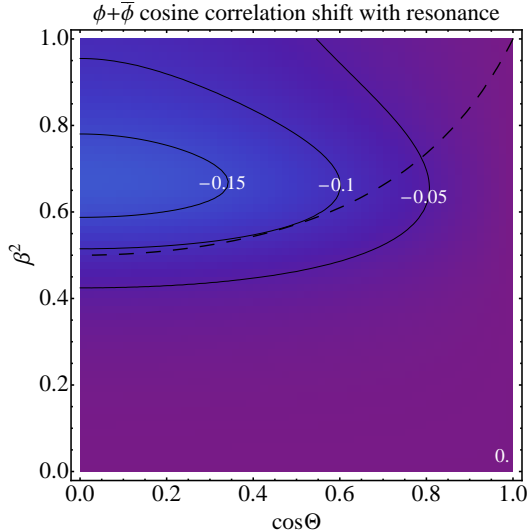}
\caption{\it Shift in the strength of the $\phi+\bar\phi$ cosine correlation in the presence of our example resonance, in $q\bar q \to t\bar t$.  Plotted versus top production angle and squared-velocity in the partonic CM frame.  (Dashed line indicates $p_T = m_t$.)} 
\label{fig:resonance1}
\end{center}
\end{figure}

\begin{figure}[tp]
\begin{center}
\epsfxsize=0.44\textwidth\epsfbox{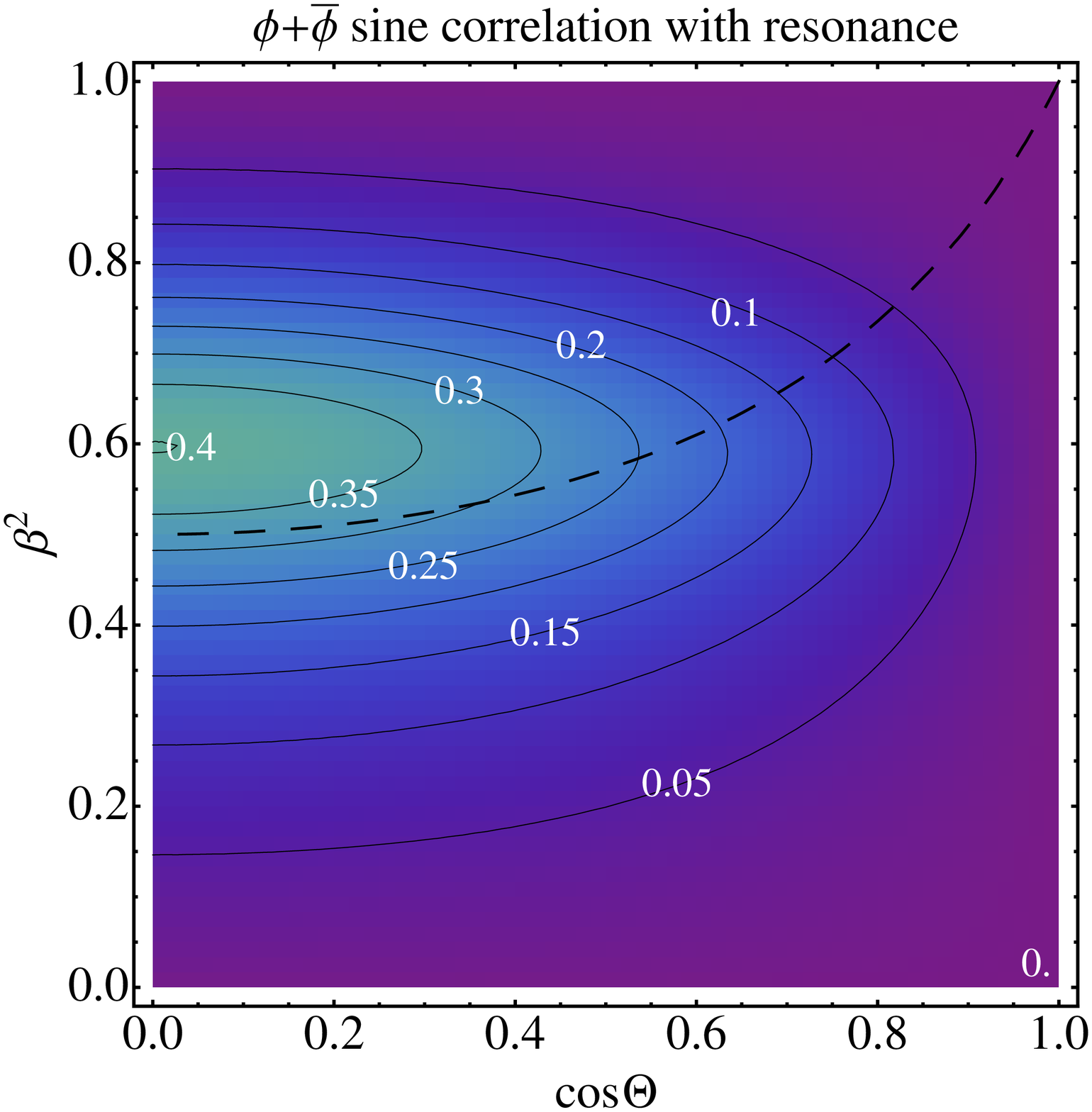}
\epsfxsize=0.44\textwidth\epsfbox{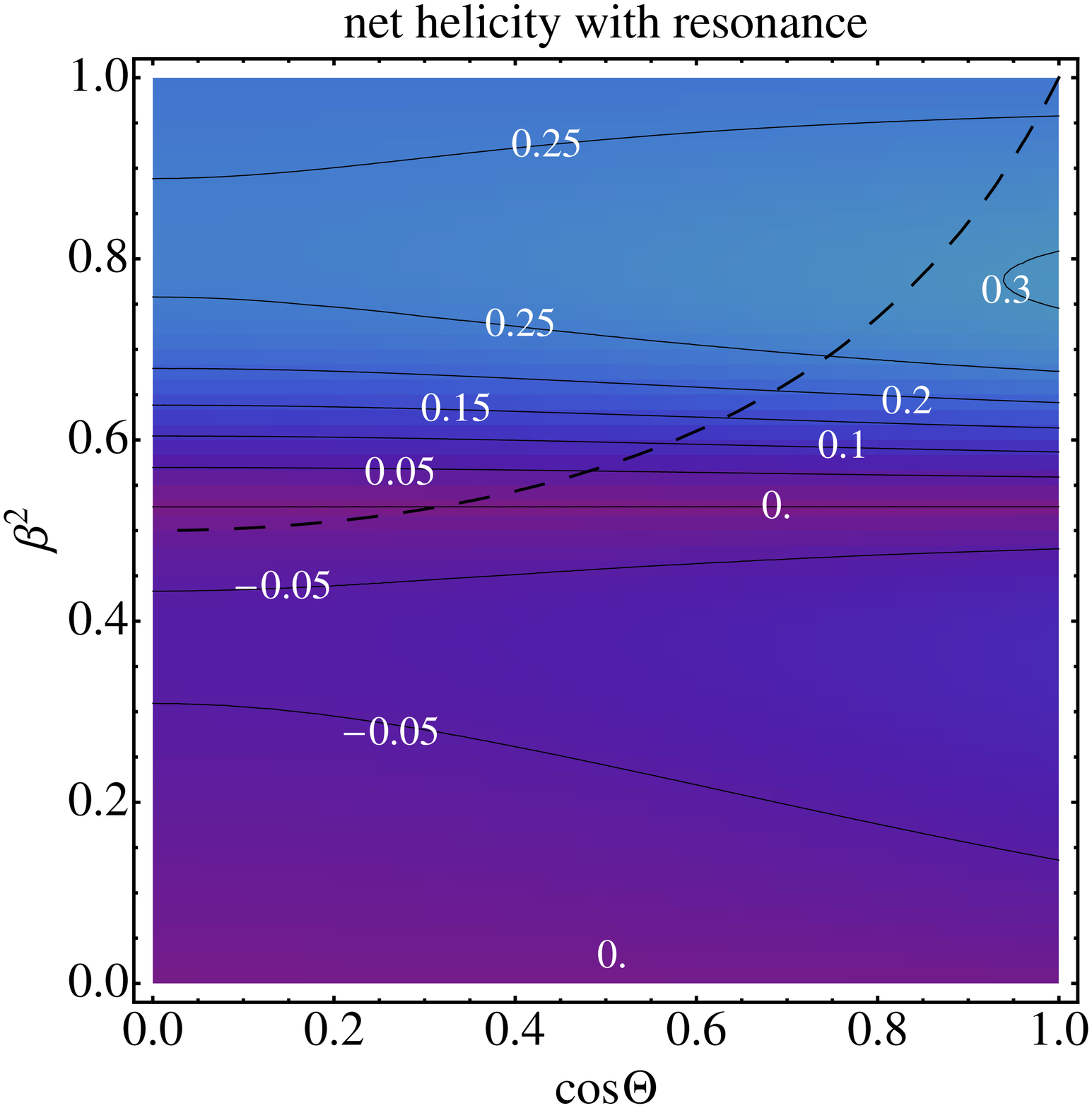}
\caption{\it Induced strength of the $\phi+\bar\phi$ sine correlation (left) and net top quark helicity (left) in the presence of our example resonance, in $q\bar q \to t\bar t$.  Plotted versus top production angle and squared-velocity in the partonic CM frame.  (Dashed line indicates $p_T = m_t$.)} 
\label{fig:resonance2}
\end{center}
\end{figure}

The small coupling to light quarks and the lack of a tightly-localized resonance peak should ensure that this particle escapes detection in dijet searches.  Our choice of parameters also keeps it well-hidden in the $t\bar t$ invariant mass spectrum at the Tevatron and LHC~\cite{CDFresonance,Abazov:2011gv}.  The relative modifications to the $q\bar q \to t\bar t$ rate are modest, vanishing near threshold and with a gradual turn-on that peaks at 15\% near 600~GeV.  Nonetheless, the effects on individual top spins and spin correlations are significant.  By far the dominant contributions appear in the $xy$-block of the correlation matrix and in the $z$-polarization.  The $zz$ entry, which leads to the standard polar-polar correlation, is nowhere modified by more than 0.1 from the SM, and typically less than 0.05.  The main effects on $t\bar t$ decay kinematics are a decrease of the $\phi+\bar\phi$ cosine-modulation, the appearance of the advertised $\phi+\bar\phi$ sine-modulation, and linear biases in the polar decay angles $\cos\theta$ and $\cos\bar\theta$.  We plot the strengths of these three effects in Figs.~\ref{fig:resonance1} and~\ref{fig:resonance2} (as above, normalizing to distribution asymmetries multiplied by two).  Notably, the net polarization flips sign as we pass through the resonance, while the azimuthal correlation effects do not.

While the P-violation from this model can be seen in the top quark polarization at
high-energy, the presence of a comparable (in many regions larger) effect from
P-violating spin correlations presents us with an interesting opportunity.  We study
the prospects for measuring these correlations at the LHC in the next section.  We
note that they may also be observable at the Tevatron, though we have not undertaken
a dedicated study here.

The inclusive polarization induced in $q\bar q \to t\bar t$ at the LHC is only 6\%, and is highly diluted by the much larger unpolarized $gg \to t\bar t$ contribution.  This small size allows the model to evade existing constraints from polarization measurements at the LHC~\cite{CMSspin,ATLASspin}.  The inclusive effects at the Tevatron would be much weaker than 6\%.  In either case, the polarization might be revealed by using harder cuts.

\section{Azimuthal Spin Correlation Measurement}
\label{sec:measurement}

We now turn to the question of whether the azimuthal sum/difference spin correlations from QCD can be measured at the LHC.  We find that these correlations can be revealed with quite good accuracy even using fairly simple event reconstruction strategies.  We demonstrate this point in a set of proof-of-principle measurements carried out on simulation data, showing that the evolution from low-$p_T$ production to high-$p_T$ production should be clearly observable with the current LHC data set.  We also discuss the observability of the new physics effects from color-dipoles or a broad parity-violating resonance.

Our simulations incorporate many realistic effects such as parton showering and hadronization, jet reconstruction, $b$-tagging and mistagging efficiencies, and finite energy/directional resolution (described in detail in Appendix~\ref{sec:details}).  While we cannot accurately represent all of the effects that afflict particle detection and reconstruction, we believe that our simple model gives an adequate first approximation.

Using \MadGraph5~\cite{Alwall:2011uj} interfaced with \PYTHIA~\cite{Sjostrand:2006za}, we generate simulated samples of $t\bar t$ at LHC8, and samples of the dominant backgrounds (also detailed in Appendix~\ref{sec:details}).  The $t\bar t$ simulations include separate samples where the top decays are either handled individually by \PYTHIA\ or using the full 6-body production/decay matrix elements via \MadGraph5's decay-chain functionality.  These furnish ``uncorrelated'' and ``correlated'' samples which we compare.

We study the effects of the correlations in both dileptonic and $l$+jets decay channels.  There are various advantages and disadvantages to each.  In principle, dileptonic is ideal for spin correlation measurement due to the leptons' maximal spin analyzing powers.  However, the presence of two neutrinos significantly complicates kinematic reconstruction, and the overall rate is low due to the 5\% dileptonic branching fraction.  The $l$+jets channel is naively easier to fully reconstruct, and comes with a higher branching fraction of 30\%.  The former can especially be useful if we are interested in some very specific region of production kinematics, such as near a resonance.  But we pay a penalty in analyzing power:  the best that we accomplish on the hadronic side is to either pick the $b$-jet ($\kappa = -0.4$), or the softest non-$b$ jet in the top's rest frame ($\kappa = 0.5$).  The promise of complete kinematic reconstruction and much higher statistics are also not so immediately delivered upon, due to imperfect $b$-tagging, combinatoric ambiguities, jet overlaps, and plentiful jet-like contamination in the events.  Nonetheless, as we will find, the dileptonic and $l$+jets channels both yield observable effects with high statistical significance after complete event reconstruction.

\subsection{Dileptonic}

For this channel, the central challenge is to reconstruct the individual top momenta.  The two neutrinos make this process nontrivial, as four degrees of freedom are unmeasured by the detector, and we need to recover them without inducing spurious kinematic correlations.  We must also devise an approach for deciding how to pair $b$-jets with the leptons.

Our aim is to provide a sufficiently accurate and unbiased reconstruction such that we can clearly observe the correlations in $\phi\pm\bar\phi$.  As a proof of concept, we have taken a straightforward approach inspired by the methods in~\cite{Baringer:2011nh,Choi:2011ys}.  More involved procedures using detailed kinematic constraints~\cite{Sonnenschein:2006ud} might be able to achieve greater sensitivity than what we claim here.  Nonetheless, the three-body nature of the top decay provides us with a large portion of the relevant kinematic information in fully visible particles, and complete reconstruction of top decay angles is not necessarily very sensitive to how we treat the neutrinos.  Moreover, azimuthal decay angle measurements are highly forgiving; for example, they are independent of the absolute top velocities in the reconstructed CM frame.  Though we do not exploit it here, the necessary kinematic reconstructions can also be much simpler in the case of highly-boosted tops~\cite{Baumgart:2011wk}.

For our present study, we select events with exactly two oppositely-charged leptons, assumed to come from the top decays, and at least two jets.  To help control backgrounds (mainly $Z^{(*)}$+jets), we reject events with a same-flavor lepton pair with $m(l^+l^-) = [80,100]$~GeV, and also demand at least one $b$-tag and \met\ $> 30$~ GeV.  We chose as our $b$-jet candidates either the two hardest $b$-tagged jets or, if only one jet is tagged, the $b$-tagged jet and the hardest untagged jet.  We reconstruct the individual top systems by computing \mttwo~\cite{Lester:1999tx,Barr:2003rg}, constructed out of the two leptons, two $b$-jets, and \vecmet.  We consider both possible assumptions of pairings for the $b$'s and leptons for this calculation.  The solution of \mttwo\ presents us with an educated guess for the individual neutrino $\vec{p}_T$'s, and we further provide a guess for their $p_z$'s by matching each neutrino solution's rapidity with that of the four-vector of its associated $b$+lepton system.\footnote{We can alternatively use kinematic constraints to solve for $p_z(\nu)$ and $p_z(\bar\nu)$ after estimating the neutrinos' transverse momenta through \mttwo.  We find that this induces extra bias into the azimuthal angle distributions, at the few percent level.  Using the full method of \cite{Choi:2011ys}, including \mttwo\ for the $W$'s induces biases at the 10--20\% level.}  This choice minimizes the full invariant mass of each top candidate.  To decide which of the two jet-lepton pairings to choose, we use the following procedure, moving on to the next step if the previous one is inconclusive:
\begin{enumerate}
	\item If one pairing has $m(bl) >$ 151 GeV, and the other does not, take the latter, as in~\cite{Baringer:2011nh}.
	\item The value of \mttwo\ is strictly bounded by $m_t$.  Thus if one pairing has \mttwo$>m_t$, but the other does not, take the latter.
	\item Lastly, take the pairing that minimizes the quantity $(m(bl^+\nu) - m_t)^2 + (m(\bar bl^-\bar\nu) - m_t)^2$.
\end{enumerate}
If either reconstructed top has a mass exceeding 200~GeV, we throw out the event.  (This last cut removes only about 5\% of events.)  In our simulation sample, 60\% of the selected events were paired correctly.  Of the 40\% of incorrect pairings, in 11\% one or both $b$-jets failed basic reconstruction, and in 7\% both $b$-jets were present but not correctly picked as our candidates.  Therefore, in cases where our procedure was given the two correct $b$-jets as input, it successfully paired them with a rate better than 70\%.  We observe that the success rate increases with truth-level top $p_T$.

\begin{figure}[tp]
\begin{center}
\epsfxsize=0.44\textwidth\epsfbox{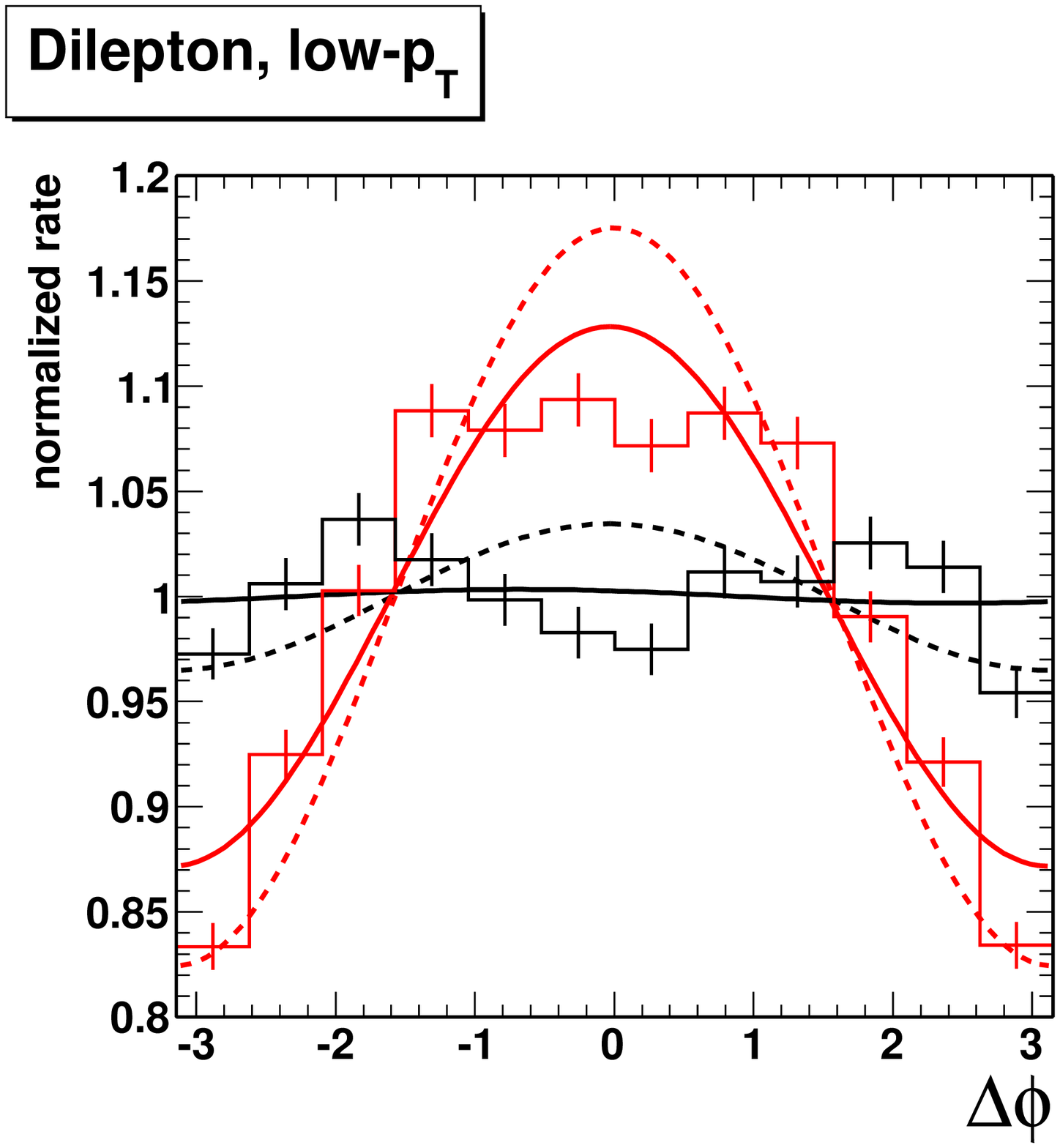}
\epsfxsize=0.44\textwidth\epsfbox{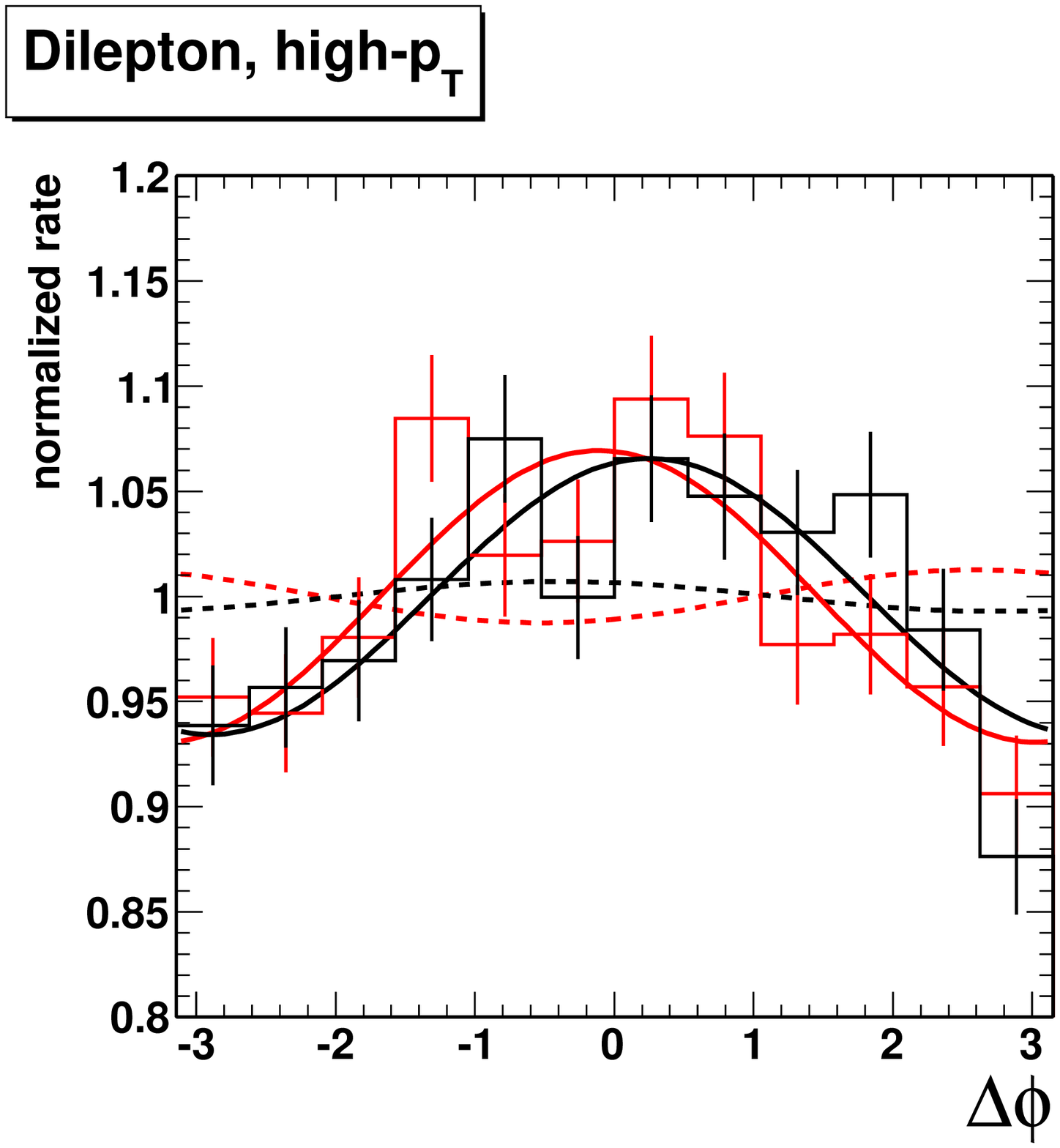}
\epsfxsize=0.44\textwidth\epsfbox{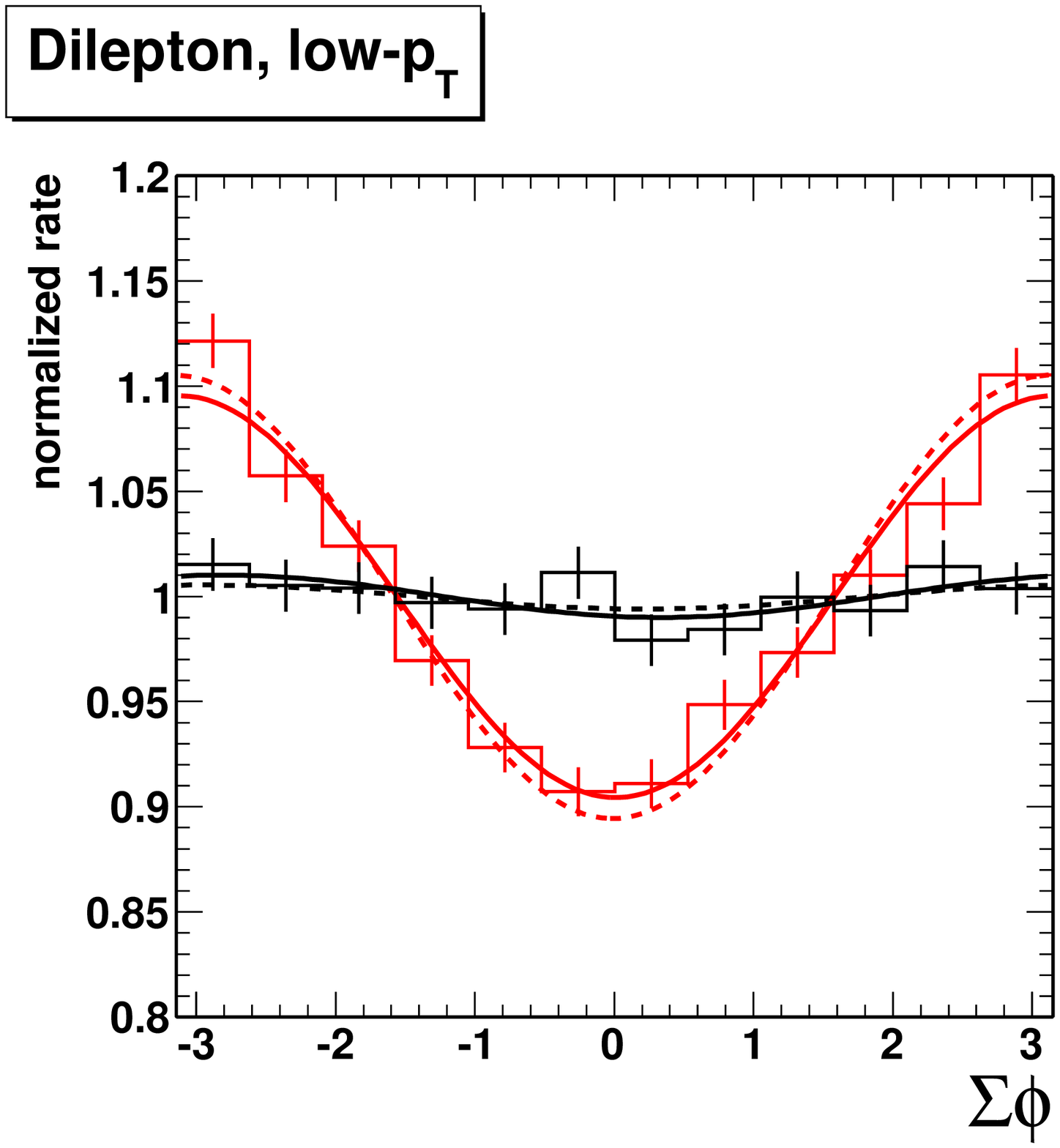}
\epsfxsize=0.44\textwidth\epsfbox{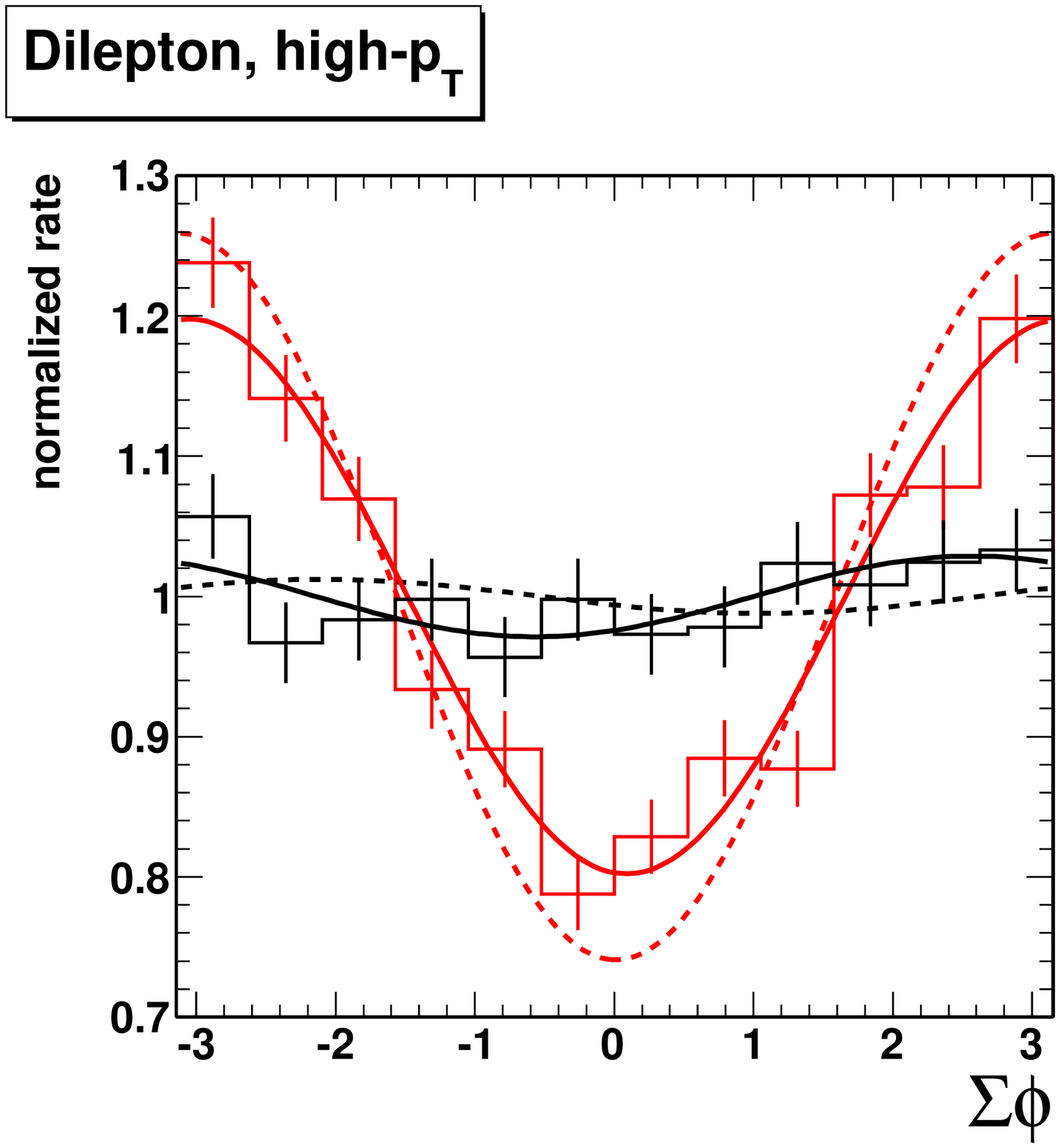}
\caption{\it Distributions of the azimuthal difference and sum for the dileptonic channel at LHC8: low-$p_T$ $\phi-\bar\phi$ (upper-left), high-$p_T$ $\phi-\bar\phi$ (upper-right), low-$p_T$ $\phi+\bar\phi$ (lower-left), high-$p_T$ $\phi+\bar\phi$ (lower-right).  The solid black and red histograms are uncorrelated and correlated simulations with full reconstruction.  (Error bars are monte carlo statistics, with an effective sample luminosity of 34~fb$^{-1}$.)  The solid curves are two-parameter fits using a flat distribution plus independent sine and cosine modulations.  The dashed curves are fits of the parton-level samples with identical acceptance cuts.}
\label{fig:llModulations}
\end{center}
\end{figure}
\begin{table}[tp]
 \centering
\begin{tabular}{l|r|c|c|c}
     LHC8, 20 fb$^{-1}$      & \; \# events \; & \; stat error \; & \; $\phi-\bar\phi$ amp uncorr/corr \; & \; $\phi+\bar\phi$ amp uncorr/corr \; \\ \hline
low-$p_T$  & 47,600 \; & 0.7\%      &  0.3\% / 12.8\% (18$\sigma$)  &  -1.0\% /  -9.6\% (12$\sigma$) \\
high-$p_T$ &  8,400 \; & 1.6\%      &  6.4\% / 6.9\% (0.3$\sigma$)  &  -2.4\% / -19.7\% (11$\sigma$)
\end{tabular}
\caption{\it Cosine modulation amplitudes for azimuthal correlation observables, and expected number of events, statistical error, and significances for LHC8 at 20~fb$^{-1}$.  Sine modulation components would have the same statistical errors, and central values are consistent with zero.  The two $p_T$ regions are determined by whether $|p_{T}(bl^+\nu)| + |p_{T}(\bar bl^-\bar\nu)|$ is less than or greater than $2m_t$.}
\label{tab:llResults}
\end{table}

We plot the distributions and fits for the azimuthal sum/difference variables in Fig.~\ref{fig:llModulations} and compare these to parton-level events with perfect reconstruction but subject to basic acceptance cuts.  We have divided the final-state phase space into two regions based on the scalar-summed $p_T$ of the two reconstructed tops:  a ``low-$p_T$'' region with $|p_T(bl\nu)|+|p_T(\bar bl\bar\nu)| < 2m_t$, and a ``high-$p_T$'' region with $|p_T(bl^+\nu)|+|p_T(\bar bl^-\bar\nu)| > 2m_t$.  The expected azimuthal modulations are fairly faithfully preserved, with most of the effect in $\phi-\bar\phi$ at low-$p_T$, and in $\phi+\bar\phi$ at high-$p_T$.  The biases induced by our simple reconstruction are typically modest, though a roughly 6\% modulation is induced in $\phi-\bar\phi$ at high-$p_T$.  In Tab.~\ref{tab:llResults}, we summarize the fitted amplitudes and their expected statistical errors given 20~fb$^{-1}$ of data at LHC8.  The effects of the correlation are in principle measurable with very high statistical significance, in excess of 10$\sigma$.

We also study the mixed polar-azimuthal correlations, involving the matrix elements $C^{13}$ and $C^{31}$.  We expect and find that the polar distributions suffer from significant acceptance bias, as the polar decay angles are strongly correlated with lepton $p_T$.  The individual $\phi$ and $\bar\phi$ distributions are similarly highly distorted by basic acceptance criteria (as we will illustrate in more detail for $l$+jets in the next subsection).  We therefore do not attempt sinusoidal fits of the $\phi'$ and $\bar\phi'$ distributions of Eq.~\ref{eq:PolarAz}, but instead default to measuring their asymmetries between the regions $[0,\pi/2]$ and $[\pi/2,\pi]$.  The asymmetry difference between correlated and uncorrelated samples is roughly 2\% for both $p_T$ regions, with expected statistical uncertainties of 0.4\% (low-$p_T$) and 1\% (high-$p_T$) at LHC8.  The reconstruction-induced asymmetry biases are percent or smaller.\footnote{We can also study the traditional polar-polar correlations, probing $C^{33}$, by measuring asymmetries in $\cos \theta \cos \bar \theta$.  We find that the correlations induce asymmetries of -3.9\% and 3.7\% on top of biases of 9.3\% and 2.7\%, for low-$p_T$ and high-$p_T$, respectively.  The statistical uncertainties are the same as for the asymmetries in $\cos \phi'$, and the effects are therefore highly significant.  However, we emphasize that our own analysis was not optimized to measure these effects, and alternative reconstruction strategies might obtain better sensitivity.}  It may therefore be possible to establish the presence of even this small correlation effect in the low-$p_T$ region with high significance, or perhaps more importantly to place tight limits on possible anomalous contributions.

Fully realistic measurements must contend with systematic uncertainties and backgrounds.  We are not in a position to fully understand systematics or the quality of background subtraction, but we can at least check that the effects of the backgrounds are small.  We include $tW$, $W^+W^-$+jets, $l^+l^-$+jets, and $\tau^+\tau^-$+jets.  (Simulations are described in detail in Appendix~\ref{sec:details}.)  Already our basic reconstruction cuts are sufficient to largely eliminate $tW$ and $W^+W^-$+jets, and our additional selections on $m(l^+l^-)$, \met, and $b$-tagging highly reduce $l^+l^-$+jets and $\tau^+\tau^-$+jets.  The largest surviving background is $l^+l^-$+jets.  It will be crucial to keep this background under control, as the tendency of the leptons to be back-to-back in the transverse plane leads to an $O(30\%)$ bias in its reconstructed $\phi - \bar \phi$ distribution in our low-$p_T$ region.  However, after all cuts the overall rate for this background is only 3\% of $t\bar t$, and the final effect on the amplitude is expected to be only $O(1\%)$.

We have also checked the stability of the difference/sum modulations to physics modeling.  We compare the modulations, both correlated and uncorrelated, in $t\bar t$ samples generated in \MadGraph5\ from simple two-body production followed by parton showering, as well as production matched up to two additional jet emissions via $k_T$-MLM.  The fitted distributions all agree at the 1\% level.

Lastly, we comment on how these measurements can be used to probe for color-dipole moments or broad parity-violating resonances, as discussed in Section~\ref{sec:NP}.  We measure the dipole moments using the $\phi-\bar\phi$ distribution.  We find good sensitivity to the CMDM using an inclusive sample, and good sensitivity to the CEDM with a cut of $m(t\bar t) > 500$~GeV ($\beta^2 > 0.5$).  The induced cosine- and sine-modulations have respective amplitudes of $0.40(\mu\times m_t)$ and $0.30(d\times m_t)$.  With 20~fb$^{-1}$ of data at LHC8, we find 2$\sigma$ sensitivity to $\mu\times m_t \simeq 0.03$ and $d\times m_t \simeq 0.05$.  In the next run of the LHC, with an expected beam energy near 13~TeV and integrated luminosity above 100~fb$^{-1}$, these limits would extend down to 0.01 or smaller.  For our example resonance model of~\ref{sec:resonance}, we fit the induced sine-wave amplitude in $\phi+\bar\phi$, taking events with $|p_{T}(t)| + |p_{T}(\bt)| \,>\, 250$ GeV and $|\cos \Theta| \,<\, 0.5$.  The amplitude is 3.4\%, with a significance of 2.5$\sigma$.

\subsection{$l$+jets}

For this analysis we demand, in addition to our basic acceptance cuts, that the event contain at least four jets, at least one of which is $b$-tagged, and that leptonic and hadronic tops can be fully reconstructed.  To reconstruct the tops, we iterate over all possible partitions of the lepton and jets into a leptonic top ($l\nu j$) and a hadronic top ($jjj$), including separately each of the two possible neutrino solutions.\footnote{If $m_T(l,$\met$) > m_W$, we use a single solution $\eta(\nu) \equiv \eta(l)$ and rescale $p_T(\nu)$ so that $m_T(l,\nu) \equiv m_W$.}  Each candidate set of the six objects must contain at least one $b$-tagged jet, and in sets with more than one $b$-jet we require that the leptonic and hadronic top individually contain at least one.  We choose the partitioning that minimizes $(m(l\nu j)-m_t)^2 + (m(jjj)-m_t)^2$.  Within the hadronic top, we further resolve the kinematics.  If the hadronic top contains one $b$-tagged jet, we take the two remaining jets to reconstruct the hadronic $W$.  Otherwise, we pick the two jets that best reconstruct the $W$ mass, and identify the third as the ``$b$-jet."  To ensure good quality reconstruction and to reduce backgrounds, we further demand that $m(jjj) = [130,200]$~GeV, $m((jj)_W) = [60,100]$~GeV, and $p_T(l\nu jjjj)/m(l\nu jjjj) < 0.1$.

To construct our spin correlation observables, we use the lepton from the leptonic side and the less energetic of the two $W$ jets in the hadronic top's rest frame.  However, we note that we get similar results (albeit with reversed modulations and somewhat smaller amplitudes) by picking the hadronic top's $b$-jet.  In a real measurement, these two choices, which have similar sensitivity, can serve as excellent cross-checks of one another.

\begin{figure}[tp]
\begin{center}
\epsfxsize=0.44\textwidth\epsfbox{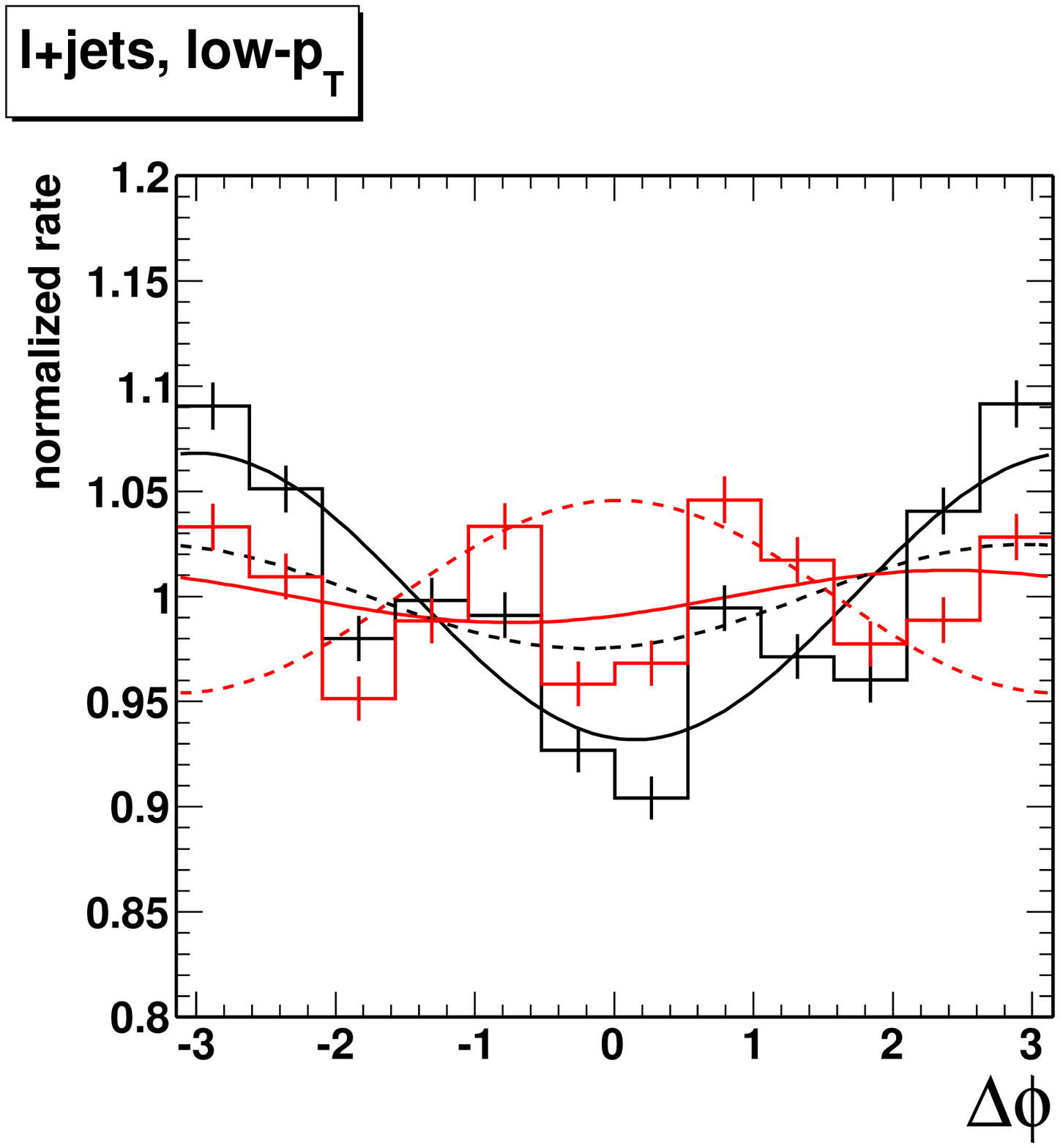}
\epsfxsize=0.44\textwidth\epsfbox{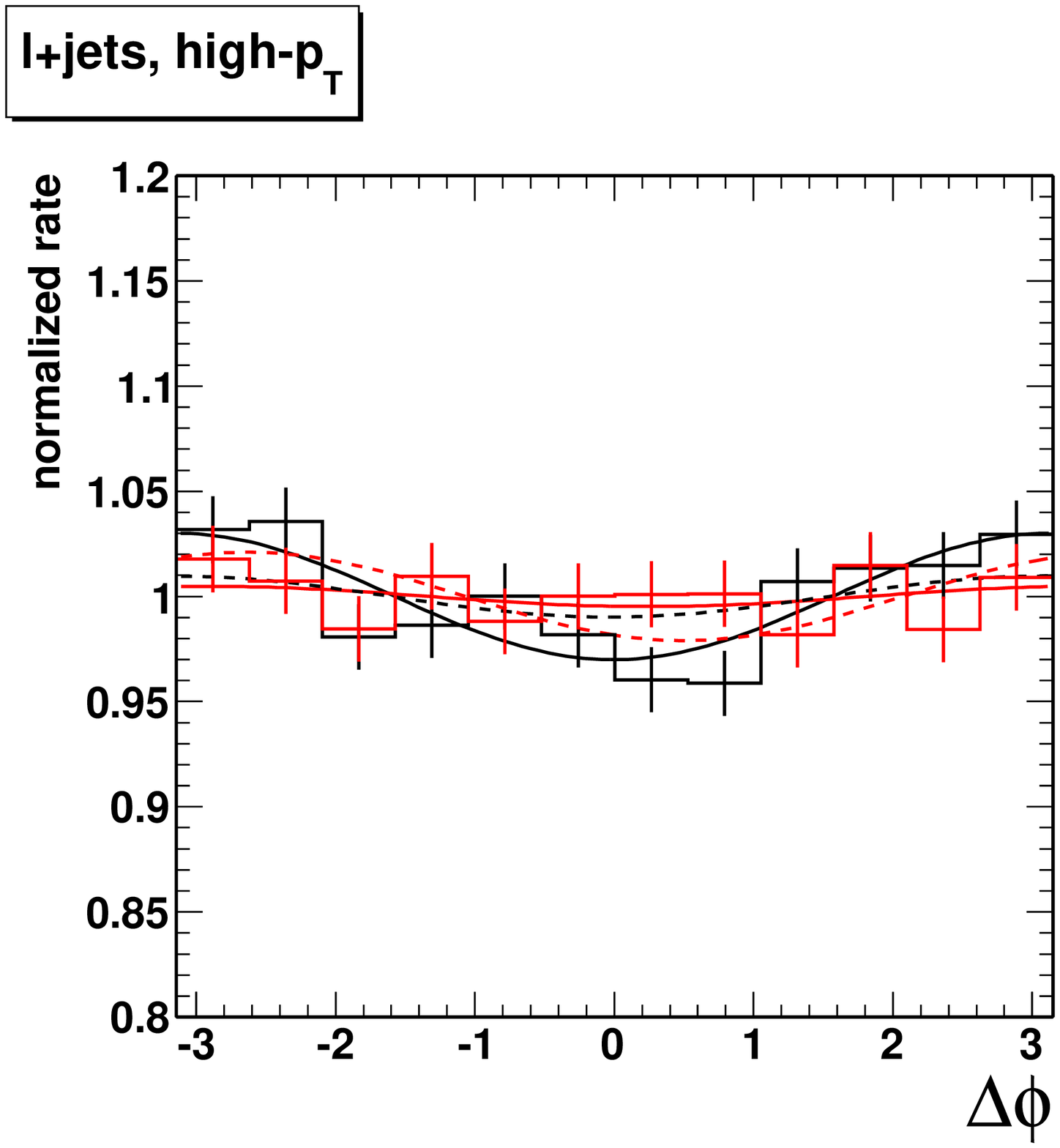}
\epsfxsize=0.44\textwidth\epsfbox{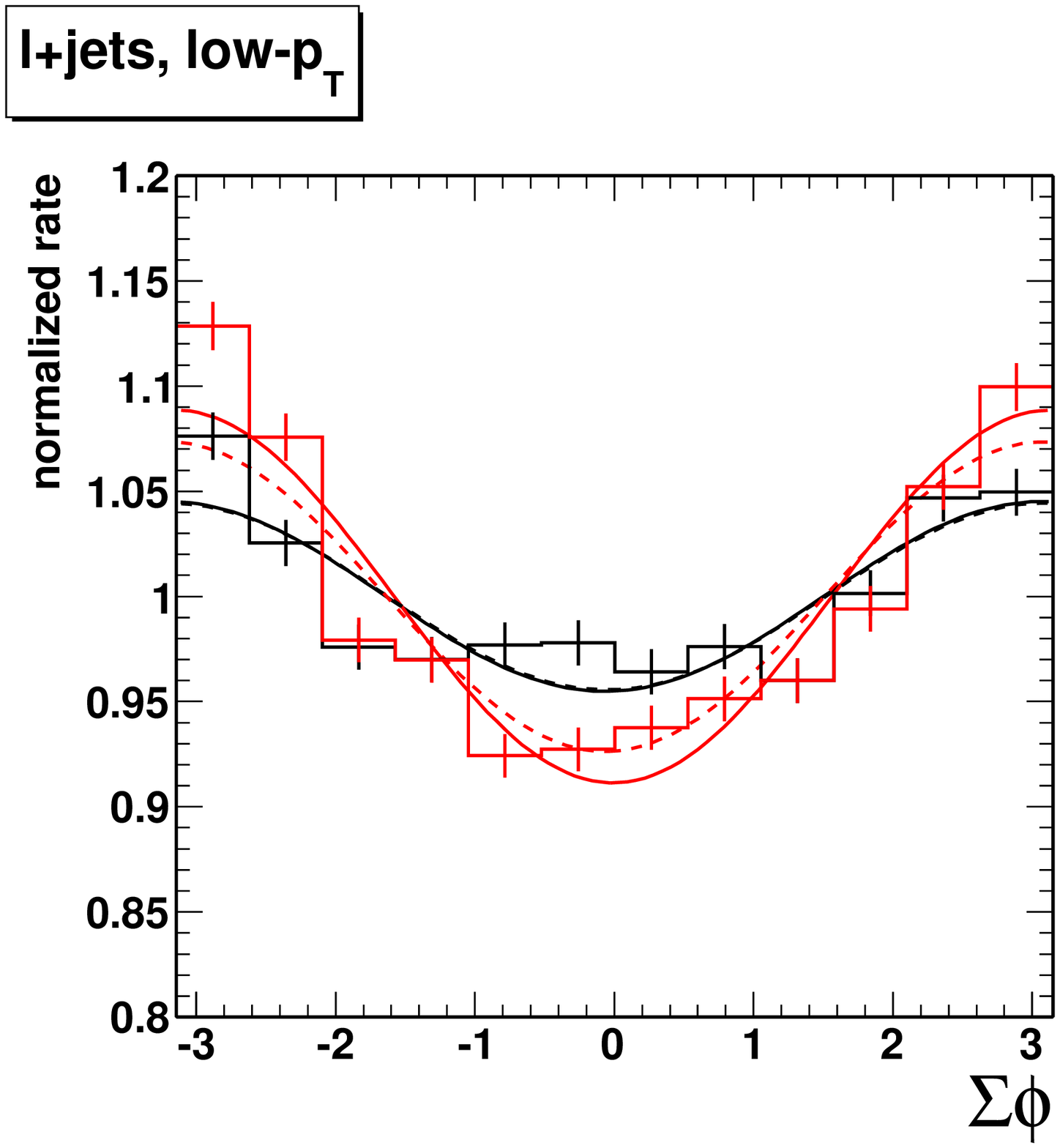}
\epsfxsize=0.44\textwidth\epsfbox{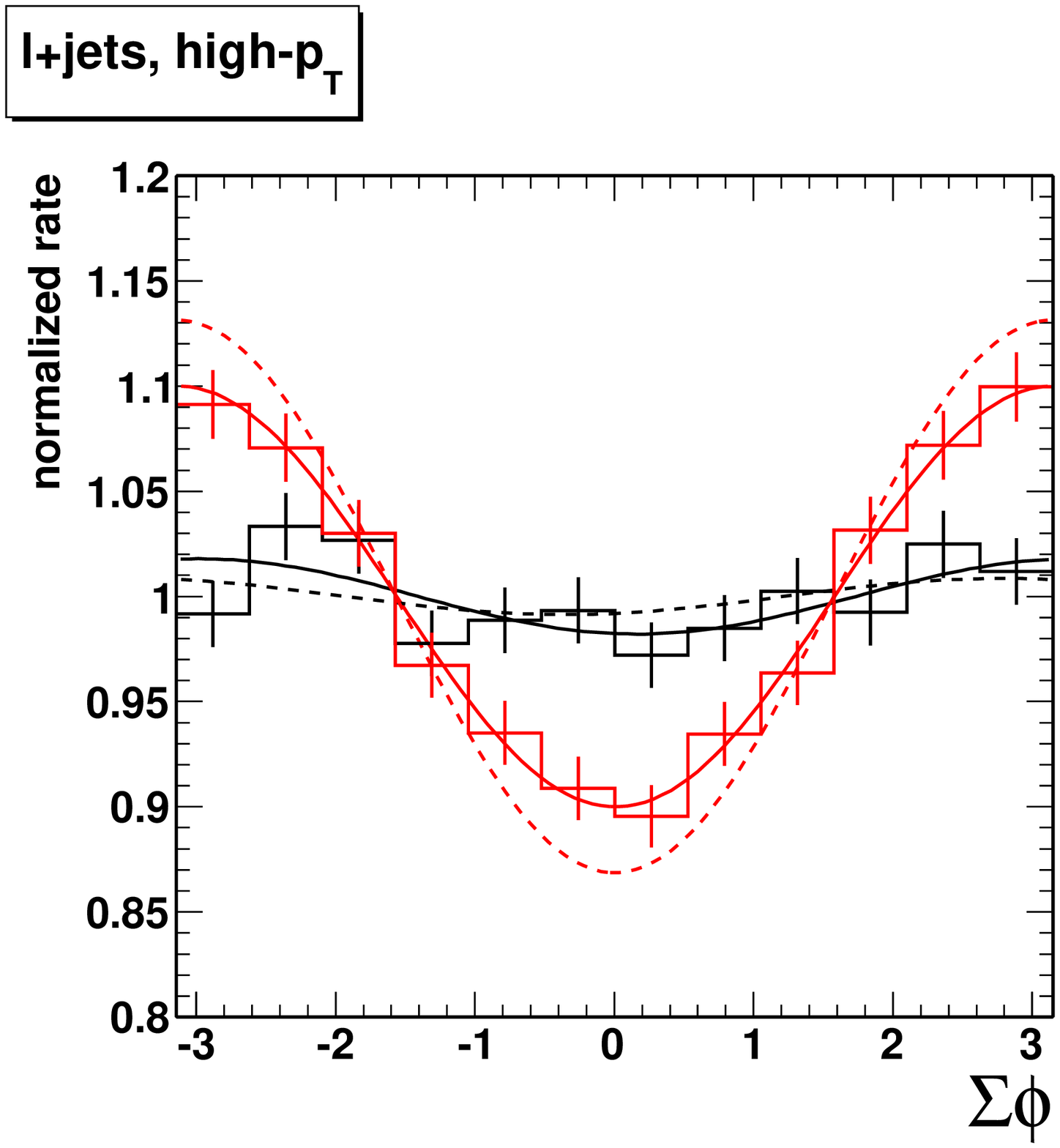}
\caption{\it Distributions of the azimuthal difference and sum for $l$+jets at LHC8: low-$p_T$ $\phi-\bar\phi$ (upper-left), high-$p_T$ $\phi-\bar\phi$ (upper-right), low-$p_T$ $\phi+\bar\phi$ (lower-left), high-$p_T$ $\phi+\bar\phi$ (lower-right).  The solid black and red histograms are uncorrelated and correlated simulations with full reconstruction.  (Error bars are monte carlo statistics, with an effective sample luminosity of 29~fb$^{-1}$.)  The solid curves are two-parameter fits using a flat distribution plus independent sine and cosine modulations.  The dashed curves are fits of the parton-level samples with identical acceptance cuts.}
\label{fig:ljetsModulations}
\end{center}
\end{figure}

We split the production phase space into ``low-$p_T$'' and ``high-$p_T$'' regions, respectively defined by $p_T(jjj) < 150$~GeV and $p_T(jjj) > 150$~GeV.  In Fig.~\ref{fig:ljetsModulations}, we show the $\phi\pm\bar\phi$ distributions from our $t\bar t$ samples.\footnote{Note that for $l^+$+jets the variables are $\phi(l^+)\pm\bar\phi({\rm softer\;}j_W)$, and for $l^-$+jets they are $\phi({\rm softer\;}j_W)\pm\bar\phi(l^-)$.  We assume CP-conservation in top decay, and do not study these two cases individually.}$^,$\footnote{The composition of these samples in $gg\to t\bar t$ ($q\bar q\to t\bar t$) is 80\% (20\%) for low-$p_T$ and 70\% (30\%) for high-$p_T$.}  For comparison, we also display fits to parton-level results with perfect reconstruction but identical acceptance cuts.   (These include a criterion, $\Delta R > 0.4$ for all $lj$ and $jj$ pairings, to reproduce the effects of lepton isolation and jet clustering.)  As expected, the effects of the correlations are smaller than in the dileptonic channel.  There are also some clear acceptance biases introduced by the cuts in the low-$p_T$ region.  Our jet-level kinematic reconstruction largely traces this bias for $\phi+\bar\phi$, but can deviate by almost 10\% for $\phi-\bar\phi$.  Still, the separation between correlated and uncorrelated tops is largely preserved.

\begin{figure}[tp]
\begin{center}
\epsfxsize=0.44\textwidth\epsfbox{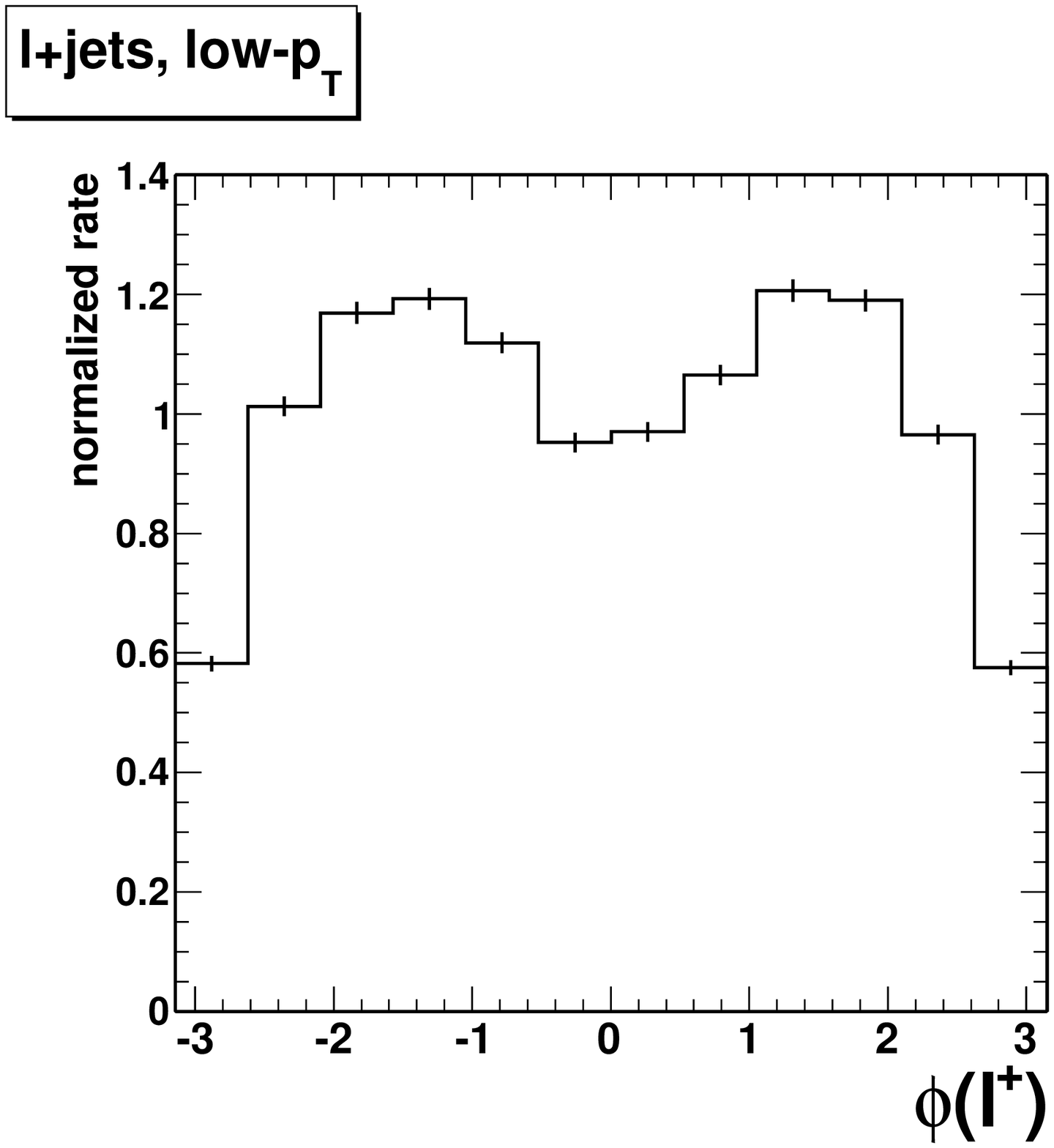}
\epsfxsize=0.44\textwidth\epsfbox{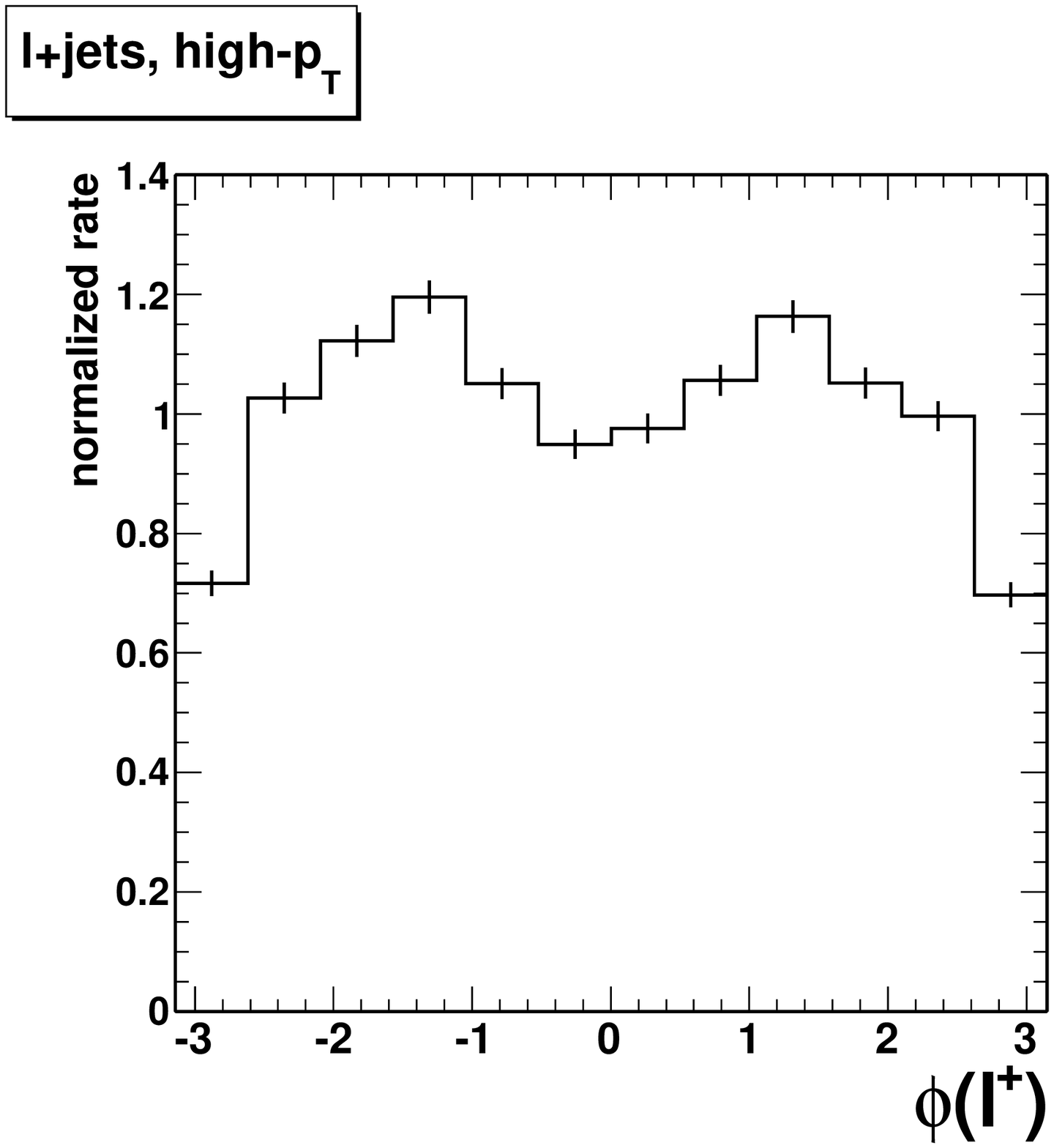}
\caption{\it Azimuthal angle distributions for positively-charged leptons in $l$+jets after all acceptance cuts: low-$p_T$ (left) and high-$p_T$ (right).  (Error bars are monte carlo statistics.)} 
\label{fig:ljetsPhil}
\end{center}
\end{figure}

We also illustrate the reconstruction bias induced on the positively-charged leptons' raw $\phi$ distributions in Fig.~\ref{fig:ljetsPhil}.  Negatively-charged leptons display similar distributions with $\phi\to\phi+\pi$, as our orientation of the $x$ and $y$-axes is always by definition tied to the positively-charged top.  Individual jets from the hadronic top also display similar distributions, which also depend on the sign of the leptonic top.\footnote{Generally, decay particles emitted within the production plane ($\phi = 0,\pi$) are less likely to be detected than particles emitted perpendicular to the production plane ($\phi = \pm\pi/2$).  For centrally produced tops, this is easy to understand:  particles emitted perpendicular to the plane are always in the central detector, whereas particles emitted within the plane can end up at very forward angles with small $p_T$.  For tops that are not exactly central, particles emitted toward $\phi=0$ are being shot back into the central part of the detector, whereas those emitted toward $\phi = \pi$ are being shot more toward the beams.}  Despite the rather severe reshaping of these otherwise flat distributions, the bias largely cancels out when forming $\phi\pm\bar\phi$ and combining positive and negative charges.  We take this as some indication that these modulation effects may not be very sensitive to detailed detector acceptances.  However, the distributions of our other correlation-sensitive azimuthal observables, $\phi'$ and $\bar\phi'$, are more directly reshaped in this manner.

The results of our sinusoidal fits to the $\phi\pm\bar\phi$ distributions are summarized in Table~\ref{tab:ljetsResults}, as well as the expected number of events and statistical errors for 20~fb$^{-1}$ integrated luminosity at LHC8.  Modulations should be resolvable with errors at the sub-percent level with respect to the total rate.  Without systematic errors, we predict roughly $10\sigma$ discrimination between correlated and uncorrelated distributions in low-$p_T$ $\phi-\bar\phi$ and high-$p_T$ $\phi+\bar\phi$.

\begin{table}[tp]
 \centering
\begin{tabular}{l|r|c|c|c}
     LHC8, 20 fb$^{-1}$      & \; \# events \; & \; stat error \; & \; $\phi-\bar\phi$ amp uncorr/corr \; & \; $\phi+\bar\phi$ amp uncorr/corr \; \\ \hline
low-$p_T$  & 73,900 \; & 0.5\%      &  -6.7\% / -0.9\% (11$\sigma$)  &  -4.5\% /  -8.9\% (9$\sigma$) \\
high-$p_T$ & 35,300 \; & 0.8\%      &  -3.0\% / -0.5\% (3.1$\sigma$)  &  -1.8\% / -10.0\% (10$\sigma$)
\end{tabular}
\caption{\it Cosine modulation amplitudes for azimuthal correlation observables, and expected number of events, statistical errors, and significances for LHC8 at 20~fb$^{-1}$.  Sine modulation components would have the same statistical errors, and central values are consistent with zero.}
\label{tab:ljetsResults}
\end{table}

We can also consider measurement of the mixed polar-azimuthal correlations via $\phi'$ and $\bar\phi'$ ({\it cf.~}Eq.~\ref{eq:PolarAz}).  The acceptance reshaping makes their distributions less amenable to sinusoidal fits, and the small size of the correlation also means that discrimination of correlated versus uncorrelated is much more difficult.  If we default back to taking asymmetries, the difference between correlated and uncorrelated is about 1\% for low-$p_T$ and 2\% for high-$p_T$ (with acceptance biases of -2\%), and the statistical resolutions are roughly 0.4\% and 0.5\%.  The small correlation effect might therefore be seen up to respective significances of roughly 2.4$\sigma$ and 4$\sigma$.\footnote{By way of comparison, we can also apply the same analysis style to the polar-polar correlations.  The statistical errors are identical.  The differences in low-$p_T$ and high-$p_T$ asymmetries are 3\% and 2\% (with opposite signs, and acceptance bias of 5\%).  While these asymmetries are both small, they would also be visible in principle.  We note that the helicity-basis polar correlations are much more sensitive to our division of top production phase space, and that in particular our dividing line at $p_T \simeq m_t$ sits near the minimum.}

As in the dilepton channel, we again consider the impact of the major backgrounds, which include $W$+jets and single-top (mainly $tW$), as well as $t\bar t$ with leptonic $\tau$ decays.  These each contribute an additional 5--10\% on top of the $t\bar t$ signal.  We have checked that the azimuthal difference/sum modulations in each of these background is at the 10\% level or less relative to their individual rates, and therefore when combined with $t\bar t$ contribute additional modulations at the sub-percent level.  We reach analogous conclusions for the contributions to the $\phi'$ and $\bar\phi'$ asymmetries.  The relative contribution of the backgrounds can be further reduced by requiring at least two $b$-tags, though at a cost of $O(50\%)$ of the signal.  We have also compared our results between matched and unmatched $t\bar t$ simulation samples.  This is a highly nontrivial cross check for $l$+jets, which uses a large number of jets in the reconstruction.  We nonetheless find percent-level agreement between the reconstructed modulations obtained in the different simulations, which suggests that higher-order corrections to our observables may be modest.

Next, we consider the possible impact of new physics.  The CMDM and CEDM respectively introduce cosine- and sine-wave modulations in $\phi-\bar\phi$.  Our simulations indicate that the amplitudes shift by $0.29(\mu\times m_t)$ and $0.27(d\times m_t)$ in our inclusive sample (combining low-$p_T$ and high-$p_T$ subsamples).\footnote{The effects grow somewhat with a top $p_T$ cut, but the inclusive analysis provides better statistical significance.}  This would give us 2$\sigma$ sensitivity to $\mu\times m_t$ and $d\times m_t$ of roughly 0.03 by the end of 2012.  These are comparable to the limits that we obtained with the dileptonic channel above, and a combined measurement over both channels would be warranted.  For our parity-violating resonance example model, we study a subsample with $p_T > 100$~GeV and $|\cos\Theta| < 0.5$, and observe a 2.7\% sine-wave modulation in $\phi+\bar\phi$.  This corresponds to about 4.5$\sigma$ at LHC8.  We thus find that for this scenario, given our particular choices of cuts and reconstruction methods, $l$+jets outperforms dilepton (in which we found the effect to be visible with 2.5$\sigma$ significance).    For both types of new physics, it may also be useful to fold in the independent $l$+jets measurements obtainable by correlating a lepton with a $b$-jet.

Somewhat longer-term, for a projected 13~TeV LHC with 100~fb$^{-1}$, the statistical significances will improve over our 2012 estimates by roughly a factor of four.  For example, dipole strengths weaker than 0.01 will become constrained.  A combined dilepton and $l$+jets spin correlation measurement might push down to roughly $5\times10^{-3}$, which for the CEDM is approaching the scale of the indirect neutron EDM constraint~\cite{Kamenik:2011dk}.  Clearly, by that point it also makes sense to further subdivide the production phase space, which will allow very detailed maps of the correlations if systematic errors can be controlled.   In particular, effects such as those from our example resonance might be cleanly localized with high precision as in Fig.~\ref{fig:resonance2}.  With high statistics, any effects of interest in specific production regions can also be further enhanced by applying cuts to the top decays, for example selecting only leptonic tops that decay with $\theta \simeq \pi/2$.  The $l$+jets channel is particularly well-suited for such measurements, due to its amenability to complete kinematic reconstruction.

Given the large number of highly-boosted top quarks produced at the very relativistic energies of LHC8 and LHC13, it will also be useful to employ jet substructure techniques and alternative lepton isolation strategies for extracting the decay kinematics (see, e.g.,~\cite{Abdesselam:2010pt,Altheimer:2012mn} and references therein).

\section{Conclusions}
\label{sec:conclusions}

A systematic treatment of top quark production and decay reveals how the complete set of spin correlations imprint themselves on $t\bar t$ events, including all interference effects, and provides us with novel ways to search for new physics.  In particular, we have emphasized that sums and differences of the azimuthal decay angles of the top and antitop about their production axis encode a significant portion of the full 3$\times$3 spin correlation matrix, and that modulations in these variables exhibit a nontrivial evolution as we scan over top production phase space.

Within QCD, there is a common lore that the $q\bar q \to t\bar t$ subprocess, dominant at the Tevatron, produces top quarks whose spins are fully correlated when measured along a special off-diagonal axis that interpolates between the beam directions for slow tops and the production axis for fast tops~\cite{Mahlon:1995zn}.  We have seen that this picture does not generally capture the entire spin correlation.  In the limit of centrally-produced tops with appreciable velocities, an even larger correlation emerges due to interference between the different spin channels, leading to a modulation in the sum of the tops' azimuthal decay angles about the off-diagonal axis.  For $gg \to t\bar t$, the picture is more complicated.  Threshold production is pure $s$-wave, and can therefore be broken down into a longitudinal {\it anti}-correlation along any axis, married to an azimuthal-difference modulation around that axis.  For fast, central tops, the correlation again begins to look like that of $q\bar q \to t\bar t$.  For intermediate regions of phase space, a nontrivial crossover occurs, and in much of this crossover region the correlation is completely dominated by the azimuthal-sum modulation about the production axis.

Measurement of azimuthal correlations about the production axis, including their evolution across top production phase space, is straightforward.  Unlike polar decay angles, azimuthal sums and differences are not highly sensitive to detailed detector acceptances.  We have demonstrated these points with a set of simulation measurements set at the 8~TeV LHC, for both dileptonic and $l$+jets channels.  Dividing the phase space into ``low-$p_T$'' and ``high-$p_T$'' regions for illustration, and accounting only for statistical errors, we predict that the modulations in the two azimuthal angle combinations should be observable with significances above 10$\sigma$ with the current 2012 data set.

The presence of new physics will generally lead to modifications of the azimuthal decay angle correlations.  In \cite{Baumgart:2011wk}, we showed how these correlations can elucidate the coupling structure of heavy, relatively narrow resonances in the $t\bar t$ mass spectrum.  Here, we have considered two additional examples which have modest $S/B$ and whose effects are not necessarily well-localized in production phase space, but which lead to significant distortions of azimuthal correlations.  The first is the well-studied possibility of contributions from dimension-five color-dipole operators.  These induce azimuthal-difference modulations, with a phase governed by the relative strengths of the CMDM and CEDM.  We have argued that the azimuthal modulations capture the vast majority of the dipoles' effects on the spin correlations at the LHC and proposed measurements that will ultimately facilitate bounds that are an order of magnitude more sensitive than the current direct limits~\cite{Kamenik:2011dk}.  Our second example is a broad spin-one color-octet resonance that couples vectorially to light quarks and axially to top quarks.  Interference with QCD leads to a pronounced asymmetric modulation in the azimuthal-sum variable, representing a form of parity-violation that has not been considered before.  We studied a specific model point which would remain well-hidden from $t\bar t$ and dijet resonance searches, and whose dominant contribution is this parity-violating effect.

While we have restricted most of our discussion of measurement prospects to the LHC, the Tevatron might also be a promising venue in which to search for anomalous correlations.  Though the QCD correlation has only just been measured there, new physics can induce strong effects in unexpected places.  In particular, processes that dominantly affect azimuthal correlations in $q\bar q\to t\bar t$, such as our parity-violating resonance model, may show up more strongly at the Tevatron than at the LHC.

We have also mainly focused on correlations amongst azimuthal angles.  In conjunction with polar angle correlation measurements, this provides us with five of the nine entries of the full correlation matrix.  The other four can be obtained through dedicated polar-azimuthal cross-correlation measurements.  We have seen that these effects are small in QCD, though potentially measurable.  They also tend to be small in many new physics models.  However, we have already encountered an important counterexample, in the high-velocity limit in the presence of color-dipole interactions.  We have not undertaken a dedicated estimate of how visible these effects might be, as they are likely subject to large corrections beyond the leading order in the dipole strengths, but they would be interesting to explore in this context and in more general models.

Detailed tests of the Standard Model such as the ones that we are proposing demand precise predictions.  Our results here are entirely restricted to leading-order, but one may ask whether our statements are stable to NLO corrections.  We have made some modest efforts toward this end by cross-checking our results between matched and unmatched samples and finding good agreement.  Recently, a full NLO analysis of the azimuthal-sum correlation has been performed for QCD augmented by a neutral spin-one resonance~\cite{Caola:2012rs}.  Reshapings of the normalized distributions at the level of roughly 10\% were observed, though the residual scale variations tended to be much smaller.  For a resonance that exhibits the same correlation as QCD, and contributes $S/B \simeq 1/2$ within a specified mass window, the scale variations were smaller than the Monte Carlo resolution.  These results are encouraging indications that the QCD predictions are under good control.

As we enter the era of truly precision top physics, it will be important to have a clear understanding of what information is available to measure.  Top quark spin correlations are a rich phenomenon that merit more detailed examination beyond the standard approaches.  This paper has been aimed toward comprehensively understanding the patterns of azimuthal decay correlations exhibited by QCD and by a handful of new physics scenarios.  We hope that our observations will promote these correlations to a more prominent place in the Tevatron's and LHC's top quark physics programs.


\acknowledgments{MB was supported by NSF-PHY-0910467 and DE-FG-03-91ER40682.  MB wishes to thank the Galileo Galilei Institute for the their hospitality while a portion of this work was completed.
  BT was supported by DoE grant No.\ DE-FG-02-91ER40676 and by NSF-PHY-0969510 (LHC Theory Initiative).}


\appendix

\section{Correlation Matrices in Leading Order QCD}
\label{sec:formulas}

The matrix $C$ is defined in Eq.~\ref{eq:proddecomp}, and corresponds to the expectation value of the product of top and antitop spin operators, $\langle 4 S^i \bar S^{\ibar} \rangle$.  In this appendix, we summarize the structure of these matrices in leading-order QCD as a function of the top production velocity $\beta$ and production angle $\Theta$ in the $t\bar t$ center-of-mass frame.  Various combinations of these matrix elements are plotted above in Section~\ref{sec:QCD}.

Our new physics models induce much more complicated correlations, and we have omitted the formulas for brevity.  However, we plot many of the relevant contributions in Section~\ref{sec:NP}.

\subsection{$q\bar q \to t\bar t$}

\be
{\rm off}\mbox{-}{\rm diagonal \; basis:} \;\;\;
C & \,=\, & \left( \begin{matrix}   
\frac{\betasq\sinsqTh}{2-\betasq\sinsqTh} & 0 & 0 \\
0 & \frac{-\betasq\sinsqTh}{2-\betasq\sinsqTh} & 0 \\
0 & 0 & 1
\end{matrix} \right) \nonumber \\
{\rm helicity \; basis:} \;\;\;
C & \,=\, & \left( \begin{matrix}   
\frac{(2-\betasq)\sinsqTh}{2-\betasq\sinsqTh} & 0 & \frac{-\sin2\Theta}{\gamma(2-\betasq\sinsqTh)} \\
0 & \frac{-\betasq\sinsqTh}{2-\betasq\sinsqTh} & 0 \\
\frac{-\sin2\Theta}{\gamma(2-\betasq\sinsqTh)} & 0 & \frac{2\cossqTh + \betasq\sinsqTh}{2-\betasq\sinsqTh}
\end{matrix} \right)
\ee

\subsection{$gg \to t\bar t$}

helicity basis:
\be
C & \,=\, & \left( \begin{matrix}   
\frac{-1 + \betasq(2-\betasq)(1+\sin^4\Theta)}{1 + 2\betasq\sinsqTh - \beta^4(1+\sin^4\Theta)} & 0 & \frac{-\betasq\sin2\Theta \sinsqTh}{\gamma(1 + 2\betasq\sinsqTh - \beta^4(1+\sin^4\Theta))} \\
0 & \frac{-1 + 2\betasq - \beta^4(1+\sin^4\Theta)}{1 + 2\betasq\sinsqTh - \beta^4(1+\sin^4\Theta)} & 0 \\
\frac{-\betasq\sin2\Theta \sinsqTh}{1 + 2\betasq\sinsqTh - \beta^4(1+\sin^4\Theta)} & 0 & \frac{-1 + \betasq\left(\betasq(1+\sin^4\Theta) + (\sin^2 2\Theta)/2\right)}{1 + 2\betasq\sinsqTh - \beta^4(1+\sin^4\Theta)}
\end{matrix} \right)
\ee

\section{Simulation Details}
\label{sec:details}

We generate the LHC8 $t\bar t$ signal and its backgrounds at leading-order using \MadGraph5\ {\tt v1.3.30}~\cite{Alwall:2011uj} interfaced with \PYTHIA~\cite{Sjostrand:2006za}.  The signal samples consist of ``uncorrelated'' and ``correlated'' portions, with top decays respectively handled by \PYTHIA\ or by \MadGraph5.  Our baseline samples use simple two-body production matrix elements.  To these we apply K-factors of $O(2)$ to achieve NLO normalization.  We supplement them with samples matched up to two additional jet emissions using $k_T$-MLM ($k_T = 30$~GeV).  Our backgrounds for the dilepton channel include simulations of $l^+l^-$+2$j$ (including $\tau$s), $W^+W^-$+2$j$, and $tW$+$j$.  Our backgrounds for the $l$+jets channel include simulations of $W$+4$j$, $tW$+1$j$, $t$+3$j$, and $t\bar t$ with $W\to\tau\nu$ decays.  All backgrounds (excepting $t\bar t$ with $\tau$s) are matched using traditional MLM with $p_T = 20$~GeV and $R = 0.4$, with the five-flavor scheme that includes $b$-quark emissions.  We have not studied in detail the azimuthal modulations of lower multiplicity samples, but we have verified, for example, that the exclusive matched $W$+3$j$ contribution is small compared to inclusive matched $W$+4$j$ after running our jet-level analysis.

We also study several new physics scenarios, including color-dipole moments and a resonance in the $t\bar t$ mass spectrum.  For these, we do not generate new Monte Carlo samples, but instead reweight our $t\bar t$ events according to the appropriate ratio of differential 6-body cross sections using the parton-level event record.  

After showering and hadronization in \PYTHIA, we process the output of the physics simulations into reconstructed leptons, jets, and \met.  We demand that leptons be isolated from surrounding activity within an $\eta$-$\phi$ cone of radius $R = 0.4$, such that the scalar-summed $p_T$ of the cone particles cannot exceed 10\% of the lepton's $p_T$.  Leptons should also have $p_T > 30$~GeV and $|\eta| < 2.5$.  (Leptons that fail these criteria are treated as ``hadrons'' and clustered into the jets.)   The remaining particles in the event we cluster into jets in \FastJet\ {\tt v2.4.2}~\cite{Cacciari:2005hq} using the anti-$k_T$ algorithm~\cite{Cacciari:2008gp} with $R = 0.4$.  We keep jets with $p_T > 30$~GeV and $|\eta| < 2.5$.  We determine whether a jet carries flavor by looking back through the \PYTHIA\ event record for the hardest bottom or charm hadron within the jet radius, not counting charm generated in bottom decay.  Each jet then gets a $b$-tag with probability 70\% for true $b$-jets, 10\% for charm-jets, and 2\% for unflavored jets.  To roughly model detector energy resolution, we smear the energies of reconstructed leptons and jets before the application of acceptance cuts:  $\sigma(E)/E = 0.02$ for electrons, $(0.1)\sqrt{E/{\rm TeV}}$ for muons, and $(0.8)\sqrt{{\rm GeV}/E} \oplus 0.04$ for jets.  We further smear the directions of the jets following~\cite{CMSpf}, by $0.025$ separately in $\eta$ and $\phi$.  (This jet energy/direction smearing adds to effects already introduced by parton showering and jet reconstruction.)  We define \vecmet\ to balance the vector-summed transverse momentum of all reconstructed leptons and jets, which gives resolutions similar to those in~\cite{CMSpf}.


\bibliography{lit}
\bibliographystyle{apsper}

\end{document}